\documentclass[11pt,a4paper]{article}

\usepackage{amsmath}
\usepackage{amsfonts}
\usepackage{amssymb}

\usepackage{amssymb}
\usepackage{amsmath}
\usepackage{graphicx}
\usepackage[a4paper,top=2cm,bottom=2cm,left=2cm,right=2cm]{geometry}
\usepackage[margin=10pt, font=small, labelfont=bf, format=hang, font=up ]{caption}

\usepackage{url}
\usepackage[bookmarks, pagebackref=false]{hyperref}
\usepackage{color}
	\definecolor{rossoCP3}{cmyk}{0,.88,.77,.40}
		\definecolor{graa}{rgb}{0.8,0.8,0.8}
		\definecolor{blaa}{rgb}{0.2,0.2,0.6}
		\definecolor{gron}{RGB}{0,150,0}
		\hypersetup{
			colorlinks, 
			bookmarksopen, 
			bookmarksnumbered,
			citecolor=blaa, 		
			linkcolor=black, 
			urlcolor=blaa,			
			}

\newcommand{\ea}[1]{
\begin{align}
#1
\end{align}
}

\newcommand{\ie}{{\it i.e.\ }}
\newcommand{\eg}{{\it e.g.\ }}
\newcommand{\ap}{{\alpha'}}

\def\be{\begin{equation}}
\def\ee{\end{equation}}
\def\ba{\begin{eqnarray}}
\def\ea{\end{eqnarray}}
\def\bea{\begin{eqnarray}}
\def\eea{\end{eqnarray}}

\newmuskip\pFqmuskip

\newcommand*\pFq[6][8]{%
  \begingroup 
  \pFqmuskip=#1mu\relax
  \mathchardef\normalcomma=\mathcode`,
  \mathcode`\,=\string"8000
  \begingroup\lccode`\~=`\,
  \lowercase{\endgroup\let~}\pFqcomma
  {}_{#2}F_{#3}{\left(\genfrac..{0pt}{}{#4}{#5}\Big | #6\right)}%
  \endgroup
}
\newcommand{\pFqcomma}{{\normalcomma}\mskip\pFqmuskip}

\numberwithin{equation}{section}

\begin{document}

\large
 \begin{center}
 {\Large \bf  String Memories ... Lost and Regained}

 \end{center}

 \vspace{0.1cm}



  \begin{center}
    
{\large Andrea Addazi$^{1,2}$, 
 Massimo Bianchi$^{3}$, Maurizio Firrotta$^{3}$, Antonino Marcian\`o $^{4,5}$.}
\\
\vspace{0.5cm}
{\it $^1$ Center for Theoretical Physics, College of Physics Science and Technology \\ Sichuan University, 610065 Chengdu, China}
\\
{\it $^{2}$ INFN sezione di Roma {\it Tor Vergata}\\
Via della Ricerca Scientifica 1, 00133 Roma, Italy}
\\
{\it $^{3}$ Dipartimento di Fisica, Universit\`a di Roma {\it Tor Vergata}\\ 
Via della Ricerca Scientifica 1, 00133, Roma, Italy}
\\
{\it $^{4}$  Department of Physics \& Center for Field Theory and Particle Physics\\ 
Fudan University, Shanghai 200433, China}
\\
{\it $^{5}$ Laboratori Nazionali di Frascati INFN \\ Via Enrico Fermi 54, 
Frascati (Roma), Italy}
\end{center}

\vspace{1cm}
\begin{abstract}
\large
\noindent
We discuss stringy $\ap$ corrections to the gravitational wave signal generated in the merging of two black holes. We model the merging with two BPS compact massive objects in the heterotic string, described by standard vertex operators or coherent states. Despite the expected cubic suppression in $\ap$ w.r.t. the General Relativity result, at tree level the string corrections seem to leave a footprint or a memory on the gravitational wave signals within the sensitivity region of aLIGO/VIRGO and future interferometers. Including loop effects that broaden and destabilise the string resonances suggests a sort of lost stringy memory effect that can be regained through the analysis of the quasi normal modes in the ring-down phase.  
\end{abstract}

\baselineskip = 20pt

\vspace{0.4cm}

\section{Introduction}
\label{IntroSect}

Direct detection of gravitational waves (GWs) \cite{GW1,GW2,GW3,GW3,GW4,GW5} has opened a new chapter in the (theoretical) exploration of Black Hole physics \cite{Barack:2018yly, Cardoso:2019rvt}. 
Many modified gravity models --- including higher-derivative gravity and massive gravity --- as well as models that violate fundamental symmetries, such as Lorentz and CPT invariance, have been highly constrained by the GW data hitherto collected \cite{Abbott:2018lct,L1,L2}. Thus, GW physics is running at the frontier of high precision tests of the gravitational sector of fundamental interactions. In this context, a natural question that one can ask is: {\it ``Can we probe quantum gravity with gravitational waves?''}

UV issues in quantum gravity have been debated for more than half a century, now. GWs may provide precious information on the quantum nature of the gravitational field in the IR region that can help discriminating among competing theories\footnote{Another direction of the quantum gravity phenomenology that may be coded in GWs is the possible hints for the existence of {\it Exotic Compact Objects} (ECOs) \cite{Maselli:2017cmm} such as {\it Fuzzballs} \cite{Mathur:2005zp} that may be exposed by searching for departures from the zero Love number bound, set by fundamental uncertainty principle limits on the BH radius \cite{Addazi:2018uhd}, or from corrections to the ring-down modes 
\cite{Brustein:2017koc, Bianchi:2020des}.}.  
An exciting perspective is to search for possible  hints of String Theory (ST) inside the GW data. 
String corrections typically produce higher-derivative terms in the low-energy effective Lagrangians that are suppressed by the Regge slope $\ap$ and by the string coupling $g_s$. 
Interesting constraints on the String Tension, the size of the internal Compactification Volume and limits on Large or Warped Extra Dimensions may be achieved in the near future.
Moreover, string theory predicts the presence of stringy resonances (Regge recurrences) as well as Kaluza-Klein (KK) excitations, which can both leave imprints in the GW signal. Last but not least, the BH information may be stored in the form of quantum, possibly soft, hairs, as originally suggested in \cite{QH1,QH2,QH3,QH4, Hawking:2016msc}. This may result into a multi-polar structure much richer than the one appearing in General Relativity\footnote{For recent work on fuzzball multi-poles see e.g. Refs.~\cite{Bena:2020see, Bianchi:2020bxa, Bena:2020uup, Bianchi:2020Comp}.} (GR) \cite{Geroch:1970cd, Hansen:1974zz, Berti:2009kk}. The correspondence between strings, branes, their bound-states and `large' BHs has a long story that we cannot review here, see e.g. Refs.~\cite{Horowitz:1996nw, Maldacena:1996ds, Damour:1999aw, Chialva:2009pf, Brustein:2016msz} for early and recent work with different perspectives. 

In the attempt to expose $\ap$ effects in GWs beyond GR, we use a toy model for BH merging within the framework of high-energy string scattering. Despite the complexity of the problem, we identify a simple case that may offer some insights in the dynamics of BH mergings in ST.  We will consider the tree-level merging of two very massive states, which we assume to retain a compactness comparable to BH's or neutron stars. We will further assume that the incoming states have the same quantum numbers as extremal 2-charge BHs and that the scattering produces a non{-}BPS charged BH state plus (soft) gravitons. The computation of the (tree-level) amplitude for this scattering process turns out to be remarkably simple in the heterotic string framework, where BPS micro-states just correspond to vertex operators or coherent states thereof\footnote{Picturing BHs as coherent states seems also to be supported by recent studies of BH formation/evaporation in highly inelastic, ultra-planckian string collisions \cite{Dvali:2014ila, Addazi:2016ksu}.}. Closed-string coherent states can be powerfully treated within the DDF approach \cite{DelGiudice:1971yjh, Ademollo:1974kz, Hindmarsh:2010if, Skliros:2011si, Skliros:2016fqs}, using similar techniques as in the case of open (super)strings \cite{Bianchi:2019ywd, Aldi:2019osr}. This provides a simplified framework for our tree-level calculations, which enables us to obtain an exact result for the relevant scattering amplitudes. Our main purpose is to explore $\ap$ deviations from the GR (or Supergravity) results.  After showing the complete agreement of our final expressions with GR at the leading order, we confirm that the first stringy corrections arise at cubic order in $\ap$ and as such would give too rapid a decay with the distance. Yet, the collective effect of all string resonances leads to a sizeable correction to the GW signal\footnote{Our result may be compared with the recent analysis in \cite{DiVecchia:2019myk, DiVecchia:2019kta, Parra-Martinez:2020dzs}, which is obtained from the extremal BH scattering in the $\mathcal{N}=8$ supergravity perspective, and in which $(\ap)^{3}$ effects should be related to the $G_N^3$ effects.}.



Heuristic estimates show that forthcoming gravitational interferometers are actually close to access a non-negligible part of the parameter space of $\ap$ corrections. Thus the logic behind our calculation is the following. As a first step, we will show that the process of two BPS BH-like `compact' states merging into a non{-}BPS BH-like `compact' state, with concomitant release of GW, can be computed in the context of heterotic string theory\footnote{Although the in-coming BHs can be taken to be both BPS, the kinematical conditions would trivialize the scattering amplitude in the soft limit, if the third BH were BPS too.}. Then, by `stripping off' the graviton polarization, we will use the `truncated' scattering amplitude as a source for the gravitational wave equation. The GW profile is obtained from the convolution of the string source with the graviton propagator. This allows us to study how $\ap$ corrections from the stringy-BH sources alter GW emission. 

Not surprisingly, our results confirm gravitational memory, based on the celebrated Weinberg's soft graviton theorem \cite{Weinberg}, recently extended\footnote{These results are in agreement with the ones obtained from the eikonal scattering perspective  \cite{Ciafaloni:2018uwe,Addazi:2019mjh,DiVecchia:2019kta}.} in \cite{He:2014laa, Strominger:2014pwa, Cachazo:2014fwa, Bern:2014vva, He:2014bga, Bianchi:2014gla, Bianchi:2016viy, DiVecchia:2015oba, DiVecchia:2015srk, Bianchi:2015yta, Guerrieri:2015eea, Bianchi:2015lnw, Bianchi:2016tju} and \cite{Sen:2017nim, Laddha:2017ygw, Laddha:2018vbn, Sahoo:2018lxl, Laddha:2019yaj,Saha:2019tub}, and suggest an interpretation of the $\ap$ effects as a sort of {\it string memory}. However taking into account their finite width beyond tree level, the GW profiles from string resonances are exponentially decaying in time and the effect seems to be {\it lost}. Nevertheless, the {\it lost} memory may be {\it regained}, if one looks at the quasi-normal modes (QNM's) \cite{Berti:2009kk, Bianchi:2020des} that govern the ring-down phase of the produced stringy BH. This effect may be enhanced for some string states that can be very long-lived \cite{Chialva:2004xm}. This suggests that some string information may have milder decay in time, opening a (small) window for searches of the string (lost and regained) memory effect. 

In order to further motivate our analysis, let us estimate the order-of-magnitude of the putative string corrections and argue that some portion of the parameter space of the high-energy phenomena can indeed be accessed and thus constrained by forthcoming gravitational interferometers\footnote{We will work in the perturbative regime 
whereby $g_s=e^{\langle \phi_{\rm dil} \rangle}$ is small, although the expectation value of the dilaton field can only be fixed by including fluxes and (non)perturbative effects.}. At tree-level, scattering amplitudes acquire stringy corrections with respect to standard quantum field theory results in the form of powers of $\nu\equiv \ap k \cdot p_{a}$, where $k$ is the graviton four-momentum and $p_{a}$ the BH four-momenta.  
Considering BHs of mass around $M_{BH}\simeq 20\, M_\odot\approx 20 {\cdot}\, 10^{30}\, {\rm Kg}$ and a LIGO/VIRGO frequency of $10\div 100\, {\rm Hz}$, barring very special kinematics, one can reach values of $\nu \simeq O(1)$ even assuming a relatively high string tension (around the Grand Unification Scale or beyond). Such a surprising amplifier is provided by the BH mass entering the expression for $\nu$, provided the massive (non)BPS states involved retain the required compactness. 

The rest of the paper is organized as follows. 
In Section \ref{GWprodBHmerg}, we discuss our toy model for (BPS) BH mergers and GW production in heterotic string and compute the relevant amplitude at tree level. We briefly discuss loop corrections, exponentiation (eikonal approximation) and compare our analysis with the one in \cite{Chialva:2005gt} for BH production in cosmic superstring collisions. 
In Section \ref{StringMemoSect} we analyze the role of the Regge recurrences that are stable at tree-level and produce a modulation of the GW signal w.r.t. GR. We will treat the corrections in three different schemes. First as higher-derivative corrections that individually seem to produce no effect at large distances. Second as a sum over (tree-level stable) resonances that has a sizeable effect on the GW profile in real space, coded in a power series in $u/\ell_a$, with $u=t-R$ the retarded time and $\ell_a = \ap k p_a/\omega$ for $a,b=1,2,3$. Third in the high-energy limit $\ap k p_a >> 1$, both at fixed angle $\ap k p_a\approx \ap k p_b$ for all $a,b$, in Section \ref{LargeOmegaEll}, and small angle (Regge limit) $\ap k p_1<< \ap k p_2\approx \ap k p_3$, in Section \ref{ReggePlunge}, relevant for Extreme Mass Ratio Inspirals (EMRI's).   
Then in Section \ref{LostringMemoSect} we discuss how string memories look irremediably lost once quantum effects producing a finite width are included and how they can be (partially) regained either by a collective effect or by their imprints on the QNM's governing the ring-down phase of the produced non{-}BPS BH. 
Section \ref{ConcSect} contains our conclusions and final comments.
For the interested reader, four Appendices contain useful formulae and conventions on the kinematics of two-body decays (App.~\ref{KinematicsApp}) and on the vanishing of the amplitude for the 3 BPS process (App.~\ref{no3BPSApp}), on Generalized Hypergeometric Functions and Meijer G-functions (App.~\ref{GenHypFunApp}), and on the resummation for `rational' kinematics (App.~\ref{RationalKinApp}).

\section{Gravitational Wave Production from BPS-BHs merging}
\label{GWprodBHmerg}
\noindent
A remarkably simple and interesting example of GW production in a `macroscopic' high-energy inelastic string process is based on the heterotic string\footnote{On a general six dimensional compact manifold with some non-trivial 1-cycle.} scattering involving very massive, spin-less and compact BPS states such as
$${\bf BPS}\, {\bf BH}_{1}+{\bf BPS}\, {\bf BH}_{2}\rightarrow {\bf (non{-}BPS)\,BH}+{\bf GW} \,.$$
\begin{figure}[h!]
\center
\includegraphics[scale=0.5]{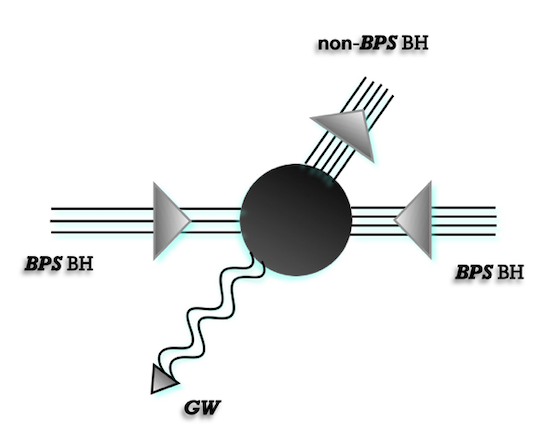}
\caption{Schematic representation of the selected scattering process}
\end{figure}
With some abuse of language we call BHs the large mass (non)BPS states with windings and momenta such that 
\be 
\begin{split}
&M_{BPS}^2 = |{\bf P}_L|^2 = |{\bf P}_R|^2 + {4\over\ap}(N_R-1)\,, \\
&M_{non{-}BPS}^2 = |{\bf P}_L|^2 + {4\over\ap}N'_L = |{\bf P}_R|^2 + {4\over\ap}(N_R-1) \,,
\end{split}
\ee
where ${\bf P}_L$ and ${\bf P}_R$ denote the internal L/R momenta and $N_L$ and $N_R$ the (mass) levels. States of this kind are also known as `small' BH's (2-charge `fuzz balls' \cite{Mathur:2005zp}) in that they have no (or `zero-area') horizon in the supergravity approximation but have a non vanishing entropy associated to the exponential growth of the degeneracy (\`a la Hardy-Ramanujan-Hagedorn), i.e.
\be
d({\bf P}_L, {\bf P}_R) = e^{S_{BH}/\kappa_B} \approx e^{2\pi \sqrt{{\ap\over 4} (|{\bf P}_L|^2-|{\bf P}_R|^2)}}= e^{2\pi \sqrt{N_R}} \,.
\ee
These are very peculiar BHs, which can carry arbitrarily high spin \cite{Bianchi:2010dy, Bianchi:2010es}. More `realistic' models for `large' BH's in string theory have been proposed over the years \cite{Horowitz:1996nw, Maldacena:1996ds, Damour:1999aw, Chialva:2009pf, Brustein:2016msz, Mathur:2005zp} that do not allow a simple analysis of the merging as the one we perform here.  

We consider one of the three BHs to be non{-}BPS since otherwise the kinematic conditions would trivialise the scattering amplitude in the soft limit,
as shown\footnote{In retrospect this might be interesting from an `academic' perspective, since the GW signal would be genuinely stringy in nature, being absent in the GR limit.} in appendix \ref{no3BPSApp}. $N'_L$ measure the level of L-excitation over the BPS ground-state with given ${\bf P}_L$. Last but not least, we assume the three massive states to be spin-less and compact, i.e. $J_a=0$ and $R_a \approx G_N M_a $. As we will see in Section \ref{CoherentSect}, compact-ness can be achieved for coherent states, though it might be harder to achieve for mass eigenstates such as the ones we consider in this Section, for which $R_a \approx \ap M_a > G_N M_a $, in the perturbative regime whereby $g_s <\!\!<1$. 

\subsection{Scattering Amplitude}

The tree-level (sphere) scattering amplitude we consider is
\begin{equation}
{\cal M}_{3+1}(k,h; p_a, \zeta_a)= g_s^4 C_{S_2} \int_{S_2} {\prod_{i=0}^{3} d^2z_i \over {\cal V}_{CKG}} \langle {\cal W}_{\cal G}(z_0,\bar{z}_0) {\cal W}_{\cal BH}(z_1,\bar{z}_1)  {\cal W}_{\cal BH}(z_2,\bar{z}_2)  {\cal W}_{\cal BH}(z_3,\bar{z}_3)\rangle  \, , 
\end{equation}
where $C_{S_2} = {8 \pi / g_s^2 \alpha' }$ is the normalisation constant of the sphere $S_2$,  ${\mathcal{V}}_{CKG}$ is the volume of its conformal Killing group (CKG) ${\mathbb{PSL}(2,\mathbb{C})}$,  $k$ and $h$ are respectively the momentum and polarization of the graviton, $p_a$ and $\zeta_a$, with $a=1,2,3$, are respectively the momentum and the polarization of each one of the three BHs. 

Setting $\alpha'=2$ as usual and using the first non-canonical $(0)$ superghost picture, the graviton vertex operator (see e.g. \cite{Friedan:1985ey}) reads
\begin{equation} \label{wb}
{\cal W}^{(0)}_{{\cal G}}(z,\bar{z}) = h_{\mu\nu} \Big(\imath \partial X_L^\mu + k_{L}{\cdot}\psi_L \psi_L^\mu\Big)e^{\imath k_{L}{\cdot}X_L}(z)\, \imath \bar\partial X_R^\nu e^{\imath k_{R}{\cdot}X_R}(\bar{z})\,,
\end{equation}
where $X_{L(R)}$ denotes the L-(R-) moving bosonic coordinates, $\psi_L$ the L-moving fermionic coordinates, $k^\mu_L=k^\mu_R= (\omega, \vec{k})$ is the null momentum of the graviton, with no internal components, i.e. ${\bf P}_L={\bf P}_R=0$. The split of the polarisation as $h=h_L\otimes h_R$ is often adopted.

In the same $(0)$ superghost picture and with a splitting of the polarisation $\zeta=\zeta_L\otimes \zeta_R^{N_R}$, the massive (BPS) BH vertex operator reads 
\begin{equation}
{\cal W}^{(0)}_{{\cal BH}}(z_2,\bar{z}_2)=\Big(\zeta_{2,L}{\cdot}\imath \partial X_L + K_{2,L}{\cdot}\psi_L \zeta_{2,L}{\cdot}\psi_L\Big)e^{\imath K_{2,L}{\cdot}X_L}(z_2)\,F_{N_2}^{\zeta_{2,R}} [\bar\partial X_R]e^{\imath K_{2,R}{\cdot}X_R}(\bar{z}_2)\,,
\end{equation}
where $K_{2L}=(p_2, {{\bf{P}}}_{2L})$, $K_{2R}=(p_2, {{\bf{P}}}_{2R})$ and $F_{N_2}^{\zeta_2,R} [\bar\partial X_R]$ is a polynomial in $\bar\partial^{n_k} X_R$, with conformal dimension $\bar\Delta_2=\sum_k n_k=N_2$.

The remaining two BH vertex operators can be expressed in the canonical $(-1)$ picture. The first (BPS) can be written as
\begin{equation}
{\cal W}^{(-1)}_{{\cal BH}}(z_1,\bar{z}_1)=e^{-\phi}\zeta_{1,L}{\cdot} \psi_{L} \,e^{\imath K_{1,L}{\cdot}X_L}(z_1)\,F_{N_1}^{\zeta_{1,R}} [\bar\partial X_R]\,e^{\imath K_{1,R}{\cdot}X_R}(\bar{z}_1)\,,
\end{equation}
where $K_{1L}=(p_1, {{\bf{P}}}_{1L})$, $K_{1R}=(p_1, {{\bf{P}}}_{1R})$ and, again, $F_{N_1}^{\zeta_{1R}} [\bar\partial X]$ is a polynomial  of conformal dimension $\bar\Delta_1=\sum_k n_k=N_1$.

For simplicity let us take the final non{-}BPS one to be  `excited'  only with `internal' bosonic oscillators {\it viz.}, namely 
\begin{equation}
{\cal W}^{(-1)}_{{\cal BH}}(z_3,\bar{z}_3)=e^{-\phi}\zeta_{3,L}{\cdot} \psi_{L} \, H^L_{N_3}[\partial^l X_R]\,e^{\imath K_{3,L}{\cdot}X_L}(z_3)\,\bar{H}^R_{\bar{N}_3}[\bar\partial^r X_R]\,e^{\imath K_{3,R}{\cdot}X_R}(\bar{z}_3)\,,
\end{equation}
where $K_{3L}=(p_3, {{\bf{P}}}_{3L})$, $K_{3R}=(p_3, {{\bf{P}}}_{3R})$, $N_{3L} = \sum_{k=1}^{k_{Max}} k m_k$ and, for later use, $n_3= \sum_{k=1}^{k_{Max}} k$ (for states in the first Regge trajectory $N_3=n_3$). 

BRST invariance requires the mass-shell conditions spelled out above 
as well as `transversality' and `traceless-ness' conditions $\zeta_{3L}{{\bf{P}}}_{3L}=0 = \zeta_{3L}H^L_{N_3} = {{\bf{P}}}_{3L} H^L_{N_3} = {\rm Tr}(H^L_{N_3})$, which we assume to be satisfied.

The issue of collinearity that afflicts the 3-BPS case, discussed in Appendix \ref{no3BPSApp}, is solved since $\ap K_1K_2= \ap (p_1p_2 +{{\bf{P}}}_{1L} {{\bf{P}}}_{2L}) = 2 N_3\neq 0$. As a result, though the incoming BH's are BPS they are not mutually BPS since ${{\bf{P}}}_{1L}$ and ${{\bf{P}}}_{2L}$ are not collinear.  

Choosing the `polarisations' of all three massive states to be along the `internal' directions, so much so that the resulting BHs have zero spin ($J_a=0$), contractions are easy to take, only a little bit more involved w.r.t. the 3-BPS case (equivalent to 3 massless in $D=10$). In particular one can show that the terms $k_1\!\cdot\!\psi h\!\cdot\!\psi$ , in the graviton vertex, and $K_2\!\cdot\!\psi \zeta_2 \!\cdot\!\psi$, in the (BPS) BH (taken to be in the 0 picture), cannot contribute. The only difference w.r.t. the 3-BPS case is the presence of $H_{3L}$ that can contract with $\exp \imath {{\bf{P}}}_{aL}X_L$, as well as with $\zeta_{2L}\partial X$. The (vanishing) 3-vector boson YM vertex is then replaced by \cite{Bianchi:2010dy, Bianchi:2010es}
\be 
\begin{split}
&{\cal V}^{(int)L}_3(\zeta_{aL}, {{\bf{P}}}_{aL}, H^{N_{3L}}_{3L}) = [\zeta_{1L}{\cdot}\zeta_{3L} \zeta_{2L} + \zeta_{2L}{\cdot}\zeta_{3L} \zeta_{1L}]{\cdot}H^{N_{3L}}_{3L}{\cdot}_{n_3{-}1}({{\bf{P}}}_{1L}{-}{{\bf{P}}}_{2L})^{\otimes (n_3{-}1)} \\
&\quad +[\zeta_{1L}{\cdot}\zeta_{2L} \zeta_{3L}{\cdot}{{\bf{P}}}_{1L} + \zeta_{2L}{\cdot}\zeta_{3L} \zeta_{1L}{\cdot}{{\bf{P}}}_{2L} + \zeta_{3L}{\cdot}\zeta_{1L} \zeta_{2L}{\cdot}{{\bf{P}}}_{3L}] H^{N_{3L}}_{3L}{\cdot}_{n_3}({{\bf{P}}}_{1L}{-}{{\bf{P}}}_{2L})^{\otimes n_3} \,.
\end{split}
\ee

Obviously, at the expenses of more involved computations, one can also replace one or both the BPS BH's with non{-}BPS states with $M_a>|{{\bf{P}}}_{aL}|$, or even consider coherent states, as we do later on along the lines of
\cite{Hindmarsh:2010if, Skliros:2011si, Skliros:2016fqs, Bianchi:2019ywd, Aldi:2019osr}. The crucial point is that in the soft limit $k\rightarrow 0$ a non vanishing 3-point `physical' amplitude be produced together with terms of higher-order in $k$ which will be our main focus.

Assembling the various pieces, the amplitude takes the following form
\begin{equation}
{\cal M}_{3+1}(h,k; p_a, \zeta_a) =4\pi^2 g_s^2 {\cal V}^{(int),L}_{3{-}YM} {\cal V}^{(int),R}_{3{-}HS} \sum_{a=1}^3  {p_a{\cdot}h{\cdot}p_a \over k{\cdot}p_a}  \, \prod_{b=1}^{3}{\Gamma(1+{k{\cdot}p_b}) \over \Gamma(1-{k{\cdot}p_b})}\,.
\end{equation}
With the stringy identification of the 4-d Newton constant\footnote{The identification of the Newton constant comes from standard matching procedure in the heterotic supergravity.}
\begin{equation}
G_N= {g_s^2 \alpha'^4 \over 64\, \pi V_{(6)}}  \,,
\end{equation}
where $V_{(6)}$ is the volume of the compactification manifold 
and the un-normalized three-point amplitude coming from the factorisation
\begin{equation}
g_s^3 C_{S_2} {\cal V}^{(int),L}_{3}({\{\zeta_{a,L}\},\{p_{a,L}\}}) {\cal V}^{(int),R}_{3}({\{\zeta_{a,R}\},\{p_{a,R}\}}) \equiv {\cal M}_{3{-}{\cal BH}}({\{\zeta_a\},\{p_a\}})\,,
\end{equation}
the result can be written as
\begin{equation}\label{ris amp1}
{\cal M}_{3+1}(h,k; p_a, \zeta_a) = 16\pi G_N\,  \widehat{{\cal M}}_{3{-}{\cal BH}} \sum_{a=1}^3  {p_a{\cdot}h{\cdot}p_a \over k{\cdot}p_a} \prod_{b=1}^{3}{\Gamma(1+{k{\cdot}p_b}) \over \Gamma(1-{k{\cdot}p_b})}\,,
\end{equation} 
where $\widehat{{\cal M}}_{3{-}{\cal BH}}=(2\pi V_{(6)}/g_s \alpha'^3) {{\cal M}}_{3{-}{\cal BH}}$ and the adimensional prefactor can be reabsorbed in the (not-directly measured) distance $R$ traveled by the GW from the source (the merging) to the detector.

Finally, using the leading soft factor, introduced by Weinberg \cite{Weinberg}, namely
\begin{equation}
S_0=\sqrt{8\pi G_N}\sum_{a=1}^3  {p_a{\cdot}h{\cdot}p_a \over k{\cdot}p_a}\,,
\end{equation}
we may recast Eq.\eqref{ris amp1} in the more compact and final form
\begin{equation}
{\cal M}_{3+1}(h,k; p_a, \zeta_a) = \sqrt{32\pi G_N}\widehat{{\cal M}}_{3{-}{\cal BH}} \,S_0 \prod_{b=1}^{3}{\Gamma(1+{k{\cdot}p_b}) \over \Gamma(1-{k{\cdot}p_b})}\,.
\end{equation}
The appearance of the soft factor $S_0$ should not be deceiving: the expression  ${\cal M}_{3+1}$ is exact, valid for any value of $\omega$.

In principle, in order to have a reliable picture of the scattering process at least at the perturbative level, one should include loop corrections to the above tree-level amplitude. Although the one-loop (torus) contribution would not be hard to compute relying on 4-point massless amplitudes in $D=10$ \cite{Green:1982sw, Bianchi:2015vsa} and tackling the two-loop amplitude should be harder but doable, addressing higher-loops looks daunting and subtle to some extent \cite{Donagi:2013dua}. 

In fact we would argue that this is not necessary for our purposes. As in GR, the process will take place in three phases: inspiralling, merger and ring-down.  Denoting by $b$ the `impact' parameter 
and by $R_1$ and $R_2$ the `sizes' (gyration radii) of the `compact'  BPS incoming BH's, in the in-spiralling phase the two massive compact objects are well separated $b>\!\!>R_1,R_2$ and exchange mostly massless quanta (gravitons), whose contribution can be resummed in the eikonal approximation\footnote{For recent work see e.g.  \cite{DiVecchia:2019myk, DiVecchia:2019kta, Parra-Martinez:2020dzs}}, leading to a computable phase shift. GW production in this phase has been studied and produces a spectrum in line with GR \cite{Ciafaloni:2018uwe, Addazi:2019mjh}.

During the merger $b<R_1+R_2$, we can use the inelastic tree-level amplitude we computed and we get a correction to the GW signal in GR that we will study momentarily. 

In the ring-down phase the non{-}BPS BH will relax to some (meta)stable configuration. We will briefly address the spectrum of quasi-normal modes (QNM) in the ring-down phase in Section \ref{LostringMemoSect} but we plan to investigate this issue more thoroughly in the future.

Before concluding this Section we would also like to briefly compare our present results with those on BH production in cosmic superstring collisions \cite{Chialva:2005gt} and the ones on pair-production of miniBH   \cite{Bianchi:2010dy, Bianchi:2010es, Bianchi:2006nf}. In \cite{Chialva:2005gt}, the incoming states are far from being compact and the splitting and joining process is suppressed by the probability that two bits of the colliding cosmic superstrings come close enough. Then the process is further suppressed by the probability that the produced string be compact enough to behave as a BH. Most of the analysis is semi-classical and reliable in the regime of \cite{Chialva:2005gt}.  In \cite{Bianchi:2010dy, Bianchi:2010es, Bianchi:2006nf} slightly different processes are considered, whereby very energetic massless initial states come so close that $b<R_S$ and a pair of mini-BH of small size (even $TeV$ scale in principle in models with very low string tension) and opposite charge is produced. Here we have focussed  on a sort of crossed channel whereby the BPS BH's (but non-mutually BPS) are in the initial state.

\section{String Memories}
\label{StringMemoSect}

Barring irrelevant constants, the 4-point amplitude ${\rm BPS_1}+{\rm BPS_2} \rightarrow {\rm non{-}BPS_3} + h_{\mu\nu}$ can be re-written as
\be
{\cal M}_{3+1}(h,k; p_a, \xi_a) = 16\pi G_N\,  \widehat{{\cal M}}_{3{-}{\cal BH}} \sum_{a=1}^3  {p_a{\cdot}h{\cdot}p_a \over k{\cdot}p_a} \prod_{b=1}^{3}{\Gamma(1+{k{\cdot}p_b}) \over \Gamma(1-{k{\cdot}p_b})}\,,
\ee
where $p_a,\xi_a$ collectively characterise momenta and polarisations of the massive (non) BPS states, with 
$M_a^2 = -p_a^2 = |{\bf P}_{L,a}|^2 +N'_L$. Note that, even though $\sum_a p_a = 0= 
\sum_a  {{\bf{P}}}_{L,a} = \sum_a {{\bf{P}}}_{R,a}$, ${\cal M}_{3}(p_a, \xi_a)$ is `physical' and non-zero since the three momenta $p_a$ are not necessarily collinear even in the soft limit $k{\rightarrow}0$, contrary to what happens for mass-less quanta.  Yet, the kinematics is rather scant in the soft limit, since $2p_1p_2 = -M_3^2 + M_1^2 + M_2^2$ and cyclic. The phase space deforms a bit due the emission of the massless graviton. See Appendix \ref{KinematicsApp} for details. 

Moreover, even for finite $k$  
\be
k(p_1+p_2+p_3)= - k^2 = 0\,,
\ee
so much so that, defining the scattering lengths 
\be
\ell_a = np_a = np_a =E_a - \vec{n}\vec{p}_a \,,
\ee
one has 
\be
\ell_1+\ell_2+\ell_3 = 0 \,,
\ee
setting $\ell_a = - \eta_a |\ell_a|$, \ie $\ell_1$ and $\ell_2$ are positive, while $\ell_3$ is negative.

\subsection{Gravitational Memory}

In GR, the GW profile $h_{\mu\nu}$, produced by a (transverse traceless) source ${\cal S}_{\mu\nu}$, obeys the following equation\footnote{This is a linearized and gauge fixed GW equation.}
\begin{equation}
\Box h_{\mu \nu}(t,\vec{x})=-16 \pi G_N {\cal S}_{\mu\nu}(t,\vec{x})\,.
\end{equation}
For the causal retarded wave propagation, the solution takes the form
\begin{equation}
 h_{\mu\nu}(t,\vec{x})=4 G_N \int d^3 x' {{\cal S}_{\mu\nu}(t-|\vec{x}-\vec{x'}|,\vec{x'})\over |\vec{x} - \vec{x'}|}\,.
 \end{equation} 
\label{GravMemo}
For the GW produced in a high-energy collision in GR, and observed at large distances $R=|\vec{x}|>\!\!> |\vec{x}'|$, one `locally' finds the plane-wave like behaviour
\begin{equation}
h_{\mu\nu}(t,\vec{x})=\tilde{e}_{\mu\nu}(\omega,\vec{x}) \,e^{-\imath\omega t} + \tilde{e}^*_{\mu\nu}(\omega,\vec{x})\,e^{\imath \omega t}\,,
\end{equation}
with polarisation tensor dictated by the Weinberg's soft theorem \cite{Weinberg}  
\be 
\tilde{e}^{\mu\nu}(\omega, \vec{x}) = {4 G_N \over \omega R}\sum_a {p_a^{(\mu} p_a^{\nu)|}  \over np_a}\, e^{\imath \omega R} \,,
\ee
where $p_a=+p_a$ for out-going particles ($\eta={+}1$ in Weinberg's notation), while $p_a=-\tilde{p}_a$ for in-coming particles ($\eta={-}1$ in Weinberg's notation), so that $\sum_a p_a = 0 = \sum_{a\in out} p_a - \sum_{a\in in} \tilde{p}_a$.
Integrating over $\omega$ and using 
\be
{1\over {\omega{-}\imath \varepsilon}} - {1\over {\omega{+}\imath \varepsilon}} = 2\pi \imath \delta(\omega)\,,
\ee
up to sub-leading terms \cite{Sen:2017nim,Laddha:2017ygw,Laddha:2018vbn,Sahoo:2018lxl,Laddha:2019yaj,Saha:2019tub}, one finds a constant term at (late) retarded time $u>0$
\be 
e^{\mu\nu}(t, \vec{x}) = -{4 G_N \over R}\left\{
\sum_{a\in {\rm out}} {p_a^{(\mu} p_a^{\nu)|}  \over np_a} - 
\sum_{a\in {\rm in}} {\tilde{p}_{a}^{(\mu} \tilde{p}_a^{\nu)|}  \over n\tilde{p}_a} \right\} \,, 
\ee
which is known as `gravitational memory'. Recall that $n=(1,\vec{n}) = (1,\vec{x}/R)$.

\subsection{String memories}

Including stringy corrections, the GW profile ${e}_{\mu\nu}(t,\vec{x})$ is determined by the solution to the equation
\begin{equation}
 \Box {e}_{\mu\nu}(t,\vec{x})=- { \delta {\cal M}_{3{+}1}(t,\vec{x}) \over \delta h^{\mu\nu} } \equiv -{\cal M}_{3{+}1}^{(\mu\nu)|}(t,\vec{x})\, ,
 \end{equation} 
where $(\mu\nu)|$ denote the symmetric, trace-less component. In the low-energy limit the Shapiro-Virasoro factor in the heterotic string amplitude, namely  
\be
{\cal F}(k=\omega(1,\vec{n}),p_a) = \prod_b{\Gamma(1+{kp_b})\over  
\Gamma(1-{kp_b})} = \prod_b{\Gamma(1+\omega\ell_b)\over  
\Gamma(1-\omega\ell_b)} = {\cal F}(\omega,\ell_a)\,,
\ee
produces corrections to the GR results that can be written as a power series in $\ap k{\cdot}p_a$ that starts at cubic order. Fourier-transforming in $\omega$ would produce corrections decaying faster than $1/R$ at large distances that would be totally negligible. On the other hand, including the contribution of the infinite tower of string resonances turns out to produce a sizeable effect.  Indeed, starting from  
\be
\tilde{e}^{\mu\nu}(\omega, \vec{x})= \int d^3y {e^{\imath \omega|\vec{x}{-}\vec{y}|} \over 4\pi|\vec{x}{-}\vec{y}|} 
    \widetilde{\cal M}^{(\mu\nu)|}_{{3{+}1}}(\omega,\vec{y}; p_a, \xi_a) 
\approx {e^{\imath \omega R} \over 4\pi R} 
{\cal M}^{(\mu\nu)|}_{{3{+}1}}(\omega,\vec{k}=\omega\vec{n}; p_a, \xi_a)\,,
\ee
where the last approximation is valid at large distances $R=|\vec{x}{-}\vec{y}|>\!\!> L$, so that 
\be 
\tilde{e}^{\mu\nu}(\omega, \vec{x})= {4 G_N}  \, \widehat{{\cal M}}_{3}(p_a, \xi_a) 
 {e^{\imath \omega R}\over \omega R} \sum_a {p_a^{(\mu} p_a^{\nu)|}  \over  np_a} {\cal F}(\omega,\vec{n}; p_a) \,,
\ee
or, barring the overall and largely irrelevant non-zero factor $\widehat{{\cal M}}_{3}(p_a, \xi_a){\neq} 0$ and anti-Fourier transforming, one finds
\be 
\label{after}
e^{\mu\nu}(t, \vec{x})=   
{4 G_N\over R} \sum_a {p_a^{(\mu} p_a^{\nu)|} \over \ell_a} \int_{-\infty}^{+\infty} {d\omega \over 2\pi \omega} e^{- \imath \omega u} {\cal F} (\omega,\ell_a) \,,
\ee
where $u=t-R$ is the retarded time. In order to compute the $\omega$ integral it is convenient to expand 
\be
{\cal H} (\omega,\ell_a) = {1\over \omega} {\cal F} (\omega,\ell_a)
\ee
as an infinite sum of poles in $\omega$ \`a la Mittag-Leffler (ML), and obtain
\be
{\cal H} (\omega,\ell_a) = {1\over \omega} + \sum_{a=1}^3 \sum_{n_a=1}^\infty 
{(-)^{n_a} \ell_a \over {n_a!^2 (\omega\ell_a+ n_a) }} \prod_{b\neq a}{\Gamma(1- n_a \lambda_{b,a})\over  
\Gamma(1+n_a \lambda_{b,a})}\,,
\ee
where we introduced the kinematical ratios 
\be
\lambda_{b,a} = {\ell_b\over \ell_a} = {n{\cdot}p_b\over n{\cdot}p_a} = 
{k{\cdot}p_b\over k{\cdot}p_a}\,.
\ee

The pole in $\omega = 0$ reproduces the gravitational memory effect\footnote{Massless dilatons and axions do not contribute at this leading order.}. In addition to this, in (heterotic) string theory one finds genuine string corrections. Inserting the ML expansion in the integral over $\omega$ (\ref{after}),  and adopting some reasonable prescription to deform the integration path, \ie $kp_a \rightarrow kp_a -\imath \epsilon$, one finds intriguing corrections $\Delta_s e^{\mu\nu}(t, \vec{x})$ to the usual GR profile, suggesting some sort of string memory effect with a non-trivial (retarded) time dependence.
Before integration the correction reads
\begin{equation}
\Delta_s e^{\mu\nu}(t, \vec{x})= {4 G_N \over R}\sum_{b=1}^3 {p_b^{(\mu} p_b^{\nu)|}\over \ell_b} \sum_{a=1}^3 \sum_{n_a=1}^\infty 
{(-)^{n_a}  \over {n_a!^2}} \prod_{b\neq a}{\Gamma(1- n_a \lambda_{b,a})\over  
\Gamma(1+n_a \lambda_{b,a})} \int_{-\infty}^{\infty} {d\omega \over 2\pi} { \ell_a e^{-\imath \omega u} \over \omega \ell_a {+} n_a{+}\imath \epsilon}\,.
\end{equation}
\begin{figure}[h!]
\center
\includegraphics[scale=0.45]{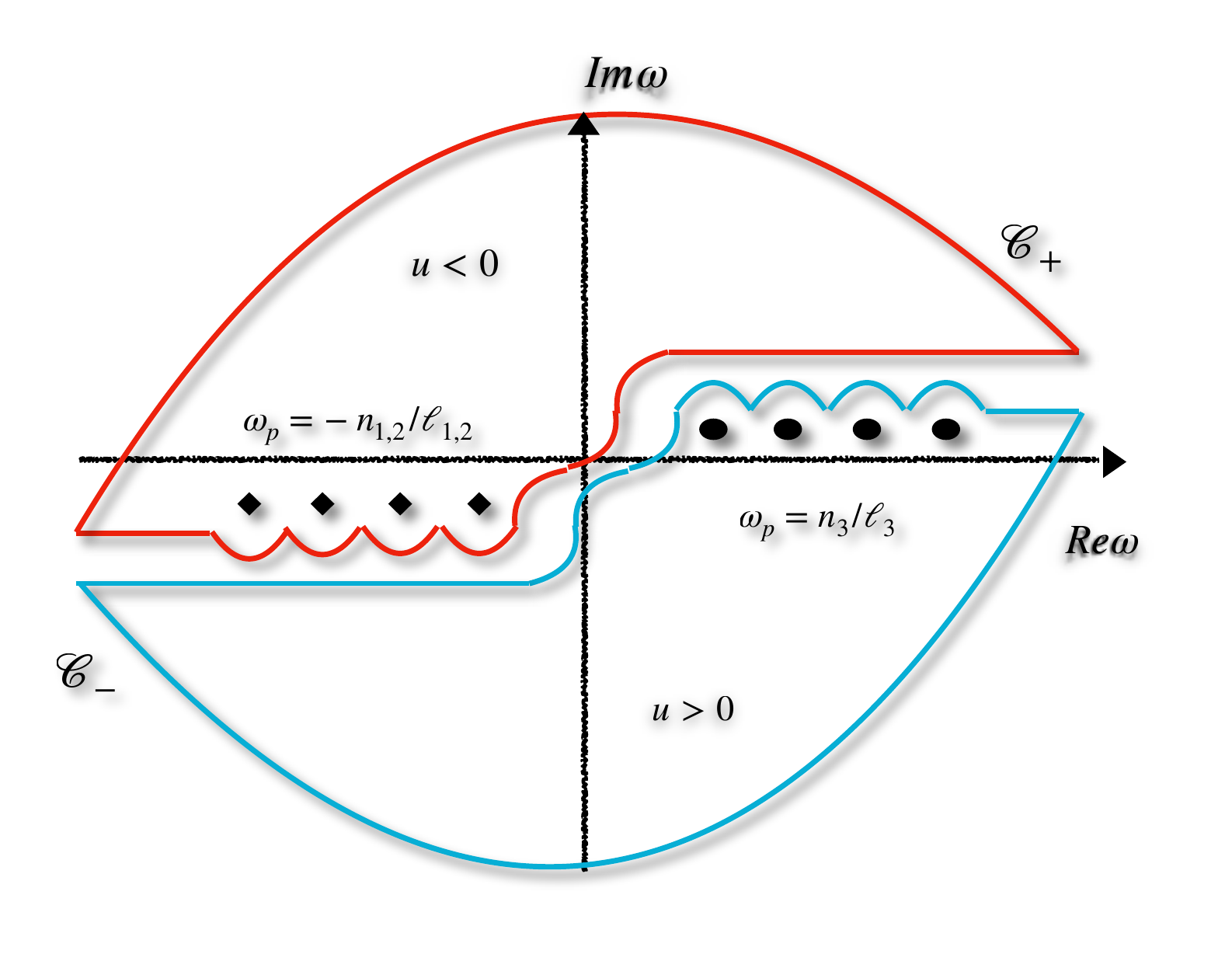}
\caption{Schematic picture of the integration contours due to the $+\imath \epsilon$ prescription. The $\omega_p$ poles are represented in the physical region of the parameter $\ell_a$. }
\end{figure}

Performing the integral, in the physical kinematic region where $\ell_{1,2}>0$ and $\ell_3<0$, for $u=t-R >0$ one finds
\be 
\Delta^{(>)}_s e^{\mu\nu}(t, \vec{x}) = -{4 G_N \over R}\sum_{b=1}^3 {p_b^{(\mu} p_b^{\nu)|}\over n{\cdot}p_b} \sum_{n_3=1}^\infty 
{(-)^{n_3} \over n_3!^2} {\Gamma(1{-}n_3\lambda_{1,3})\Gamma(1{-}n_3\lambda_{2,3})
\over  
\Gamma(1{+}n_3\lambda_{1,3}) \Gamma(1{+}n_3\lambda_{2,3})} e^{\imath n_3 u/{\ell}_3} \,,
\ee
while for $u=t-R <0$ one finds
\be
\Delta^{(<)}_s e^{\mu\nu}(t, \vec{x}) ={4 G_N \over R}\sum_{b=1}^3 {p_b^{(\mu} p_b^{\nu)|}\over n{\cdot}p_b}
\sum_{a=1,2} \sum_{n_a=1}^\infty 
{(-)^{n_a} \over n_a!^2} {\Gamma(1{-}n_a\lambda_{a{+}1,a})\Gamma(1{-}n_a\lambda_{a{+}2,a})
\over  
\Gamma(1{+}n_a\lambda_{a{+}1,a})\Gamma(1{+}n_a\lambda_{a{+}2,a})} e^{\imath n_a u/{\ell_a}} \,, 
\ee
with $a{+}3\equiv a$, recall that $n{\cdot}p_a = - \eta_a E_a(1-\vec{n}\vec{v}_a)$ with $\vec{n}= \vec{x}/R$.  

Notice that the typical time-scales of the signal are set by $\ell_a$, which we have estimated to be in the aLIGO/VIRGO range ($\omega\approx 10\div 100 Hz$) for $1/\sqrt{\ap} \approx 10^{15\div 16}GeV$ and $M_a\approx 10\div 50 M_{\odot}$. Since the amplitude is of the same order as the leading GR contribution, we expect the correction to be detectable in the near future. 



Although the resulting series cannot be resummed in general, for special values of the kinematical parameters $\ell_a=np_a$ they can we written in terms of known functions.

\subsection{`Rational' Kinematics}

Following the detailed analysis in Appendix \ref{KinematicsApp}, in the CoM frame of the system one has 
\be
E_1 = {\widetilde{M}_3^2+M_1^2-M_2^2 \over 2\widetilde{M}_3} \,, \qquad E_2 = {\widetilde{M}_3^2+M_2^2-M_1^2 \over 2\widetilde{M}_3} \,, \qquad
|\vec{p}| = {\sqrt{{\cal F}(M^2_1,M^2_2,\widetilde{M}^2_3)}\over 2\widetilde{M}_3} \,,
\ee
where $\widetilde{M}_3 = E_3{+}\omega=\omega{+}\sqrt{M_3^2{+}\omega^2} $ and $\vec{p}=\vec{p}_1=-\vec{p}_2$, while $\vec{p}_3={-}\vec{k}={-}\omega \vec{n} ={-}\omega \vec{x}/R$, and  
\be
{\cal F}(x,y,z)= x^2+y^2+z^2-2xy-2yz-2zx
\ee
is the ubiquitous `fake square', which is positive in the physical domain
\be
0<\mu_1<1 \quad , \quad 0<\mu_2<1\quad , \quad (\mu_1-\mu_2)^2 - 2(\mu_1+\mu_2) + 1>0\,.
\ee
\begin{figure}[h!]
\center
\includegraphics[scale=0.45]{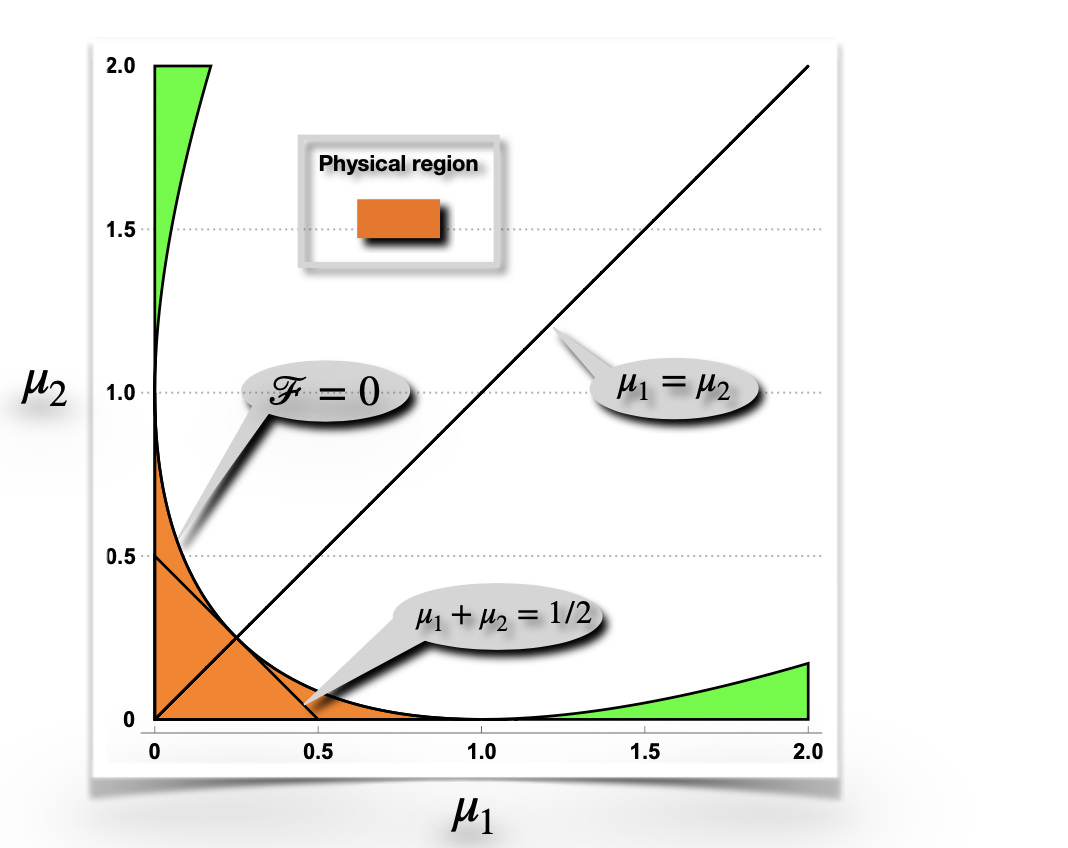}
\caption{Physical domain in the plane 
$(\mu_1=M_1^2/\widetilde{M}_3^2, \mu_2=M_2^2/\widetilde{M}_3^2)$ for generic $k=\omega(1,\vec{x}/R)$ and $\widetilde{M}_3 = E_3 +\omega = \sqrt{M_3^2+\omega^2} + \omega$.}
\label{physregionplot}
\end{figure}
In terms of  $\mu_1=M_1^2/\widetilde{M}_3^2$ and $\mu_2=M_2^2/\widetilde{M}_3^2$ and of $\cos\theta =  {\vec{x}\cdot \vec{p}\over R|\vec{p}|}$, one has 
\be
\lambda_{13} = -{1\over 2} (1+\mu_1-\mu_2) + {1\over 2} \cos\theta \sqrt{1- 2(\mu_1+\mu_2) + (\mu_1-\mu_2)^2} \,,
\ee
\be
\lambda_{23} = -{1\over 2} (1-\mu_1+\mu_2) - {1\over 2} \cos\theta \sqrt{1- 2(\mu_1+\mu_2) + (\mu_1-\mu_2)^2}\,,
\ee
within the ranges 
\be
{-}1\!<\! {-}{1{+}\mu_1{-}\mu_2\over 2}{-}{\sqrt{1{-}2(\mu_1{+}\mu_2) {+}(\mu_1{-}\mu_2)^2}\over 2} \!\le\! \lambda_{13} \!\le\!
{-}{1{+}\mu_1{-}\mu_2\over 2}{+}{\sqrt{1{-}2(\mu_1{+}\mu_2) {+}(\mu_1{-}\mu_2)^2}\over 2}\!<\!0
\ee
and 
\be
{-}1\!<\! {-}{1{-}\mu_1{+}\mu_2\over 2}{-}{\sqrt{1{-}2(\mu_1{+}\mu_2) {+}(\mu_1{-}\mu_2)^2}\over 2} \!\le\! \lambda_{23} \!\le\!
{-}{1{-}\mu_1{+}\mu_2\over 2}{+}{\sqrt{1{-}2(\mu_1{+}\mu_2) {+}(\mu_1{-}\mu_2)^2}\over 2}\!<\!0\,.
\ee
(Recall that $\lambda_{13} + \lambda_{23}=-1$.) A typical value   
is $\lambda_{13}=\lambda_{12}=-1/2$, which is reached for 
\be
\cos\theta_{(-{1\over2}, -{1\over2})}= {\mu_1-\mu_2\over \sqrt{1- 2(\mu_1+\mu_2) + (\mu_1-\mu_2)^2}}\,,
\ee
and is allowed ({\it i.e.}$|\cos\theta|\le 1$) only for $\mu_1+\mu_2\le1/2$. \\

For $\mu_1=\mu_2=\mu=M^2/\widetilde{M}_3^2$, things look simpler and one has 
\be
\lambda_{13} = -{1\over 2}+ {1\over 2} \cos\theta \sqrt{1- 4\mu} 
\, , \qquad \qquad  
\lambda_{23} = -{1\over 2} - {1\over 2} \cos\theta\sqrt{1- 4\mu} \,,
\ee
with 
\be
-1< -{1\over 2}- {1\over 2}\sqrt{1- 4\mu}\le \lambda_{13,23}\le -{1\over 2}+ {1\over 2}\sqrt{1- 4\mu} <0\,.
\ee
In particular, $\lambda_{13}=\lambda_{23}=-1/2$ is found for $\cos\theta = 0$ for all $\mu\le 1/4$ and for all $\cos\theta$ for $\mu= 1/4$. 

\begin{figure}[h!]
\center
\includegraphics[scale=0.45]{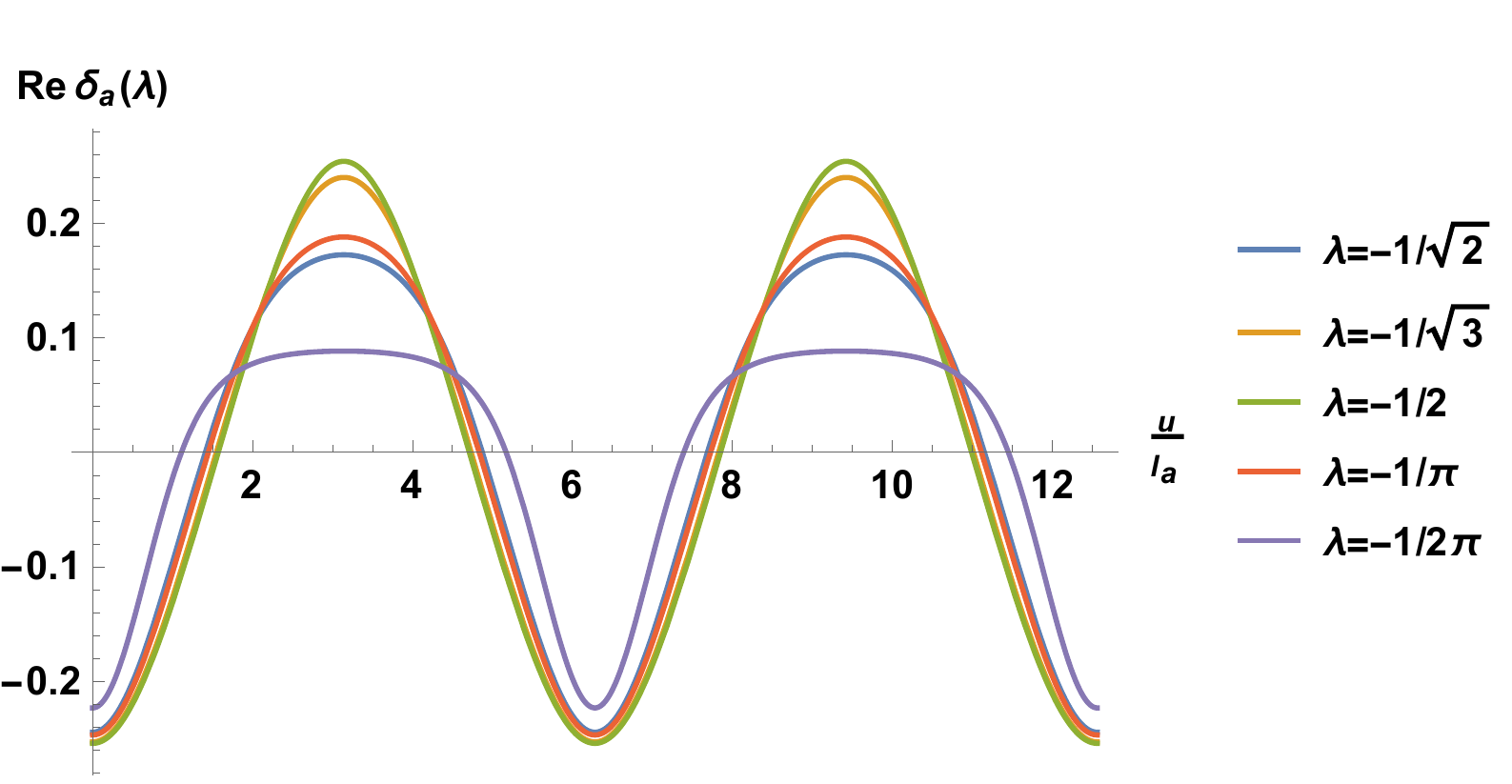}
\includegraphics[scale=0.45]{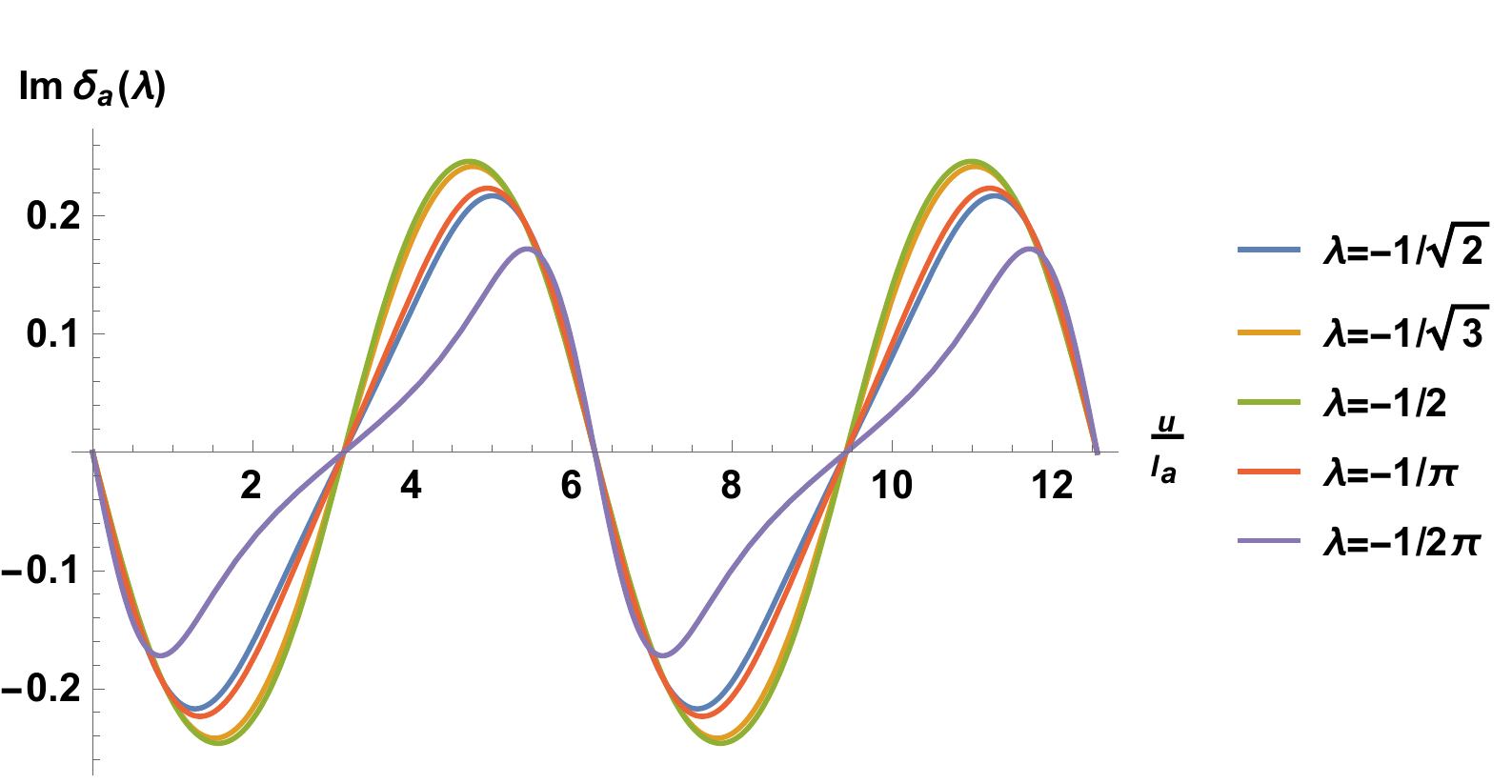}
\includegraphics[scale=0.45]{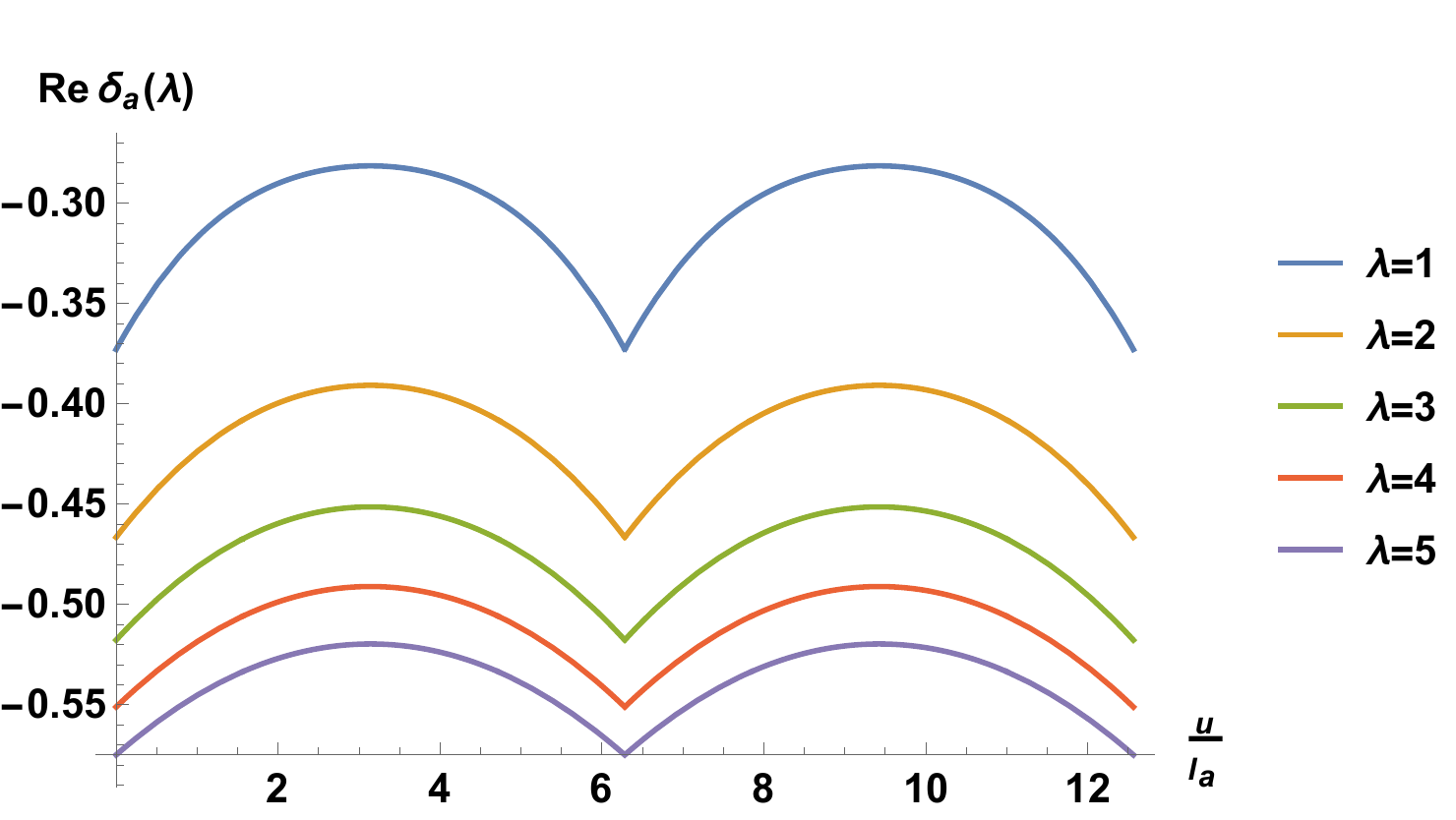}
\includegraphics[scale=0.45]{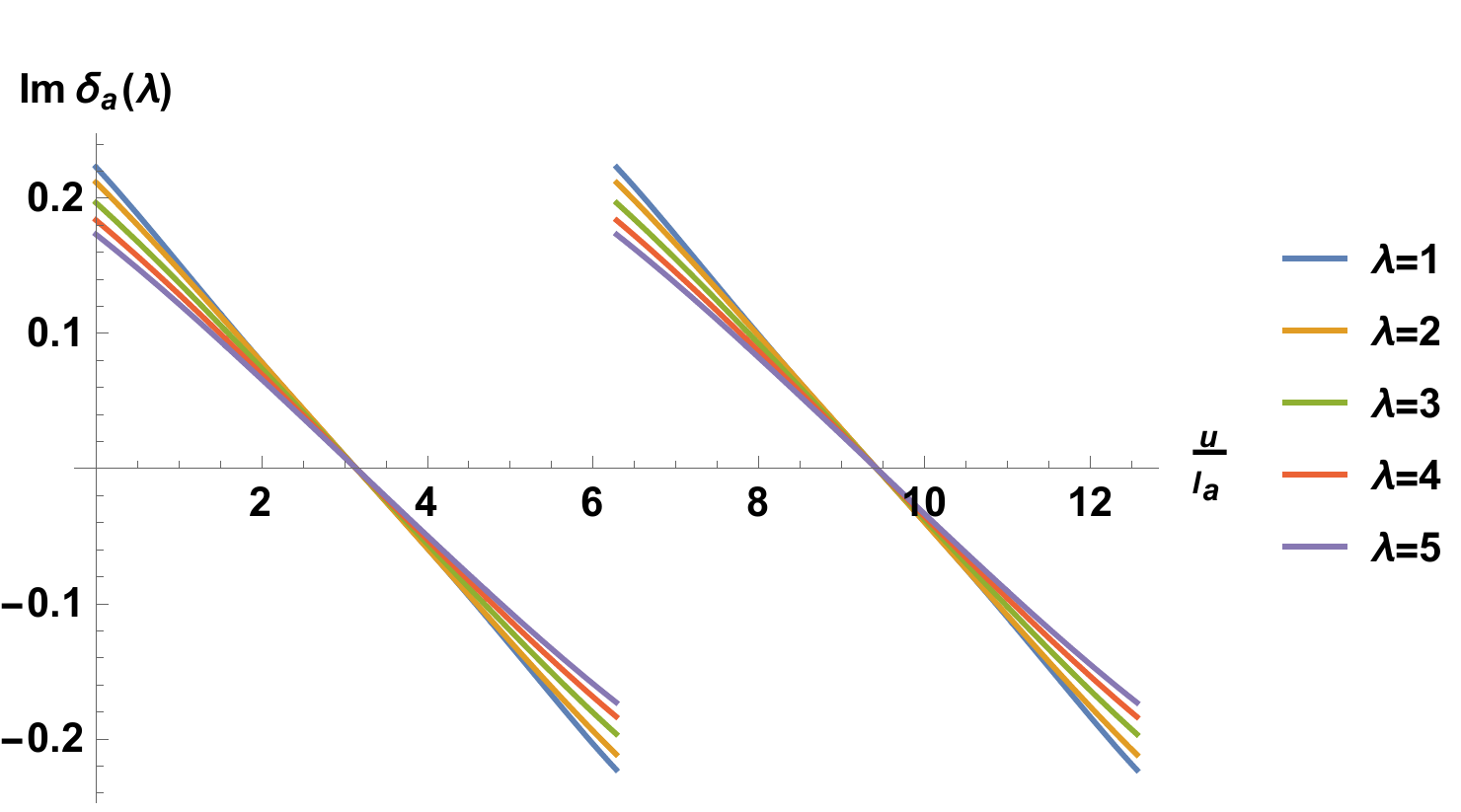}
\caption{Plots of real and imaginary parts of $\delta_a(\lambda_{a{+}1,a},\lambda_{a{+}2,a};u/\ell_a)$ for special choices of the (independent) kinematical ratio $\lambda$.}
\label{hyperplotFig}
\end{figure}

In this `symmetric' condition, the series  
\be
\begin{split}
\sum_{n_3=1}^\infty 
{(-)^{n_3} \over n_3!^2} {\Gamma\left(1{+}{n_3\over 2}\right)^2\over \Gamma\left(1{-}{n_3\over 2}\right)^2} e^{-\imath n_3 u/{\ell}_3} &= -{1\over 4} e^{- \imath n_3 u/{\ell}_3}  {}_2F_1\left({1\over 2},{1\over 2};1; {e^{-2\imath n_3 u/{\ell}_3}\over 16}\right) \\
&= -{1\over 2\pi} e^{- \imath n_3 u/{\ell}_3}{\cal K} \left({e^{-2\imath n_3 u/{\ell}_3}\over 16}\right) 
\end{split}
\ee
becomes a complete elliptic integral of the first kind. For the same kinematics, the other two series are identical, with $\lambda_{31}=-2$, $\lambda_{21}=+1$ or $\lambda_{32}=-2$, $\lambda_{12}=+1$, and yield 
\be
\sum_{n=1}^\infty 
{(-)^{n} \over n!^2} {\Gamma\left(1{+}2{n}\right) \Gamma\left(1{-}{n}\right)\over 
\Gamma\left(1{-}2n\right) \Gamma\left(1{+}{n}\right)} e^{-\imath n u/{\ell}_3} = -{1\over 2} + 
{1\over \pi} {\cal K} (16 e^{-\imath n_3 u/{\ell}_3}) \,.
\ee
Since the argument $z=16 e^{-\imath n_3 u/{\ell}_3}$ has modulus $|z|=16>1$, one has to analytically continue the series and use 
\be 
\begin{split}
{2\over \pi} {\cal K} (z) &=  {}_2F_1(1/2,1/2;1;z) \\
&= {(-z)^{-1/2} \Gamma(1) \over \Gamma(1/2)^2}\sum_{n=0}^\infty {(1/2)_n^2 \over n!^2 {} z^n}  
[\log(-z)+2\psi(n{+}1)-\psi({1\over 2}{+}n)- \psi({1\over 2}{-}n)]\,.
\end{split}
\ee
Quite remarkably, this generates a non-periodic term ($\log(-z)$), and terms ($(-z)^{-{1\over 2}-n}$) with reduced periodicity to $4\pi\ell$ . This phenomenon leads to the apparent discontinuity of the plots in the second row of Fig.~\ref{hyperplotFig}. 

For other rational values of $\lambda_{13}$ and $\lambda_{12}$ one can express the results in terms of generalised hypergeometric functions. For illustrative purposes we plot the results for the three series for 
$\lambda_{13}$ in the interval $\lambda_{min} \!=\!-0.99$, $\lambda_{Max}\!=\!-0.001$ --- corresponding to $\mu \!=\! \lambda_{min}$, $\lambda_{Max}\!=\! -\lambda_{min}(1+\lambda_{min})$ --- and $\varphi_3 \!=\! u/{\ell}_3$ in the interval $(0,2\pi)$. 

Just to make compact the notation, we adopt the definition
\begin{equation}
\sum_{n_a=1}^\infty 
{(-)^{n_a} \over n_a!^2} {\Gamma(1{-}n_a\lambda_{a{+}1,a})\Gamma(1{-}n_a\lambda_{a{+}2,a})
\over  
\Gamma(1{+}n_a\lambda_{a{+}1,a})\Gamma(1{+}n_a\lambda_{a{+}2,a})} e^{\imath n_a u/\ell_a} =\delta_a(\lambda_{a{+}1,a},\lambda_{a{+}2,a};u/\ell_a)\,.
\end{equation}
This function is symmetric under the exchange of $\lambda_{a{+}1,a}\leftrightarrow  \lambda_{a{+}2,a}$ and from the kinematical constraints
\begin{equation}\label{kinconstr}
\lambda_{a{+}1,a}+\lambda_{a{+}2,a}=-1 \,,\qquad a=1,2,3\,,
\end{equation}
fixing the values of $\lambda_{1,3}$ and $\lambda_{2,3}$ one has fixed all the allowed values of the other parameters. In this regard some examples are reported in Table~\ref{TABLE1}.

For representative kinematical ratios, in Fig.~\ref{hyperplotFig} we plot the real and imaginary parts of the function $\delta_a(\lambda_{a{+}1,a},\lambda_{a{+}2,a};u/\ell_a)$, which according to Table~\ref{TABLE1} and its properties can be represented as a function of only one kinematical parameter, the other ones being fixed as in (\ref{kinconstr}).
\begin{table}[!ht]
\center
\begin{tabular}{ || c | c || c | c || c | c ||}
 $\lambda_{13}$ &$\lambda_{23}$& $\lambda_{31}$ &$\lambda_{21}$& $\lambda_{12}$ &$\lambda_{32}$ \\
\hline
\hline
 -1/2 &-1/2& -2 &1& 1 &-2 \\
\hline
  -1/3 &-2/3& -3 &2& 1/2 &-3/2 \\
  \hline
  -1/4 &-3/4& -4 &3& 1/3 &-4/3 \\ 
  \hline
  -1/5 &-4/5& -5 &4& 1/4 &-5/4 \\ 
  \hline
  -2/3 &-1/3& -3/2 &1/2& 2 &-3 \\ 
  \hline
  -3/4 &-1/4& -4/3 &1/3& 3 &-4 \\ 
\end{tabular}
\caption{ Some examples of `rational' kinematical regimes.  }
\label{TABLE1}
\end{table}  

Note that contrary to gravitational memory, string memories have a non-trivial $u$ dependence and that their origin lies in the possibility of an excited string state to emit gravitons through Regge resonance in highly inelastic processes like the one we have considered. 

\begin{figure}[h!]
\center
\includegraphics[scale=0.33]{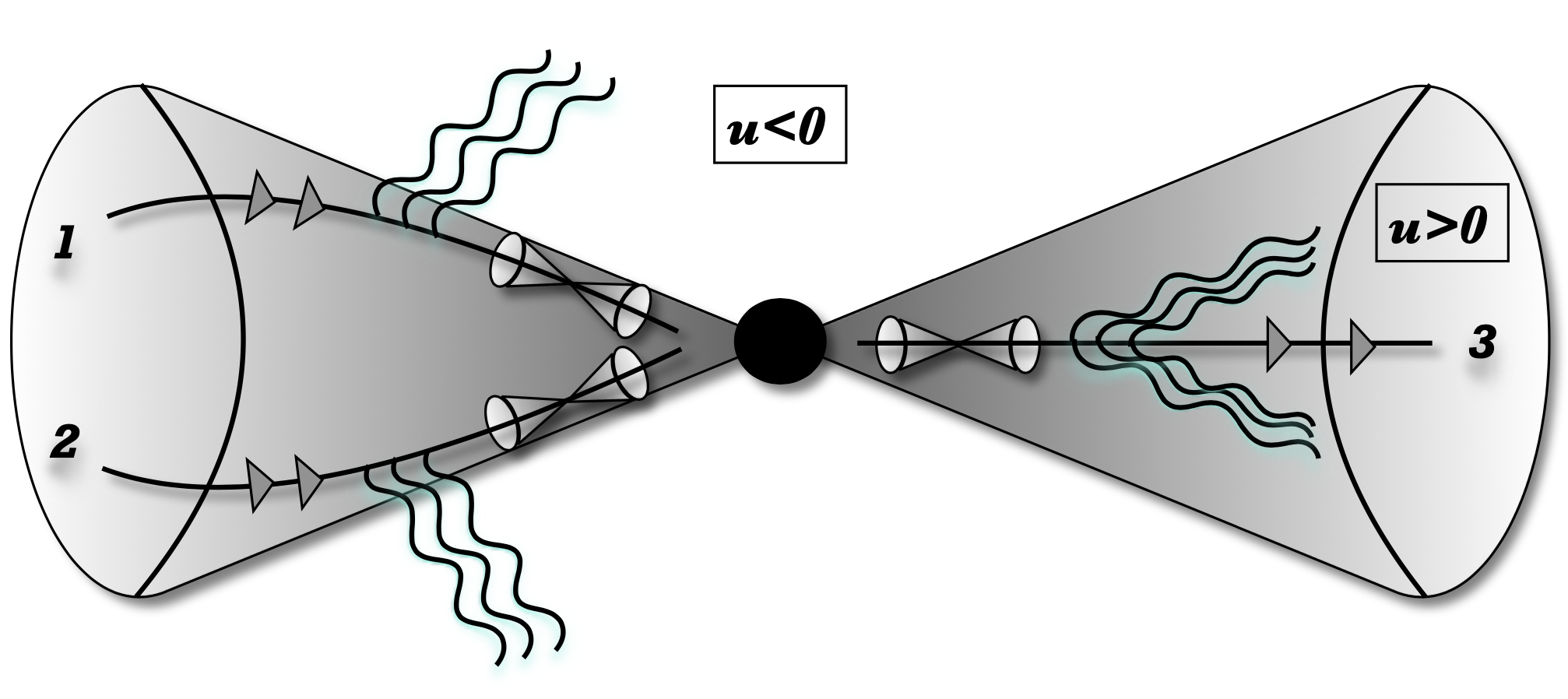}
\caption{Schematic representation of the causal structure of the process.}
\label{CausalityFig}
\end{figure}

Moreover, recall that incoming BHs produce radiation that can be detected outside the future light-cone of the merging event, i.e. for $u<0$, but also in the future $u>0$, while the produced BH emits radiation only inside the future light-cone, i.e. $u>0$, see Fig.~\ref{CausalityFig}.

In order to shed some light onto the physical implications that these $\ap$ corrections might entail, and to expose a more transparent $u$ dependence, we will focus on special/extreme kinematical regimes in the following.


\subsection{Large $\omega \ell_a$ behaviour of the String Memories}
\label{LargeOmegaEll}

Let us subtract the purely GR pole at $\omega = 0$ and consider 
\be
{\cal H} - {1\over \omega} = {1\over \omega}({\cal F} - 1)\,.
\ee
At large $\omega \ell_a$, (keep in mind that 1 and 2 are `in' and 3 is `out') the dominant behaviour is
\be
{\cal F} \approx \exp\{-\beta \omega\} \quad {\rm with} \quad -\beta = \ell_1 \log{\ell_1\over\ell_3}+\ell_2 \log{\ell_2\over\ell_3} = \ell_3 \left(\lambda_{13} \log\lambda_{13} + \lambda_{23} \log\lambda_{23} \right) \,.
\ee
{\bf $\omega \rightarrow |\omega|$} .... 
Note that $\beta$ is complex since $\ell_1, \ell_2>0$ while $\ell_3<0$ (in the physical domain). 

Plugging this in the $\omega$ integral yields the expression
\be
{{\cal E}}(u, \beta) = \int_{-\infty}^{+\infty} {d\omega\over \omega} e^{-\imath \omega u} (e^{-\beta \omega} - 1) 
\ee
that is ill-defined and requires regularisation. A reasonable choice\footnote{We thank A.~Sen and B.~Sahoo for pointing out a shortcoming in a preliminary version of this manuscript.} seems to be replacing $e^{-\beta \omega}$ with $e^{-\beta|\omega|}$ that yields 
\be 
{{\cal E}}_{reg}(u, \beta) = \int_{-\infty}^{+\infty} {d\omega\over \omega} e^{-\imath \omega u} (e^{-\beta|\omega|} - 1) = 2i \arctan{u\over \beta}
\ee
%
\begin{figure}[h!]
\center
\includegraphics[scale=0.45]{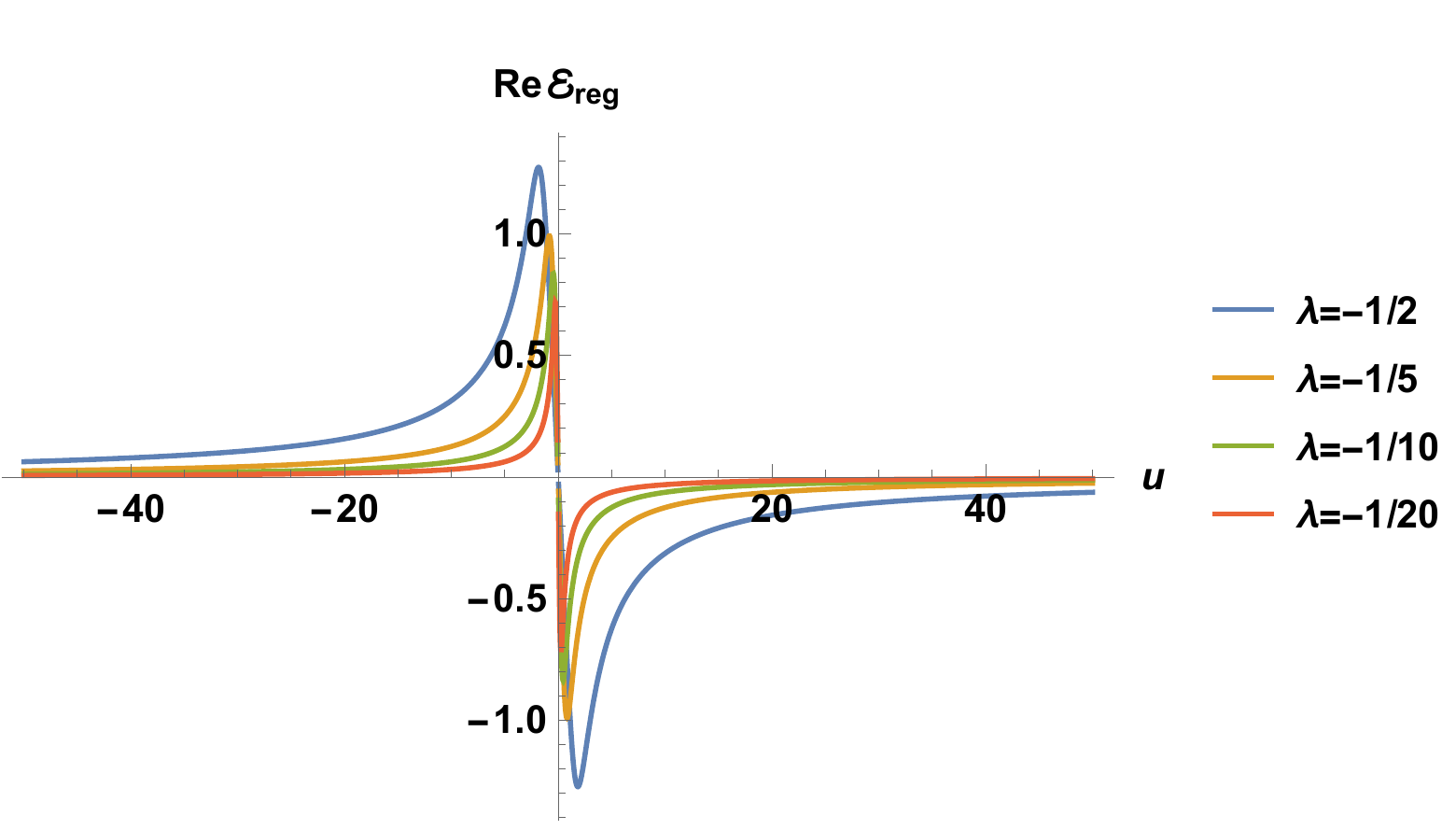}
\includegraphics[scale=0.45]{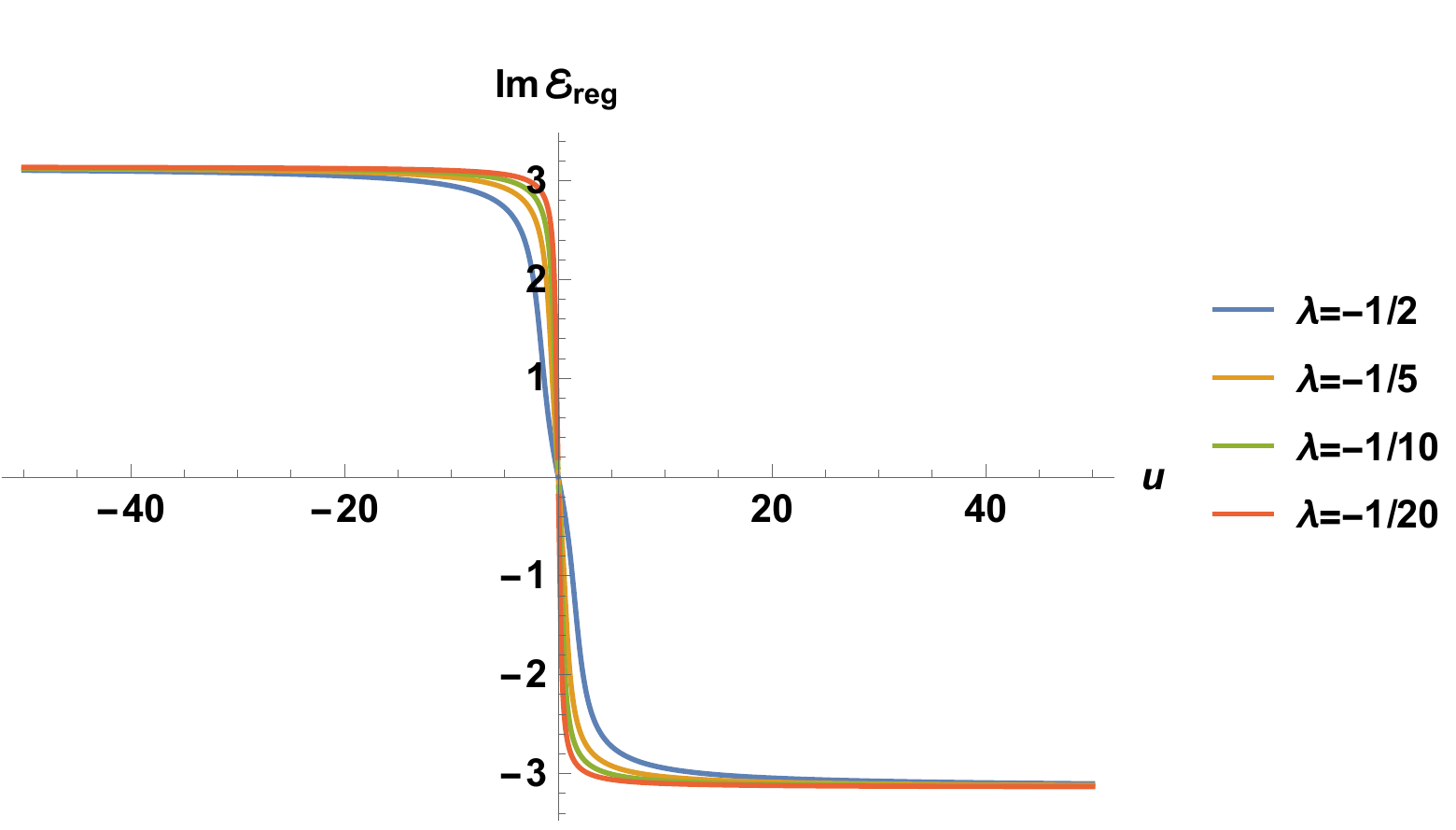}
\caption{Real and imaginary part of ${\cal E}_{reg}(u,\beta)$ as a function of $u$.}
\label{EfunctionFig}
\end{figure}
As visible from the plot in Fig.~\ref{EfunctionFig} the real part of the function ${{\cal E}}_{reg}(u, \beta)$ behaves like $1/u$, while the imaginary part tends to a constant. It is amusing to observe that a $1/u$ behaviour is similar to the one produced by the $\log\omega$ terms appearing at one-loop in the case of $D=4$, that have been identified in \cite{Saha:2019tub}, following earlier work \cite{Sen:2017nim, Laddha:2017ygw, Laddha:2018vbn, Sahoo:2018lxl, Laddha:2019yaj}. This seems to suggest that the inclusion of $\ap$-effects may emerge as a violation of the gravitational soft theorems. As previously suggested, $\log\omega$ terms may be detected through GWs. One should keep in mind that the function $\beta\omega=\ap (s\log s +t\log t+ u\log u)$ coincides with the leading log IR divergent terms appearing at one-loop in 4-graviton amplitudes in GR (or supergravity), with $G_N = \alpha' g_s^2/\hat{V}_6$ replacing $\alpha'$.

\subsection{Regge behaviour: `the plunge' $M_1<<M_2\sim M_3$}
\label{ReggePlunge}

Some kind of Regge behaviour is found in the case $\varepsilon = \ell_1<<\ell_2\approx -\ell_3 $, whereby one of the merging BPS BH's is much lighter than the other two, \ie $M_1<<M_2\sim M_3$. This process is some times called the `plunge' and leads to Extreme Mass-Ratio Inspirals (EMRIs) that will be one of the scientific goals of LISA mission. In this case one finds
\be
\begin{split}
{\cal F} &= {\Gamma(1+\omega \varepsilon) \Gamma(1+\omega \ell) \Gamma(1-\omega(\ell+\varepsilon))
 \over \Gamma(1+\omega \varepsilon) \Gamma(1-\omega \ell) \Gamma(1+\omega(\ell+\varepsilon))}\\
 &\approx (1-\omega\ell)^{-\omega\varepsilon} (1+\omega\ell)^{-\omega\varepsilon} = 
 (1-\omega^2\ell^2)^{-\omega\varepsilon} \approx (-\omega\ell)^{-2\omega\varepsilon}\,.
 \end{split}
 \ee
Plugging this into the $\omega$ integral one has \be
\int_{-\infty}^{+\infty} {d\omega\over \omega} e^{-\imath \omega u}(-\omega\ell)^{-2\omega\varepsilon}
= \int_{-\infty}^{+\infty} {\ell d\omega} e^{-\imath \omega u} e^{-(1+2\omega\varepsilon)\log\omega\ell }\,.
\ee
Setting $w = \omega\ell$, $\hat{u} = u/\ell$, $\hat\varepsilon = \varepsilon/\ell$ and performing a saddle-point approximation yields
\be
\int_{-\infty}^{+\infty} {dw} e^{-\imath w\hat{u}} e^{-(1+2w\hat\varepsilon)\log w } = C e^{-\imath w^*\hat{u}} e^{-(1+2w^*\hat\varepsilon)\log w^* }\,,
\ee
where $w^*$ satisfies the saddle-point equation
\be
-\imath \hat{u}-{1\over w^*} -2\hat\varepsilon - 2\hat\varepsilon\log{w^*}=0 \,, \qquad i.e. \qquad 
{\imath \hat{u}\over2\hat\varepsilon} - 1 = \log{w^*}\,,
\ee
where one can neglect $1/w$. So one gets
\be
-\imath \hat{u}w^* - (1+2\hat\varepsilon w^*)\log{w^*} = 1+ \log{w^*}+ 2\hat\varepsilon w^* \approx -\imath {\hat{u}\over 2\hat\varepsilon} + 2\hat\varepsilon \exp \left({\imath{\hat{u}\over2\hat\varepsilon}-1}\right) \,,
\ee
\begin{figure}[h!]
\center
\includegraphics[scale=0.3]{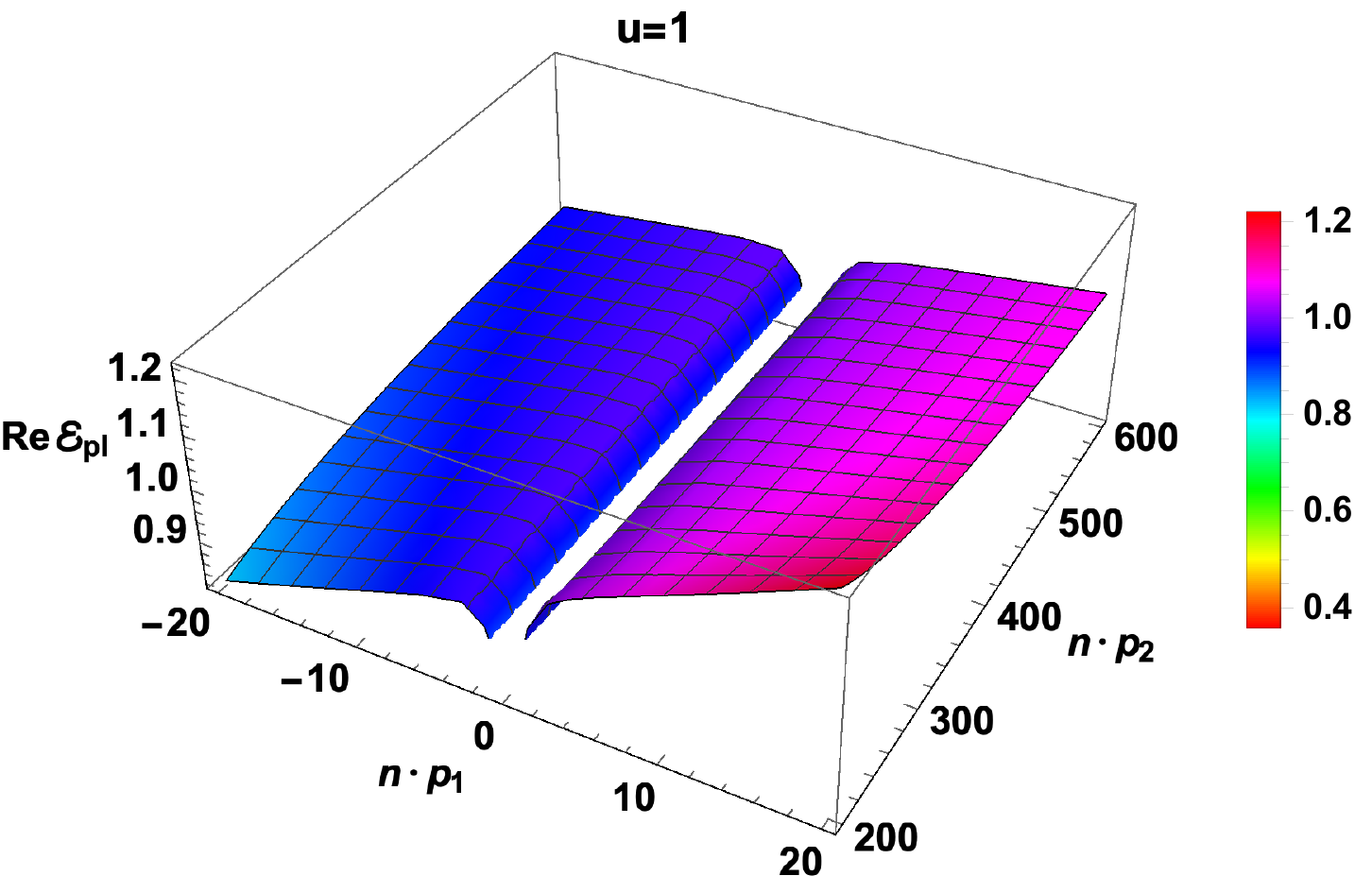}
\includegraphics[scale=0.3]{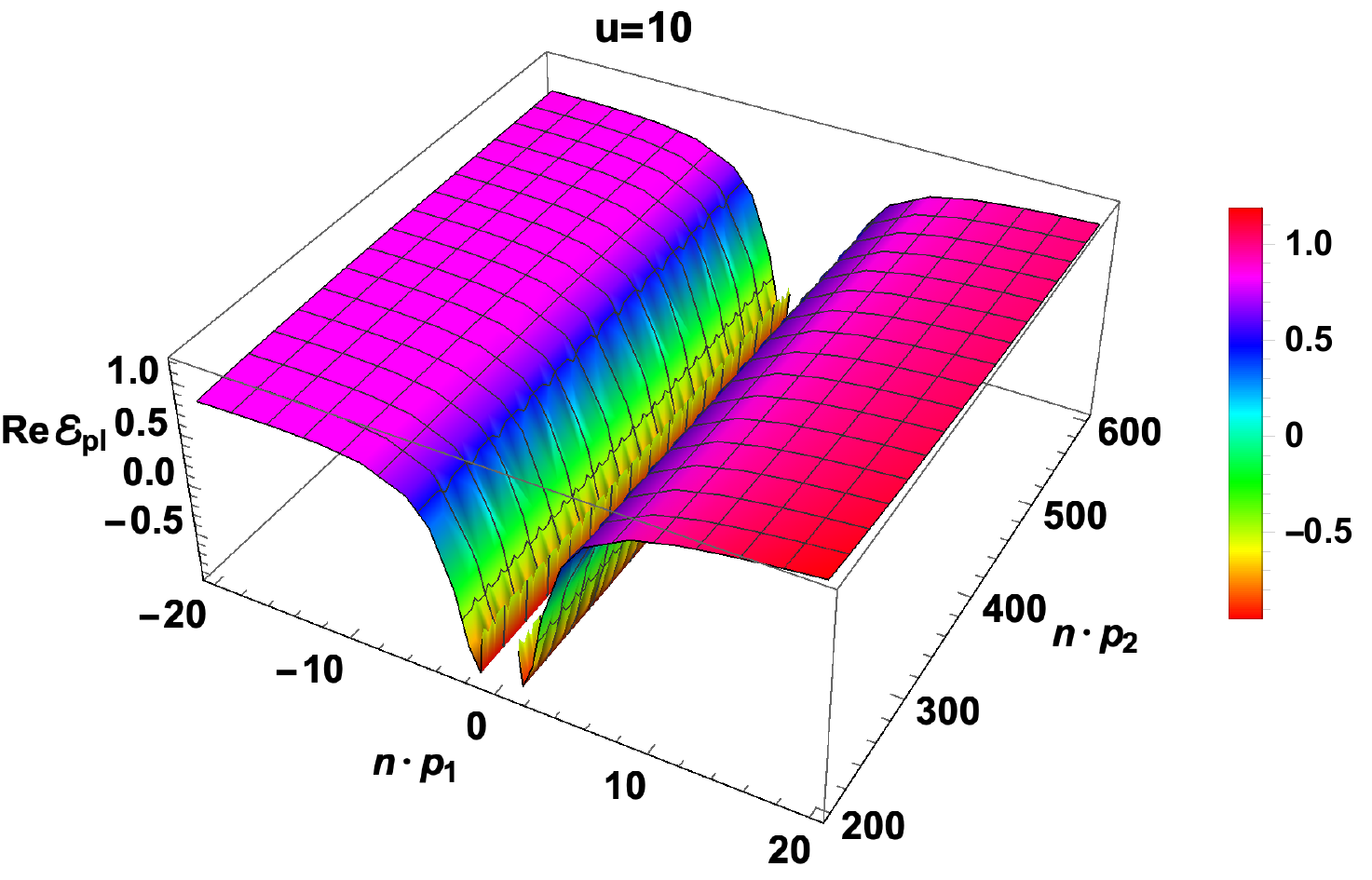}
\includegraphics[scale=0.3]{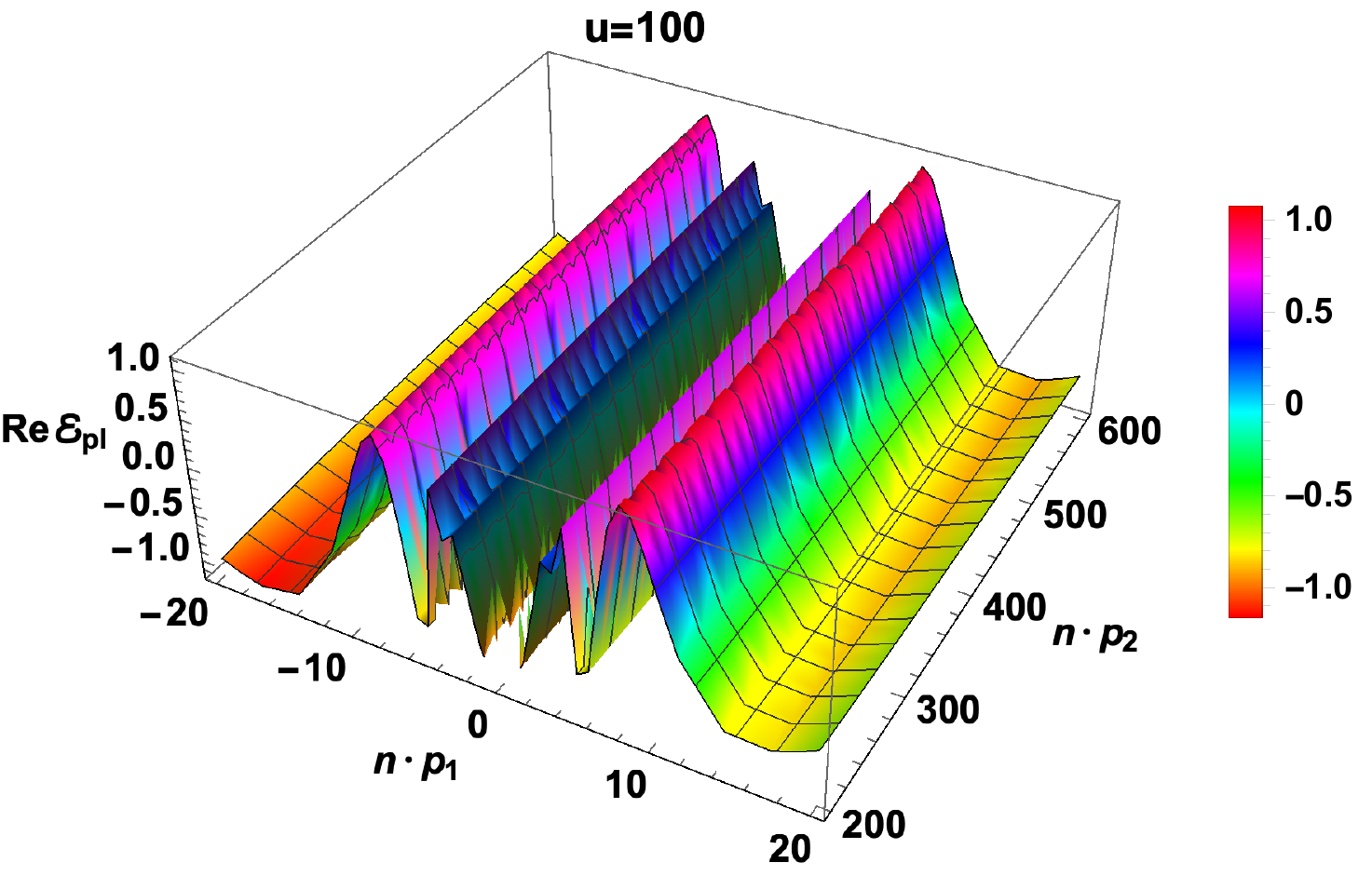}
\includegraphics[scale=0.3]{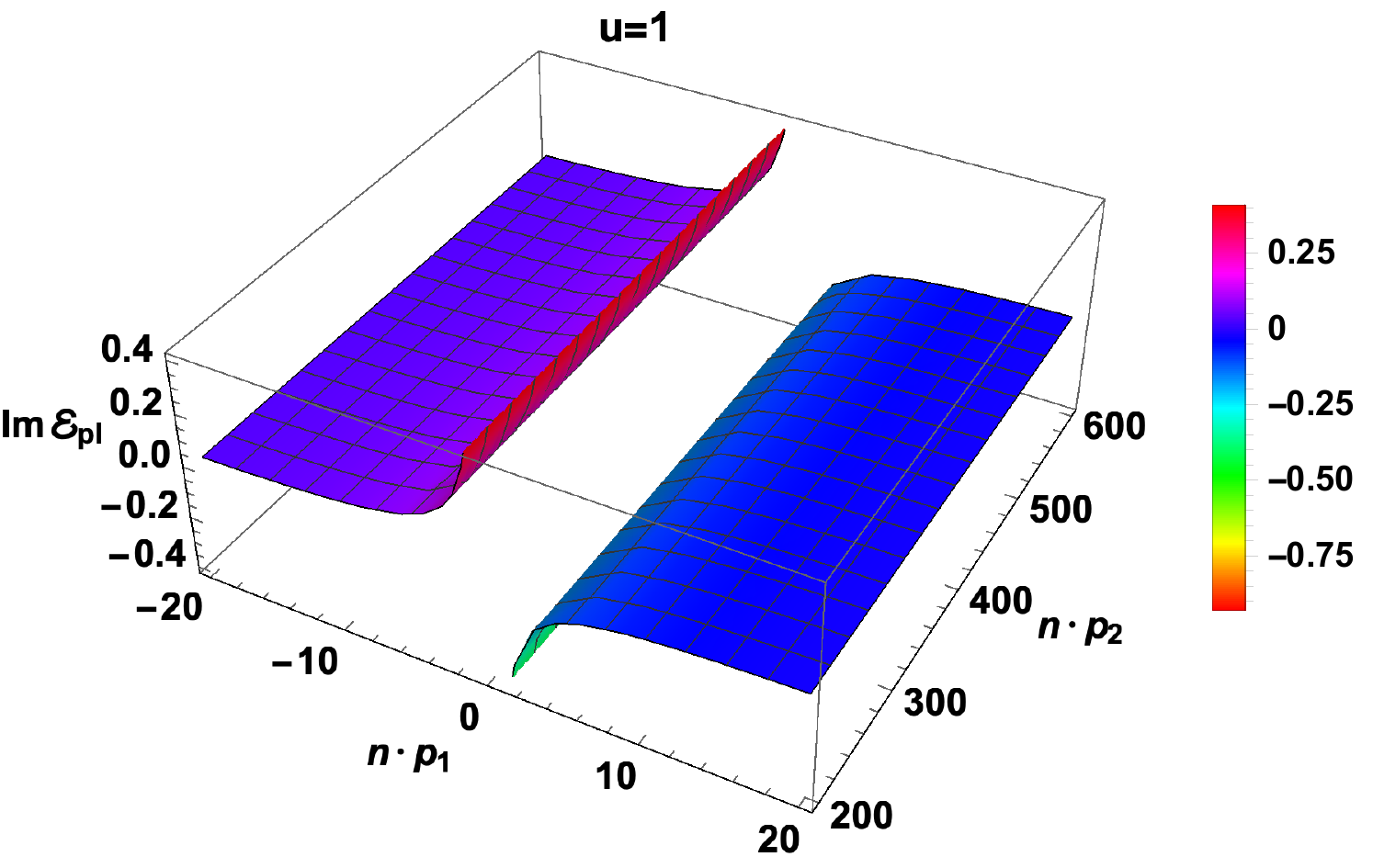}
\includegraphics[scale=0.3]{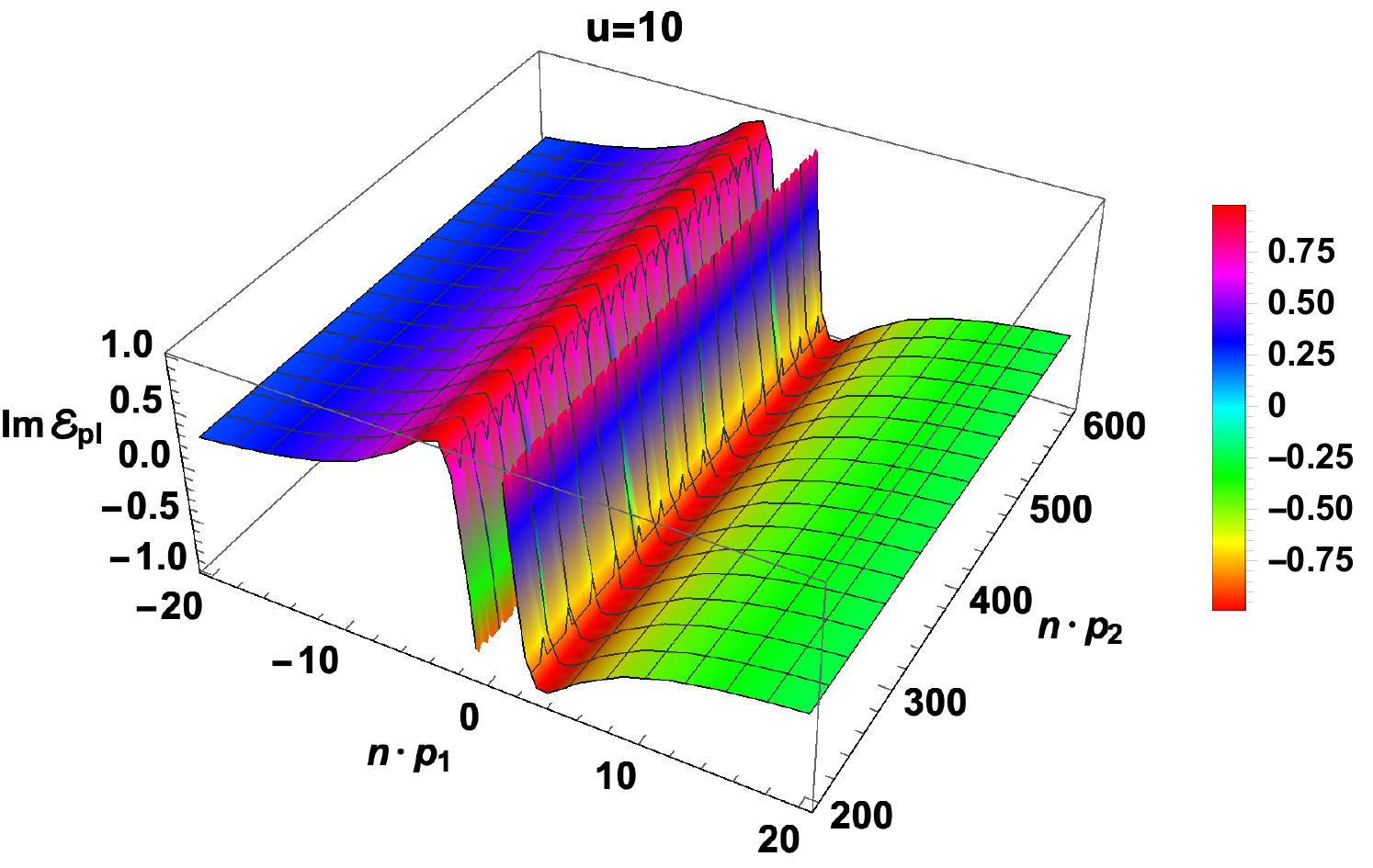}
\includegraphics[scale=0.3]{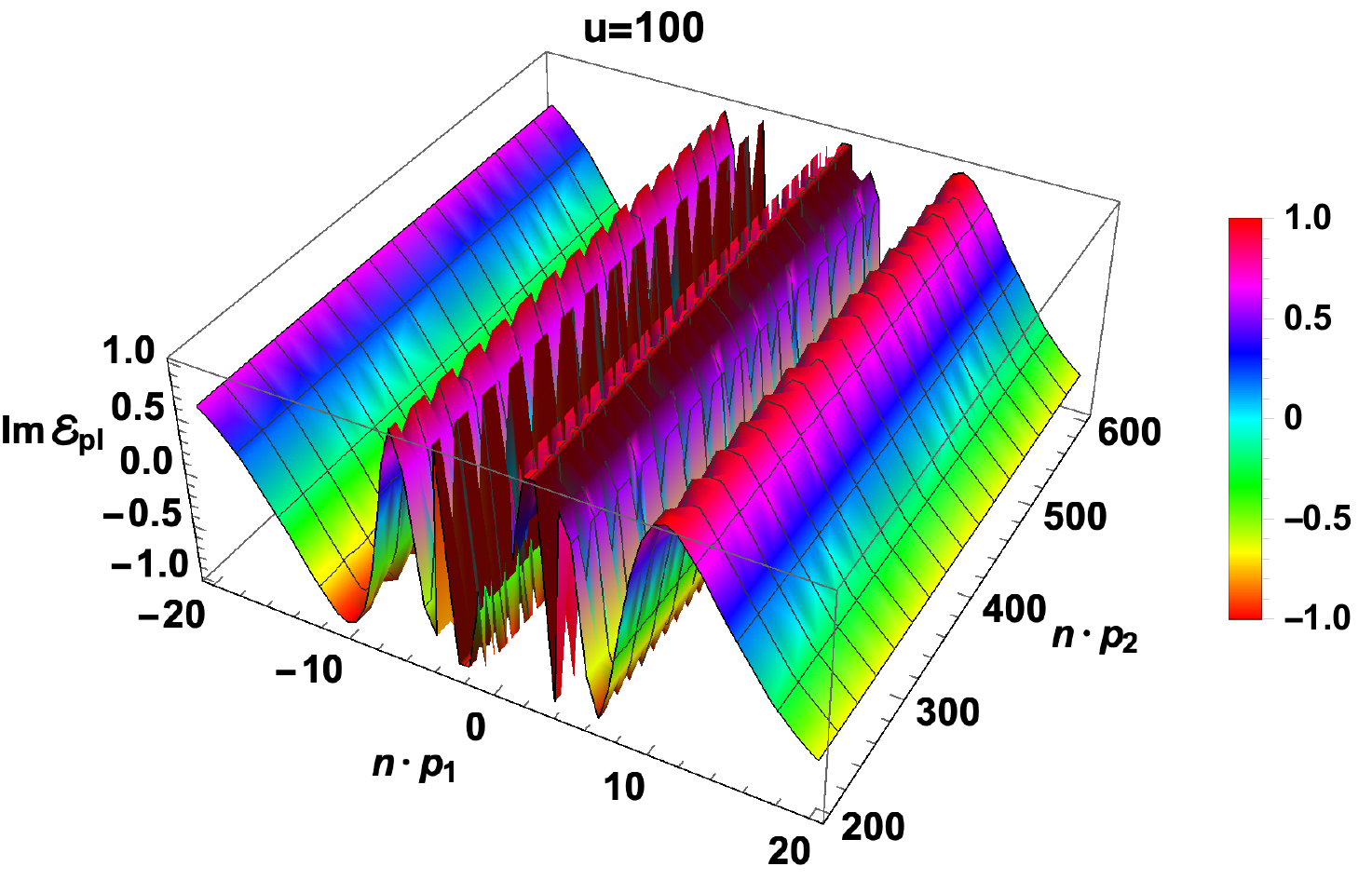}
\caption{Real  and imaginary part of $E_{pl}$ as a function of the parameters $n{\cdot}p_1$ and $n{\cdot}p_2$ for three positive reference values of $u=1,10,100$.}
\label{EplungeFig}
\end{figure}

and finally, exponentiating the result, 
\be 
{\cal E}_{pl}(np_1,np_2; u)= e^{-\imath {\hat{u}\over 2\hat\varepsilon}+\exp\left( {2\varepsilon\over \ell} \exp{\imath u\over2\varepsilon}\right)} = \exp\left( -\imath {u\over 2np_1}+{2np_1\over np_2} \exp{\imath u\over2np_1}\right) \,.
\ee
One should keep in mind that $np_1<< np_2$, the result looks different at $u>0$ from $u<0$ whereby the imaginary part gets flipped. See plots in Fig.~\ref{EplungeFig}.

\section{Lost String Memories ... regained}
\label{LostringMemoSect}

At tree level the poles in $\omega$ were located on the real axis and we have given a prescription to deform the contour of integration. However, beyond tree level massive string resonances become unstable and their masses get shifted and broadened {\it viz.}
\be
\ap M_N^2 = N \rightarrow \ap M_N^2 = N+\delta N + \imath \gamma_N \quad .\ee
Both the (adimensional) mass-shift $\delta N$ and the decay width $\gamma_N$ depend in a highly non-trivial fashion on the details of the state, \ie the vibration modes $n_k$ and polarisation tensor, and on the string coupling $g_s$. The study of this feature, that appears already at one-loop, remains beyond the purpose of our present analysis. Yet, we would like to mention that string states can be found with a very long lifetime, growing with the mass $M$ as fast as ${\cal T}\simeq g_s^{-2} M^5$ \cite{Chialva:2004xm}.  
In a semi-classical picture, these long-lived closed strings remain unbroken during their classical evolution. Emission of massless quanta provide the dominant decay channel of these and other massive string state \cite{Chialva:2003hg}. Type II superstring one-loop amplitudes were evaluated in \cite{Chialva:2009pg}, for states in the Neveu-Schwarz (NS) sector, obtaining mass-shifts and decay rates as a function of the space-time dimension and the string scale.     

In general these instabilities are experimentally relevant, encoding contributions that in non-linear optics are termed `evanescent' waves. The large number of resonances that are eventually produced in some phenomena, including super-oscillation, can actually trigger remarkable amplifier mechanisms, see e.g. \cite{superosc}. On the other hand, barring the dispersion at a given level and relying on string analyses~\cite{Chialva:2003hg, Chialva:2004xm, Chialva:2009pg} suggests power-law expressions of the decay rate such as 
\be
\label{spe}
\gamma_{N}\simeq \gamma_{0}N^{\alpha}\, , 
\ee 
with $\alpha$ real and positive. For $N\rightarrow \infty$, the high density of states allows to replace the infinite sum over the massive string resonances with an integration \ie 
\be
\label{DN}
\sum_{N=1}^{\infty}C_N e^{-\gamma_{0}N^{\alpha} u } = \sum_{N=1}^{\infty} C_N e^{-\gamma_{0}N^{\alpha} u } - C_0
\quad \longrightarrow \ee
\be \quad \int_{0}^{\infty}d\zeta\,  C(\zeta) e^{-\gamma_{0}\, \zeta^{\alpha} u } - C_0 \approx C_0 \left[
{\Gamma\left(1+\frac{1}{\alpha}\right) \over (\gamma_{0} u)^{1/{\alpha}}} - 1 \right] \,. 
\ee
This means that stringy resonances may survive in the GW signal as a cumulative effect. Yet, it seems quite unlikely that one could resolve individual peaks.

\subsection{QNM's and ring-down phase}

To make contact with the phenomenology of GWs, we may associate to $\gamma_{N}$ of Eq.~\eqref{spe} an exponential damping. This immediately calls for a consideration of quasi normal modes (QNM's) $\omega_{lmn}$\cite{Berti:2009kk} and echoes thereof \cite{Cardoso:2017njb, Cardoso:2017cqb}. QNM's represent unstable perturbations of a background metric. The real part $\mathfrak{Re}(\omega_{lmn})$ is associated to the frequency of the unstable closed orbits of a (massless) probe, while the imaginary  $\mathfrak{Im}(\omega_{lmn})$
 to the Lyapunov exponent $\gamma \approx \tau^{-1}$ that governs the chaotic behaviour of geodesics around the `photon-sphere' \cite{Bianchi:2020des}. 

In GR, the uniqueness of the frequencies and damping times is customarily related to the ``no hair'' theorem. In this sense, the detection and identification of QNM's may provide a further possible test for GR in strong-field regimes such as BHs \cite{Dreyer:2003bv}.


Indeed, while in-spiralling can be dealt with by means of a post-Newtonian analysis in GR\footnote{For recent work on the dynamics of binary systems at sixth post-Newtonian order see \eg \cite{Bini:2020nsb, Bini:2020hmy} and references therein.}, the merging phase requires analytical tools: the perturbative approach ceases to be valid as none of the two BHs can be treated as a perturbation of the other (except possibly for EMRI's \cite{BCWLisa}). In our toy model this is accounted for by the highly inelastic amplitude we used as a source for the GW signal in the merging phase. In the ring-down phase, that represent the last part of the signal, perturbation methods can be still applied to the analysis, and with satisfactory results since the signal can reliably be decomposed in QNMs \cite{Berti:2009kk}. 

Using the transverse traceless (TT) gauge, one may expand the observable amplitude $h(u)$ of the GW as
$$
h(u)\simeq \mathfrak{Re}\left[ \sum_{l,m,n} A_{l,m,n} \, e^{-\imath (\omega_{lmn} u +\phi_{lmn})} \right]\,,
$$
where the summation comprises the angular dependence of the mode amplitude and phase, captured by $l$ and $m$ with $l=2,3,...$ and $|m| \leq l$. The harmonic is taken into account through the overtone index $n=1,2\dots$.  Some of the modes $\omega_{lmn}$ of the expansion around the final BH configuration (typically a rotating BH, described by Kerr metric \cite{Kerr:1963ud}) will be predominant. 

We expect the same to take place in a complete quantum theory of gravity such as string theory. The QNM's of a stringy (nonBPS) BH, such as the ones we have considered in our toy model, or a (nonBPS) fuzz-ball are characterised by their peculiar QNM's \cite{Bianchi:2020des, Berti:2009kk}  and analysis of the ring-down signal may expose echoes \cite{Cardoso:2017njb, Cardoso:2017cqb, Cardoso:2019rvt} and novel multipolar structures \cite{Bena:2020see, Bianchi:2020bxa, Bena:2020uup, Bianchi:2020Comp} that could help discriminating between different models for the smooth horizonless compact object replacing the BH and its singular and paradoxical behaviour in GR. 

Once again this interesting analysis, that is being performed for fuzz{-}balls, is beyond the scope of our investigation and we defer it to the future.


\section{Coherent states of quasi BPS BH's}
 \label{CoherentSect}
 
For the purpose of making our computations as simple as possible, so far, we have considered mass eigen-states of heterotic string on $T^6$. We would like to generalise our analysis to quasi BPS coherent states using DDF operators \cite{DelGiudice:1971yjh, Ademollo:1974kz, Hindmarsh:2010if, Skliros:2011si, Skliros:2016fqs, Bianchi:2019ywd, Aldi:2019osr} that can be made as compact as required for the validity of our analysis.

\subsection{DDF operators for open bosonic strings}
In order to fix the notation let us introduce the DDF operators  that for the open bosonic string are defined 
as
\begin{equation}\label{DDFop}
A^i_n= \dfrac{\imath}{\sqrt{2\alpha'}}\oint {dz \over 2\pi i} \, \partial_{z}X^i(z)e^{inq{\cdot}  X(z)}
\end{equation}
where $i=1,..., D{-}2$ ($D=26$) and $q^2=0$. For convenience, we set $q^+=q^i=0$ and $q^-\neq 0$ from the start, so that 
$q{\cdot}  X = {-}q^+X^-{-}q^-X^+{+}q_iX^i = {-}q^-X^+$ with 
\begin{equation}
X^+={1\over \sqrt{2}}(X^0+X^{D{-}1})= x^+ + 2\alpha' p^+ \tau \quad .
\end{equation}
Computing the OPE and imposing 
 $ 2\alpha'{{p}}{\cdot}  q=1$
 with ${{p}}^\mu$ the zero-mode of the momentum operator, 
one finds the commutators\footnote{Note that $2\alpha'{{p}}{\cdot}  q\approx 1$ is `central' in that it commutes with $A^i_n$.}
\begin{equation}\label{DDFcomm}
[A^i_{n},A^j_{m}]=n\,\delta^{ij}\delta_{m+n,0} \, .
\end{equation}
DDF operators, though based on a choice of light-cone, can be shown to commute with the Virasoro operators $L_n$ and reproduce covariant, BRST invariant vertex operators. For closed strings one has to double the modes $A^i_{n}\rightarrow (A^i_{n,L},A^j_{n,R})$ up to subtleties with the (generalised) momentum we will deal with momentarily.

\subsection{Classical Profiles for Coherent States}

In order to illustrate the dynamical profiles of (quasi)BPS coherent states we consider simple examples that correspond to different choices of the polarizations ${\zeta}^{\mu}_{n}$ (${\tilde\zeta}^{\mu}_{n}$), or more precisely of the parameters $\lambda_n^{\mu}$ ($\tilde{\lambda}^{\mu}_{n}$).
Closed-string coherent states satisfy
\begin{equation}
A^i_n |{\mathcal{C}}(\lambda,\tilde\lambda, p)\rangle = \lambda^i_n |{\mathcal{C}}(\lambda,p) \quad , \quad 
\tilde{A}^i_n |{\mathcal{C}}(\lambda,\tilde\lambda, p)\rangle = \tilde\lambda^i_n |{\mathcal{C}}(\lambda,\tilde\lambda,p)
\rangle 
\end{equation}
Starting from the classical string profile 
\begin{equation}
\begin{split}
X^{\mu}_{(\lambda)}(\sigma,\tau)&\simeq \sum_{n=1}\left( {\lambda^{(-)\mu}_{n,L}\over n} \cos\big(n(\sigma-\tau)\big) + {\lambda^{(+)\mu}_{n,L}\over n} \sin\big(n(\sigma-\tau)\big) \right)   + \\
&\sum_{m=1}\left( {\lambda^{(-)\mu}_{m,R}\over m} \cos\big(m(\sigma+\tau)\big) - {\lambda^{(+)\mu}_{m,R}\over m} \sin\big(m(\sigma+\tau)\big) \right)\,,
\end{split}
\end{equation}
with $\lambda_{n,L,(R)}^{(\pm)}=\lambda_{n,L(R)}\pm \lambda^*_{n,L(R)} $ real polarizations, where the mass formula for a BPS state reads
\begin{equation}
{\alpha'\over 4}{{\bf{P}}}_L^2={\alpha'\over 4}M^2=\langle N_R \rangle{-}1{+}{\alpha'\over 4}{{\bf{P}}}_R^2\,, \quad {\rm with} \quad \langle N_R\rangle=\sum_{n=1}\lambda_{n,R}{\cdot}\lambda_{n,R}^*\,,
\end{equation}
while for a (quasi)BPS state one has
\begin{equation}
{\alpha'\over 4}{{\bf{P}}}_L^2{+}\langle N_L \rangle={\alpha'\over 4}M^2=\langle N_R \rangle{-}1{+}{\alpha'\over 4}{{\bf{P}}}_R^2\,, \quad {\rm with} \quad  \langle N_L\rangle=\sum_{n=1}\lambda_{n,L}{\cdot}\lambda_{n,L}^*\,.
\end{equation}
 
\begin{figure}[h!]
\center{
\includegraphics[scale=0.3]{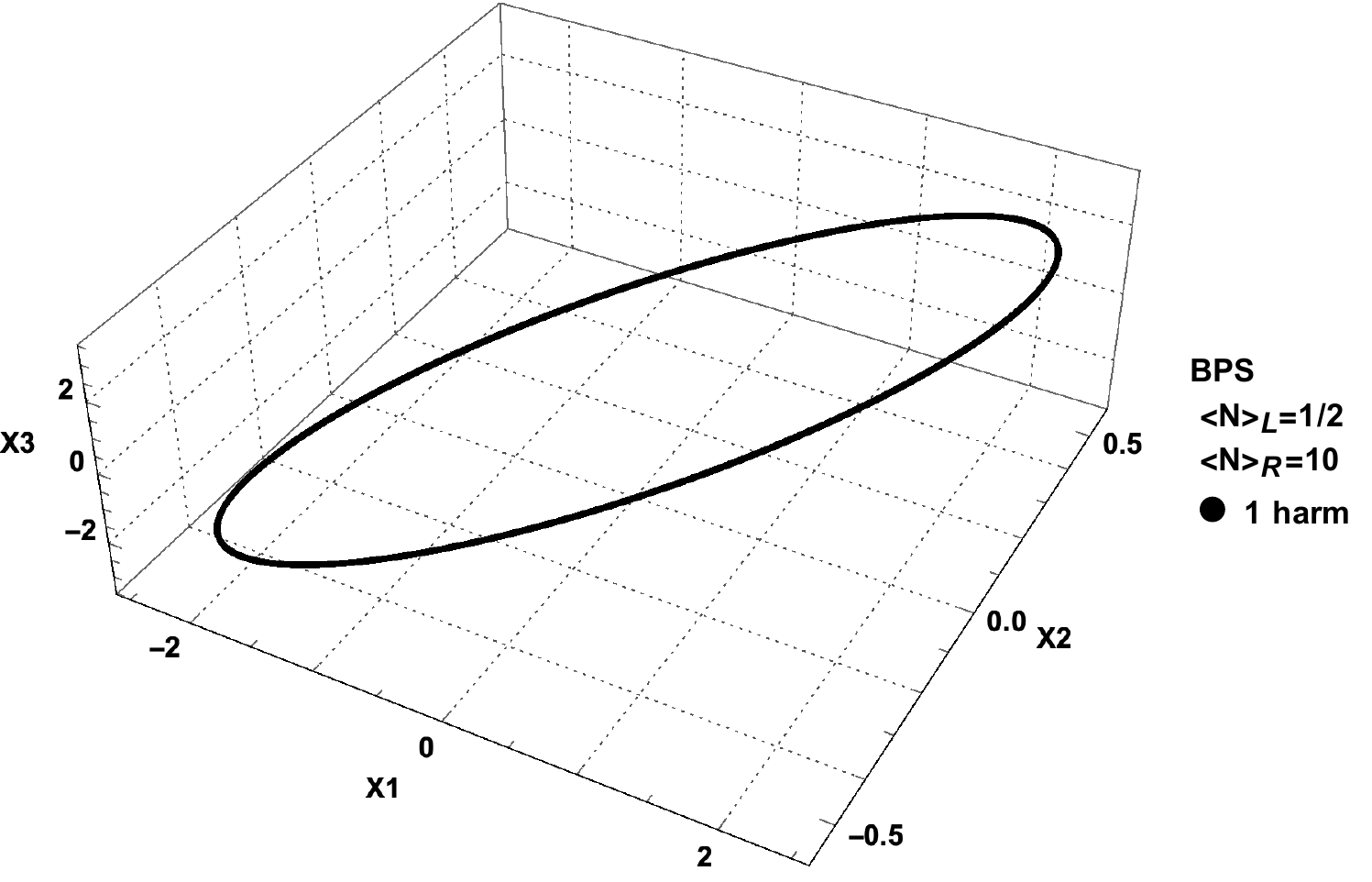}
\includegraphics[scale=0.3]{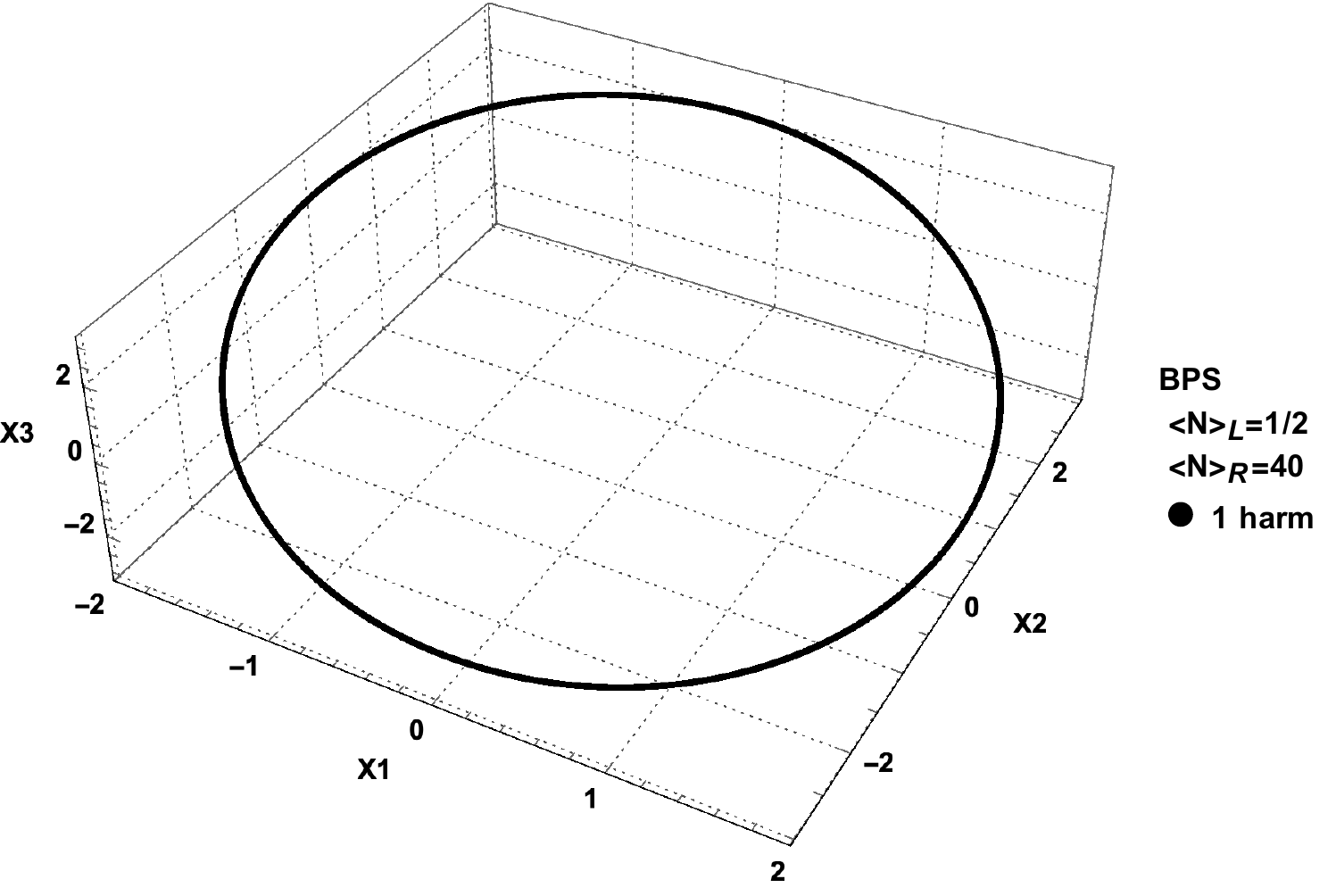}

\includegraphics[scale=0.3]{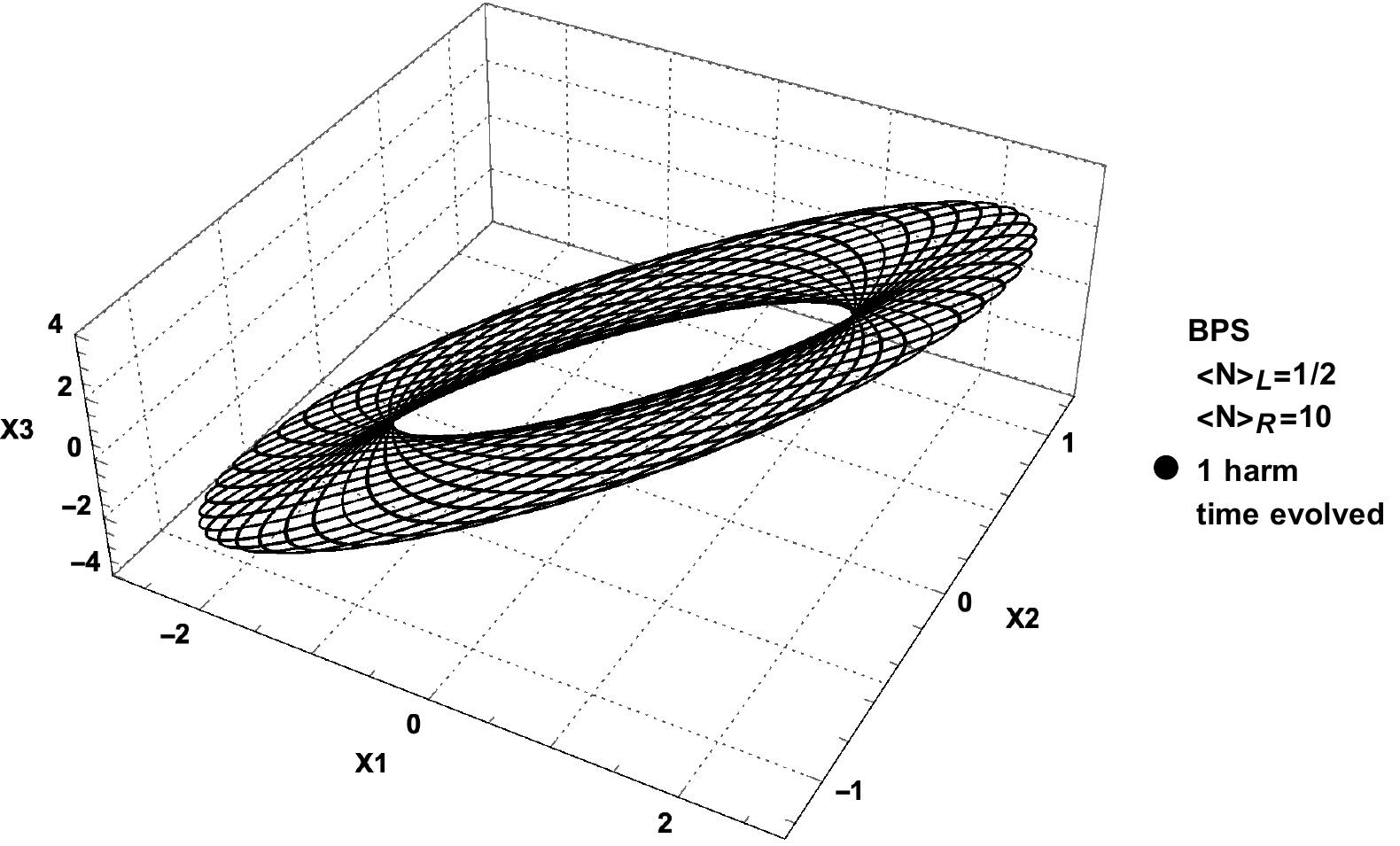}
\includegraphics[scale=0.3]{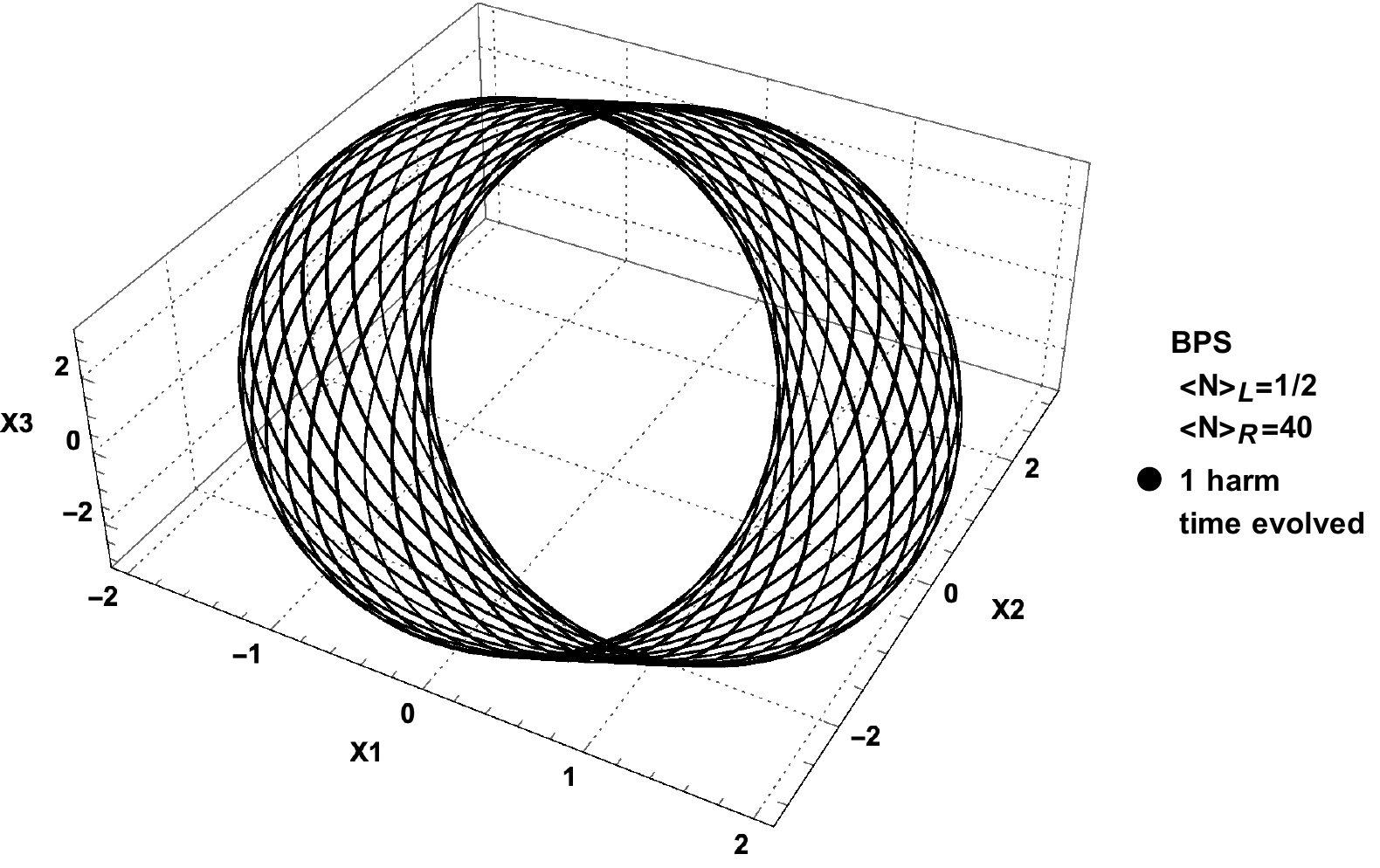}
}
\caption{BPS profiles with only one harmonic and their evolution in time.}
\label{BPS1Fig}
\end{figure}

\begin{figure}[h!]
\center{
\includegraphics[scale=0.3]{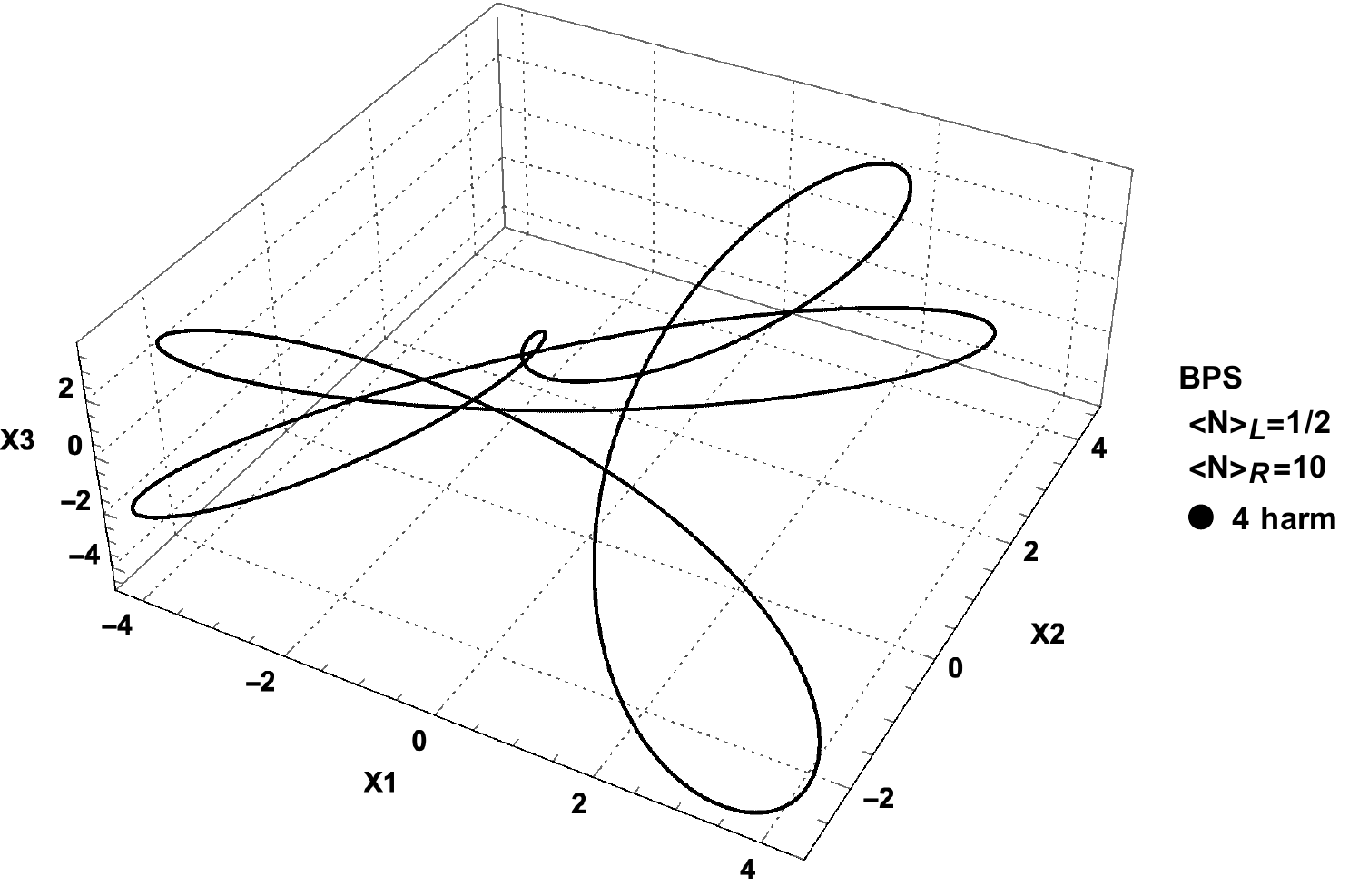}
\includegraphics[scale=0.3]{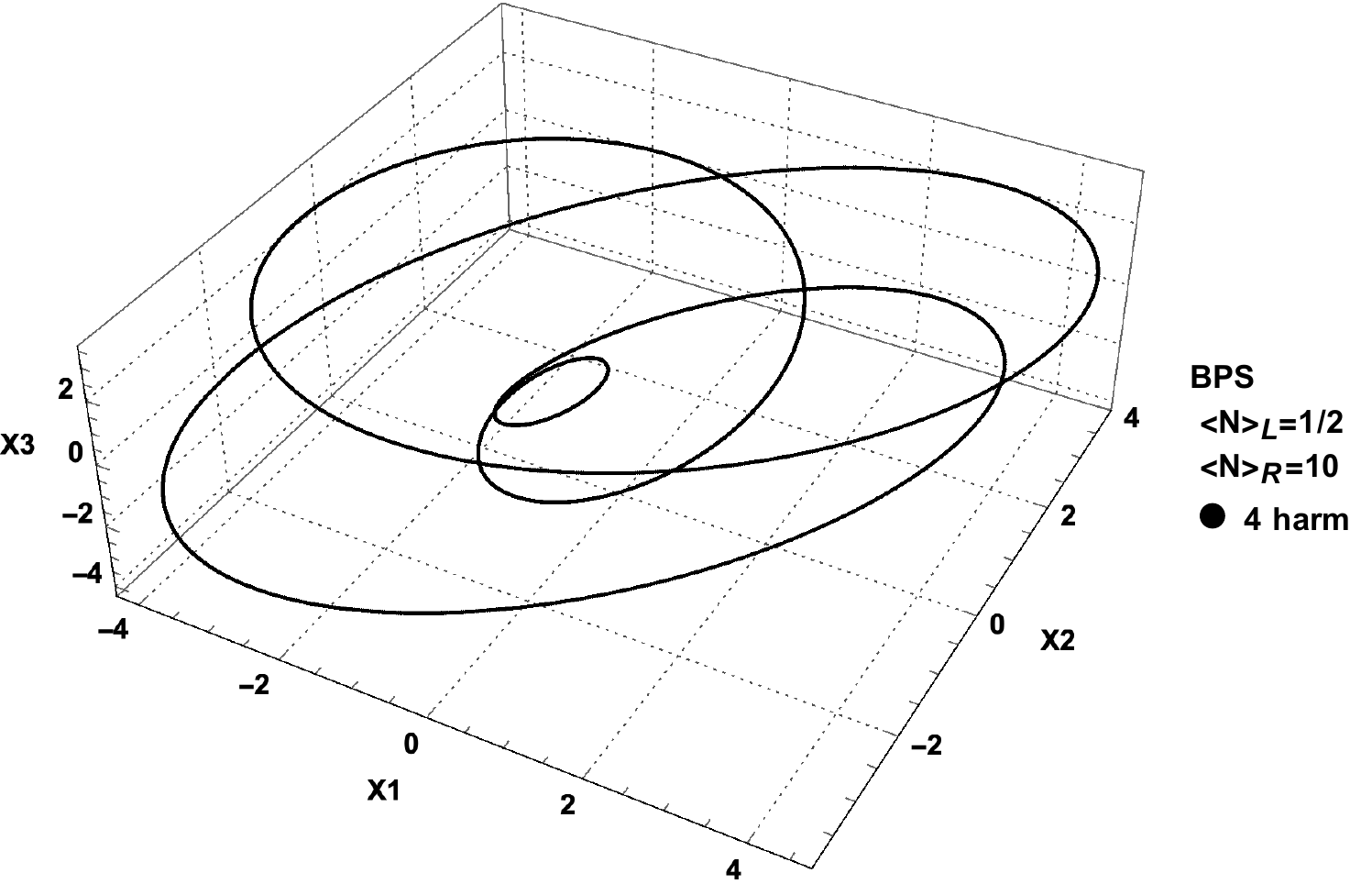}
\includegraphics[scale=0.3]{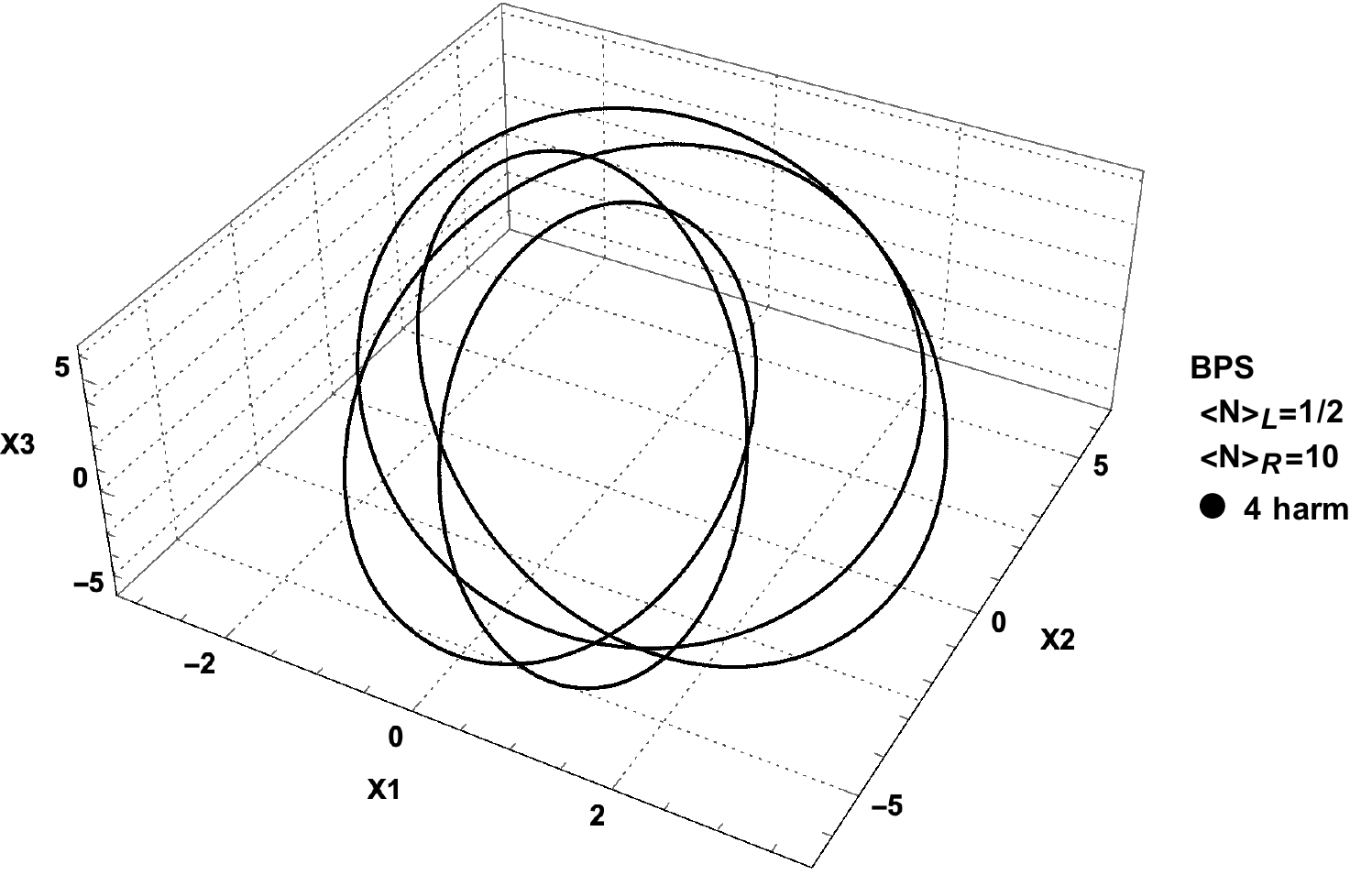}
\includegraphics[scale=0.3]{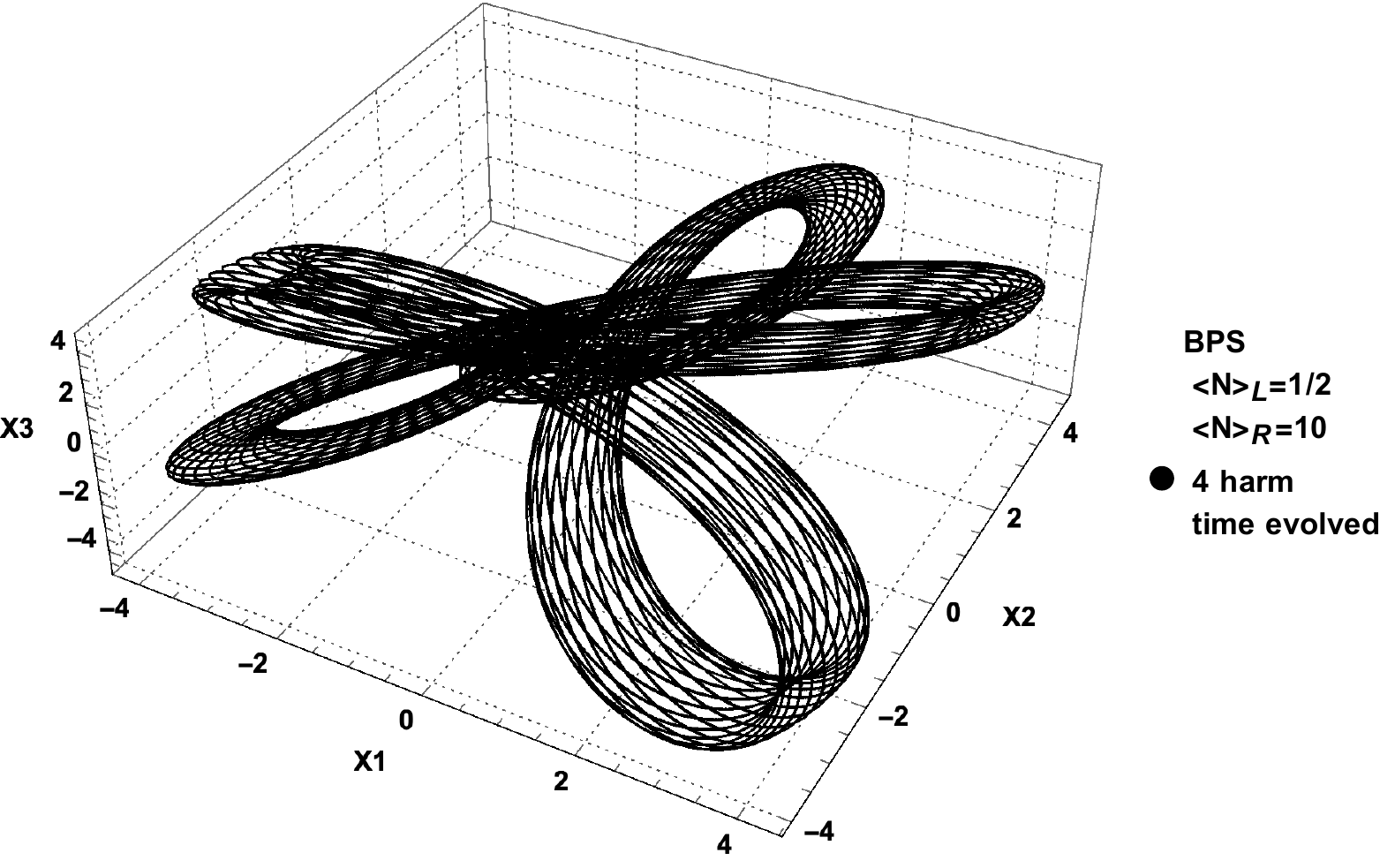}
\includegraphics[scale=0.3]{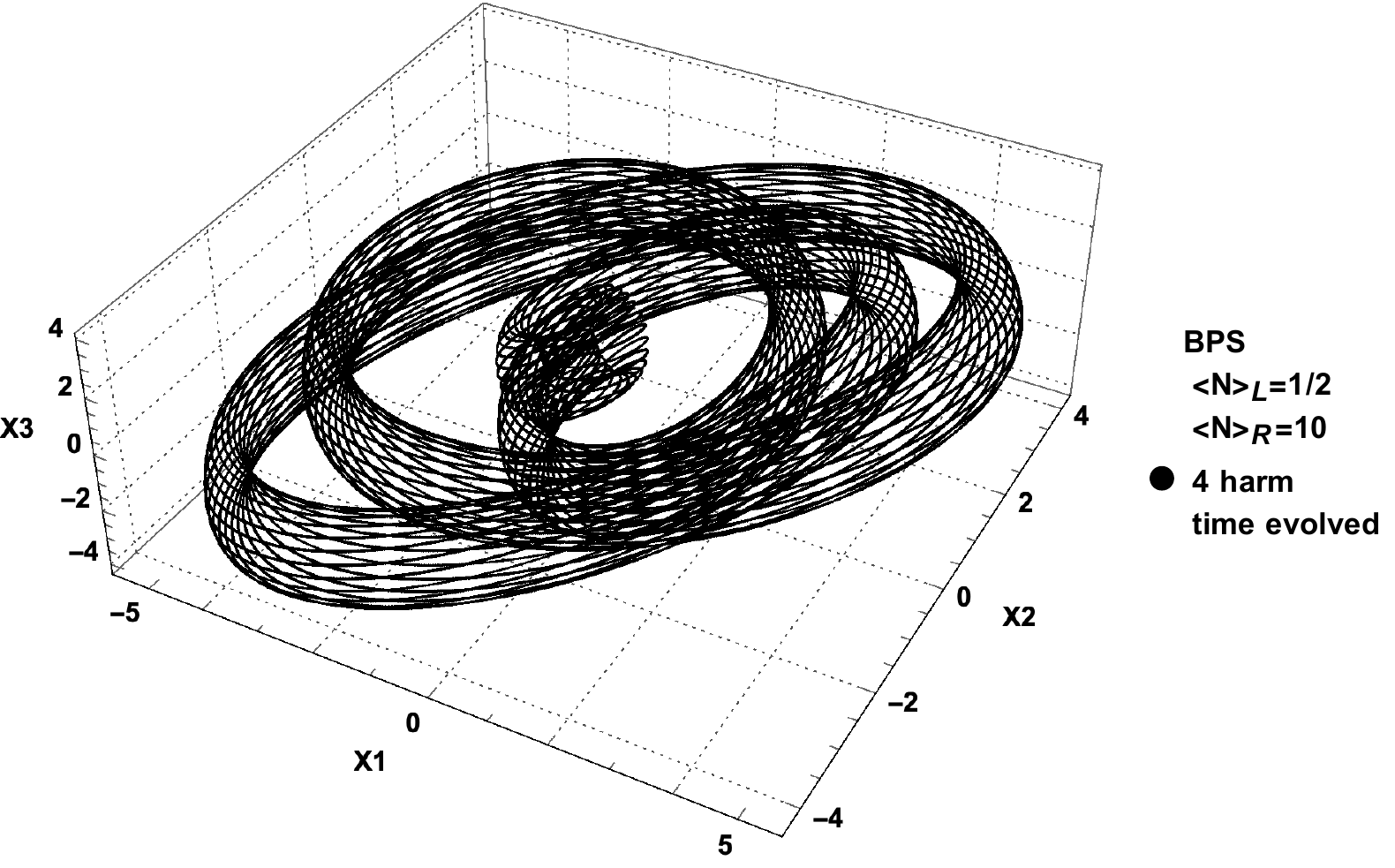}
\includegraphics[scale=0.3]{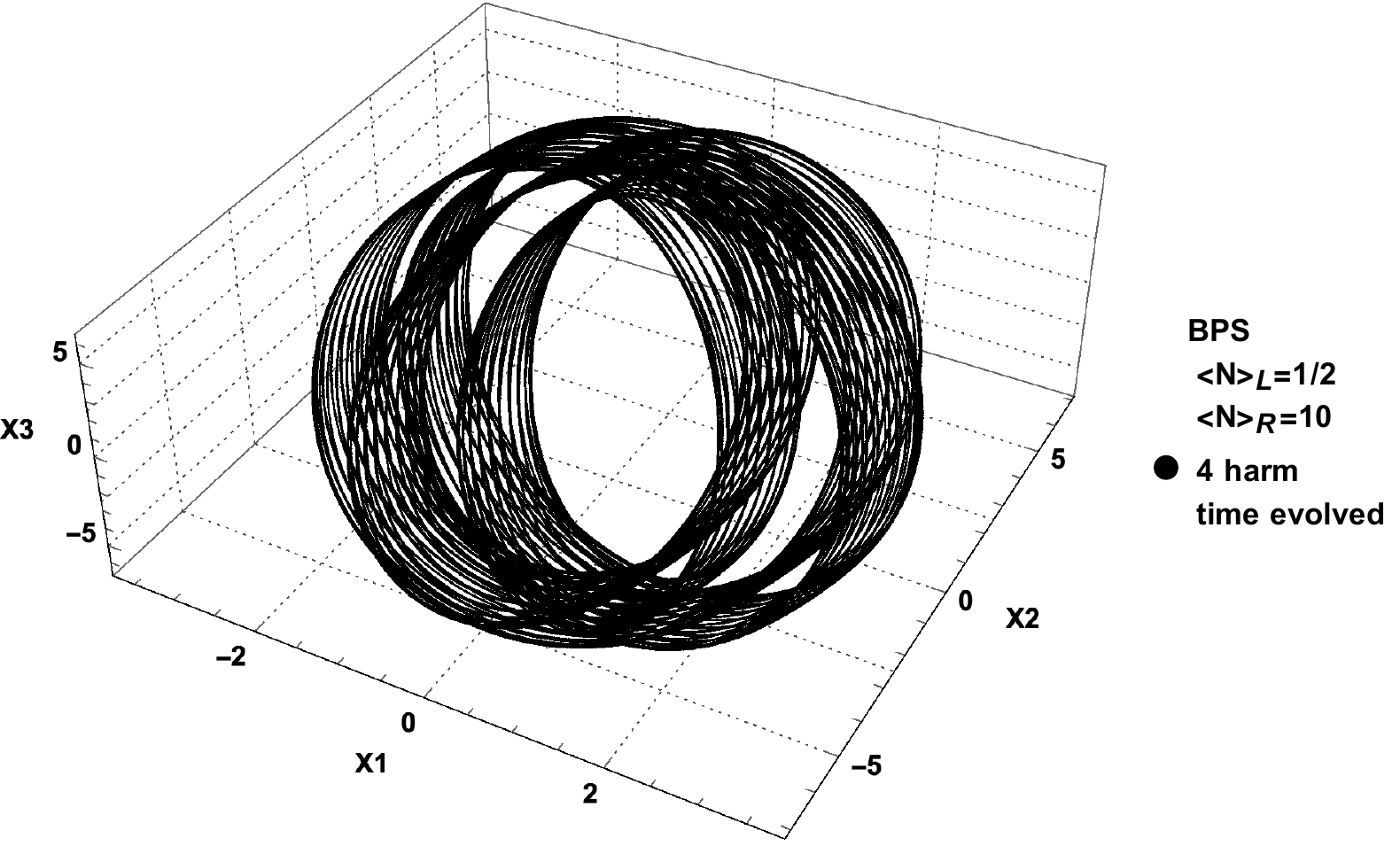}
}
\caption{BPS profiles with four harmonics and their evolution in time. The three different profiles differ for the values of the parameters $\alpha$ and $\beta$. In particular the distribution is $\alpha=1$ and $\beta=2,4,8$ respectively.}
\label{BPSFig}
\end{figure}

\begin{figure}[h!]
\center{
\includegraphics[scale=0.3]{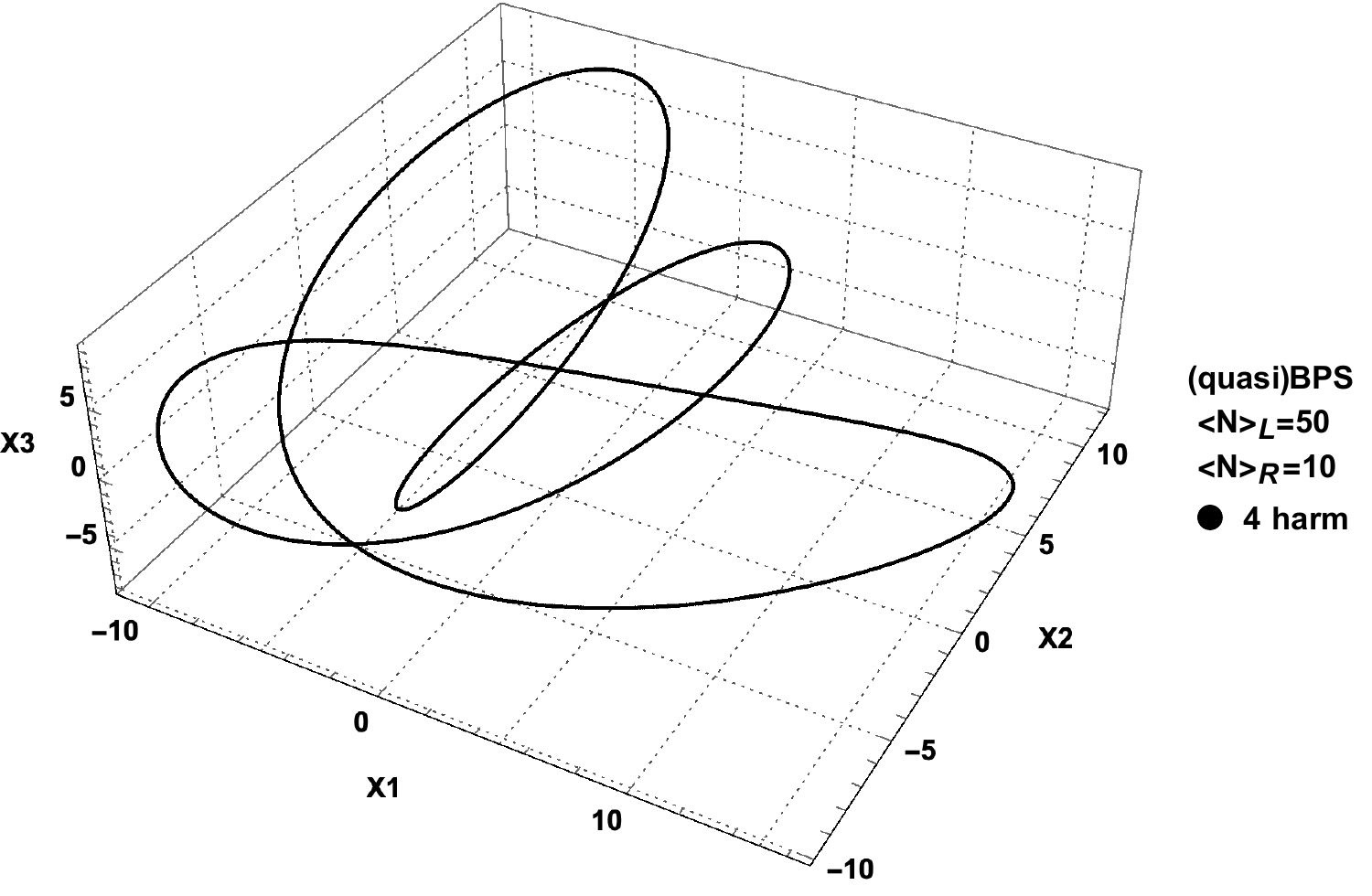}
\includegraphics[scale=0.3]{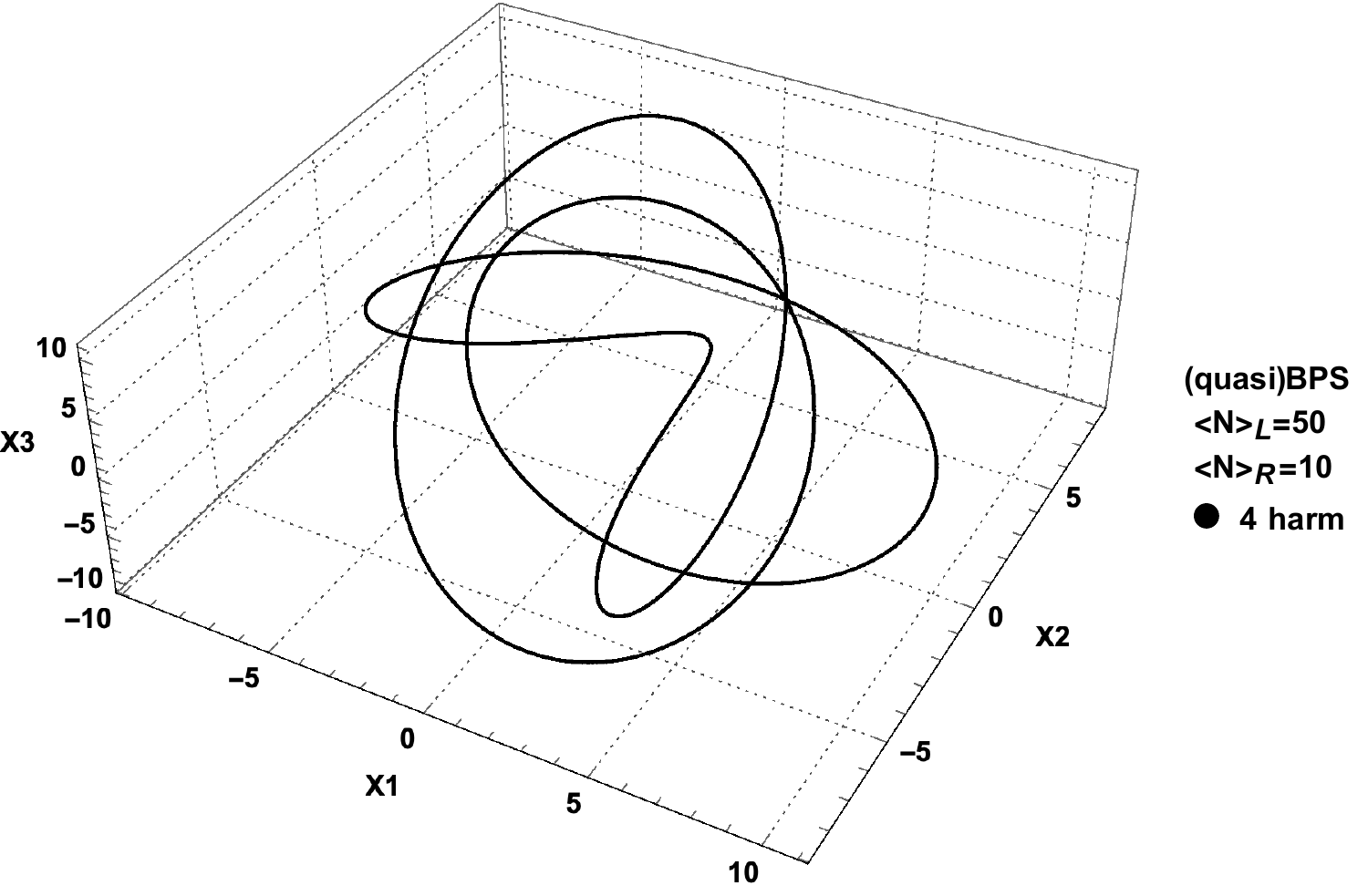}
\includegraphics[scale=0.3]{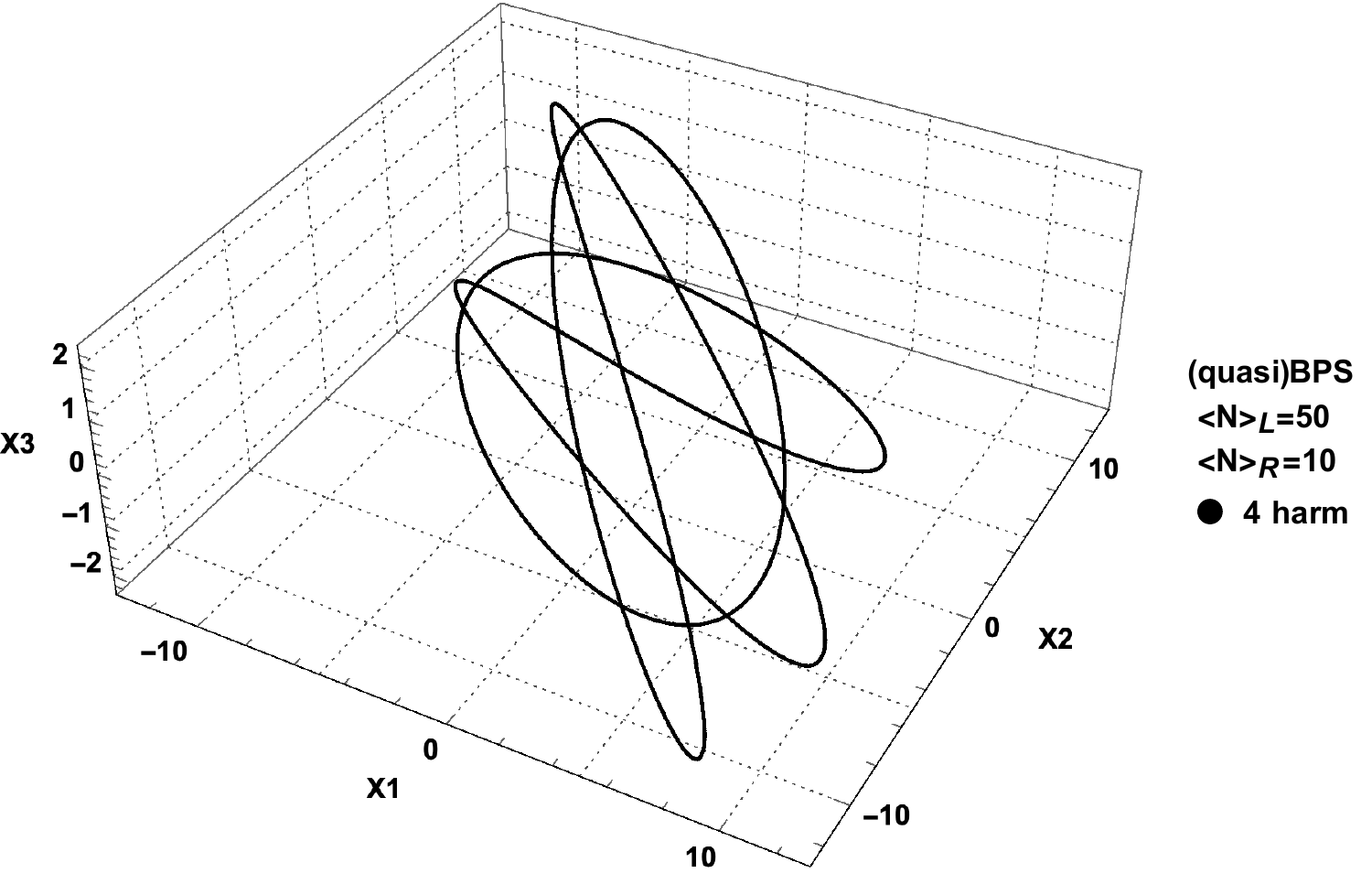}
\includegraphics[scale=0.3]{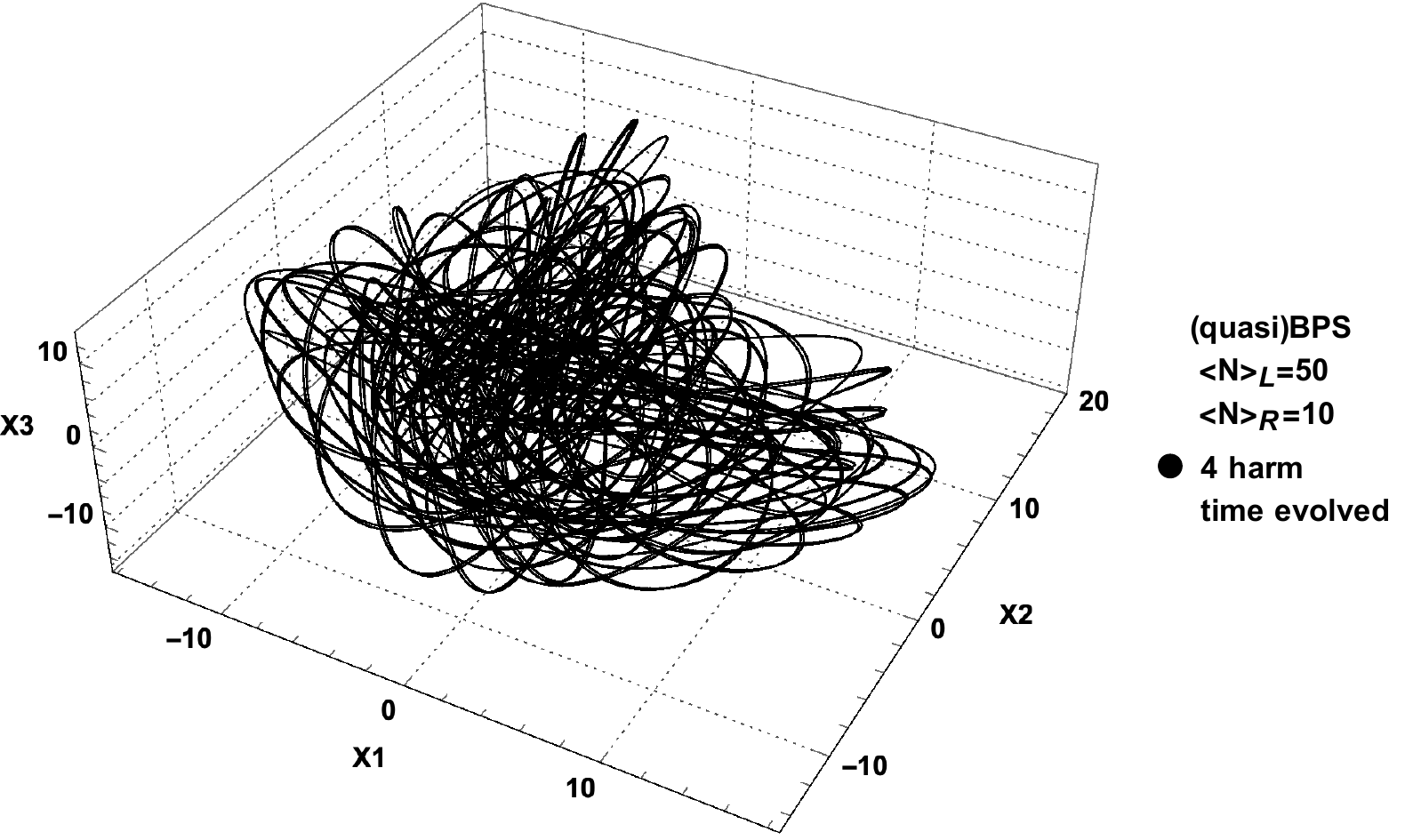}
\includegraphics[scale=0.3]{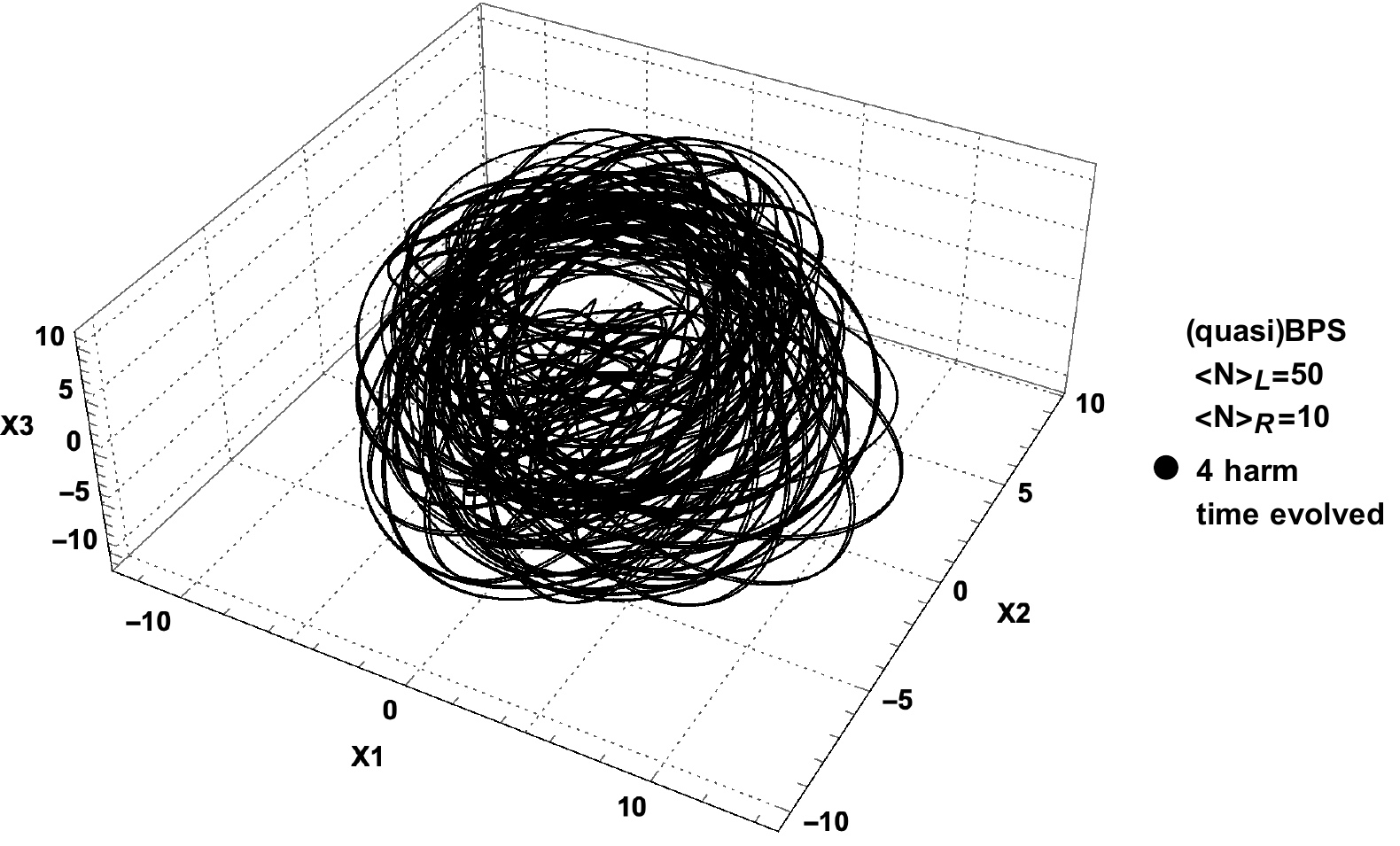}
\includegraphics[scale=0.3]{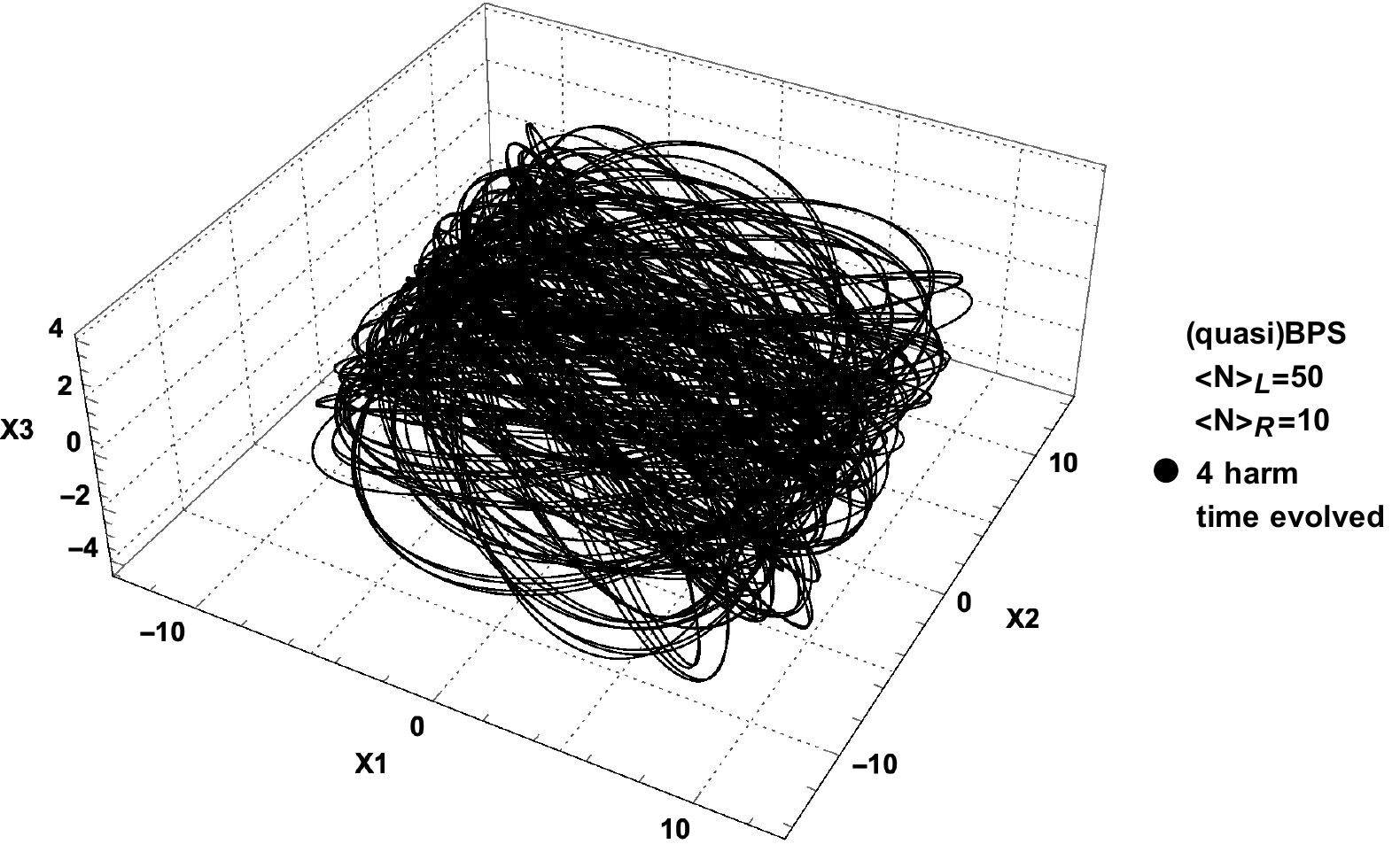}
}
\caption{(quasi)BPS profiles with four harmonics and their evolution in time. The three different profiles differ for the values of the parameters $\alpha$ and $\beta$. In particular the distribution is $\alpha=1$ and $\beta=2,4,8$ respectively.}
\label{NBPSFig}
\end{figure}

Using a simple ansatz for the coherent state polarizations  of the form $\lambda^{\mu}_n=V^{\mu}  \,e^{-\alpha n} n^{\beta}$ where $\alpha$ and $\beta$ are two free parameters, and $V^\mu$ a (null) vector the BPS and (quasi)BPS states, provided with a coherent structure, display three-dimensional profiles as the ones displayed in the plots in Fig.~\ref{BPS1Fig}, \ref{BPSFig} and \ref{NBPSFig}.

\subsection{Generalized momentum}

In order to write down vertex operators for coherent states, we start by fixing the conserved charges ${{{\bf{P}}}}_L$ and ${{{\bf{P}}}}_R$ of each massive state, and choose the reference null momentum $q$ of all states to have only the $q^-$ component \ie
 $q^+=q^I=q^i=0$, where $q^\pm= {1\over \sqrt{2}}(q^0{\pm}q^3)$ and $I=1,2$ --- space-time $(x,y)$ ---  while $i=1,...6$ (internal).
\be
p=(p^+, p^-, p^I; 0)\,, \qquad  p=(p^+, p^--{N_R{-}1\over p^+}, p^I; 0)\,,
\ee
with 
\be
-2p^+p^-+ |p^I|^2 = 2 - {{{\bf{P}}}}_L^2 \,, \qquad i.e. \qquad p^- = { |p^I|^2+ {\bf P}_L^2 -2 \over 2 p^+}\,.
\ee
The full 10-d momenta read 
\be
K_L=(p^+, {|p^I|^2+ {{{\bf{P}}}}_L^2\over 2 p^+} , p^I; {{{\bf{P}}}}_L) \, , \quad \quad K_R=(p^+, {|p^I|^2+ {{{\bf{P}}}}_L^2 \over 2 p^+} , p^I; {{{\bf{P}}}}_R)\,,
\ee
so that $K_L^2=0$, while $K_R^2={{{\bf{P}}}}_R^2{-}{{{\bf{P}}}}_L^2 = 2{-}2N_R$, as desired. Notice that the two momenta differ only in the internal part.

For quasi (or non-)BPS states, $K_L^2 ={-}2N_L' $, where $N_L'$ is the excess with respect to the BPS ground-state $N_L=\delta_L$. In this case, the momenta read
\be
K_L'=(p^+, {|p^I|^2+ {{{\bf{P}}}}_L^2+2N_L' \over 2 p^+} , p^I; {{\bf{P}}}_L)\, ,  \quad \quad 
K_R=(p^+, {|p^I|^2+ {{\bf{P}}}_R^2 +2N_R-2 \over 2 p^+} , p^I; {{\bf{P}}}_R)\,,
\ee
with ${{\bf{P}}}_L^2+ 2N_L' = {{\bf{P}}}_R^2 -2+2N_R$ such that $K^\mu_L =K^\mu_R$ along the non-compact space-time directions. We should anyway keep in mind that $K_L' = \widehat{K}_L{-}(N_L'{+}1) q$, and
$K_R = \widehat{K}_R{-}N_R q$ with the tachyonic momenta $\widehat{K}_L\neq \widehat{K}_R$ yet $\widehat{K}_L^2 =2  = \widehat{K}_R^2$. 

After this longish kinematic preamble, that should clarify issues on DLCQ raised in \cite{Hindmarsh:2010if, Skliros:2011si, Skliros:2016fqs}, for closed-string coherent states, we may proceed writing down vertex operators for the BPS or quasi-BPS BH-like coherent states using DDF operators \cite{DelGiudice:1971yjh, Ademollo:1974kz}. 

\subsection{Vertex operators}

Recalling that $K_L=\widehat{K}_L-q$ is null in $D=10$, for BPS `coherent' states we may choose
\be
W^{BPS}_C = e^{-\varphi} \zeta_L\Psi_L e^{\imath K_LX_L}  \exp\left\{\sum_{n}^{1,\infty} {\tilde\zeta_n\widetilde{\cal P}_n \over n} e^{-\imath nqX_R} +
\sum_{r,s}^{1,\infty} {\tilde\zeta_r\tilde\zeta_s \over 2rs} \widetilde{\cal S}_{r,s} e^{-\imath(r+s)qX_R} \right\} 
e^{\imath \widehat{K}_RX_R}
\ee
where the level mathcing imposes $N_R= {1\over 2} ({{\bf{P}}}_L^2-{{\bf{P}}}_R^2) + 1$ and with polarizations $\tilde\zeta_n^\alpha = \tilde\lambda^A_n(\delta^\alpha_A{-}q^\alpha \widehat{K}_A)$, $\alpha= 0, ... 25$ and $A=1,...24$ (bosonic string sector). The operatoratorial structures that appear explicitly read
\be
\widetilde{\cal P}_n^A = \sum_{h=1}^n {\imath \bar\partial X_R^A \over (h{-}1)!} {\cal Z}_{n{-}h}[{\cal U}]
\, , \quad  \widetilde{\cal S}_{n,m} = \sum_{h=1}^n h {\cal Z}_{n{+}h}[{\cal U}] {\cal Z}_{m{-}h}[{\cal U}]\,, \quad {\cal U}_{\ell,R}^{(n)} =  {-\imath \,n   q^-\bar\partial^\ell X^+_R \over (\ell{-}1)!} 
\ee
while ${\cal Z}_{n}[u_\ell] = \sum_{n_\ell: \sum_\ell \ell n_\ell = n} \prod_{\ell=1}^n {u^{n_\ell}_\ell\over n_\ell !\ell^{{n_{\ell}}}} $ are the cycle index polynomials 
\be
{\cal Z}_0=1 \,,  \quad \quad {\cal Z}_1 = u_1 \, ,  \quad \quad {\cal Z}_2 = {u_2\over 2} + {u_1^2\over 2} \, , \quad \quad 
 {\cal Z}_3 = {u_3\over 3} + {u_1 u_2\over 2} + {u_1^3\over 6} \ \  ...\,\,.
 \ee

For simplicty, and without much loss of generality, the non{-}BPS `coherent' states we consider have only `bosonic' excitations on the BPS ground-states. Modulo subtleties, addressed in \cite{Aldi:2019osr}, one has
\begin{equation}
\begin{split}
 W^{\rm non}_C =\int_0^{2\pi} {d\beta\over 2\pi}
e^{-\varphi} \zeta_L\Psi_L &\exp\left\{\sum_{n}^{1,\infty} {\zeta_n{\cal P}_n \over n} e^{-\imath n(qX_L{-}\beta)} {+}
\sum_{r,s}^{1,\infty} {\zeta_r\zeta_s \over 2rs}{\cal S}_{r,s} e^{-\imath (r{+}s)(qX_L{-}\beta)} \right\} 
e^{\imath {K}_LX_L{+}\beta} \\
&\exp\left\{\sum_{m}^{1,\infty} {\widetilde{\zeta}_m\widetilde{{\cal P}}_m \over m} e^{-\imath m(qX_R{-}\beta)} {+}
\sum_{\ell,f}^{1,\infty} {\widetilde{\zeta}_\ell\widetilde{\zeta}_f \over 2\ell f}\widetilde{{\cal S}}_{\ell,f} e^{-\imath (\ell{+}f)(qX_R{-}\beta)} \right\} 
e^{\imath {K}_RX_R{+}\beta}
\end{split}
\end{equation}
Integration over $\beta$ implements level-matching and $\zeta^\mu_n = \lambda^i (\delta^\mu_i{-}q^\mu p_i)$, ${\cal P}_n$, ${\cal S}_{r,s}$, ${\cal U}$ as for the R-movers, with $\bar\partial^\ell X_R$ replaced by $\partial^\ell  X_L$. 

Whether a coherent state is compact or not  depends on the choice of the parameters 
$\lambda^\mu_n$ and $\tilde\lambda^\mu_n$ that determine the mass $M_a$ and gyration radius $R_a$ 
\begin{equation}
M^2_a={1\over \alpha'} \sum_{n=1}^{\infty} {|\lambda^{(a)}_n|^2 } = {1\over \alpha'} \sum_{n=1}^{\infty} {|\tilde\lambda^{(a)}_n|^2 } \,,\quad R^2_a = \alpha' \sum_{n=1}^{\infty} {|\lambda^{(a)}_n|^2 \over n^2} = \alpha' \sum_{n=1}^{\infty} {|\tilde\lambda^{(a)}_n|^2 \over n^2} \: , 
\end{equation}
where the $\tilde\lambda^\mu_n$ are constrained  by level-matching.
We would like to have $R_a\simeq G_N M_a$, where $G_N= \alpha' {g_s^2 /\widehat{V}_{(6)}}$ is the 4-d Newton constant and $\widehat{V}_{(6)}= {V_{(6)}\over 64 \pi  (\alpha')^3  }$ is the adimensional compactification volume.  This means that the parameters 
$\lambda^\mu_n$ and $\tilde\lambda^\mu_n$ should satisfy 
\begin{equation}
{R_a\over M_a}= G_N \quad \leftrightarrow \quad 
\alpha' \, \left({\sum_{n=1}^{\infty} {|\lambda^{(a)}_n|^2 \over n^2} \over \sum_{n=1}^{\infty} {|\lambda^{(a)}_n|^2 }}\right)^{1/2}=\alpha' {g_s^2 \over \widehat{V}_{(6)}}  
\end{equation}
or 
\begin{equation}\label{perturb condition}
\left({\sum_{n=1}^{\infty} {|\lambda^{(a)}_n|^2 \over n^2} \over \sum_{n=1}^{\infty} {|\lambda^{(a)}_n|^2 }}\right)^{1/2} = {g_s^2\over \widehat{V}_{(6)} } \approx 10^{-4\div -3} << 1
\end{equation}
in order to fullfil the requirement on the compactness of the stringy BH's involved in the scattering process in the perturbative regime $g_s<<1$. Among a variety of choices, a possible ansatz for the coherent state polarization is as above (see Figs.~\ref{BPS1Fig}, \ref{BPSFig}, \ref{NBPSFig})
\begin{equation}
\lambda^{\mu}_n=V^{\mu}  \,e^{-\alpha n} n^{\beta}
\end{equation}
with $V^\mu$ a (null) vector and $\alpha$ and $\beta$ two tuneable parameters, the condition (\ref{perturb condition}) leads to
\begin{equation}
\left({Li_{2{-}2\beta}(e^{-2\alpha}) \over Li_{{-}2\beta}(e^{-2\alpha})}\right)^{1/2} = {g_s^2\over \widehat{V}_{(6)} } = {G_N \over \alpha'}
\end{equation}
with $Li_n(x)$ the polylog function. In the extremely simple case in which $\beta=1$ one has
\begin{equation}
G_N=\alpha' \sqrt{2}\sinh(\alpha)\big(1{-}\tanh(\alpha) \big)^{1/2}
\end{equation}
\begin{figure}[h!]
\center
\includegraphics[scale=0.4]{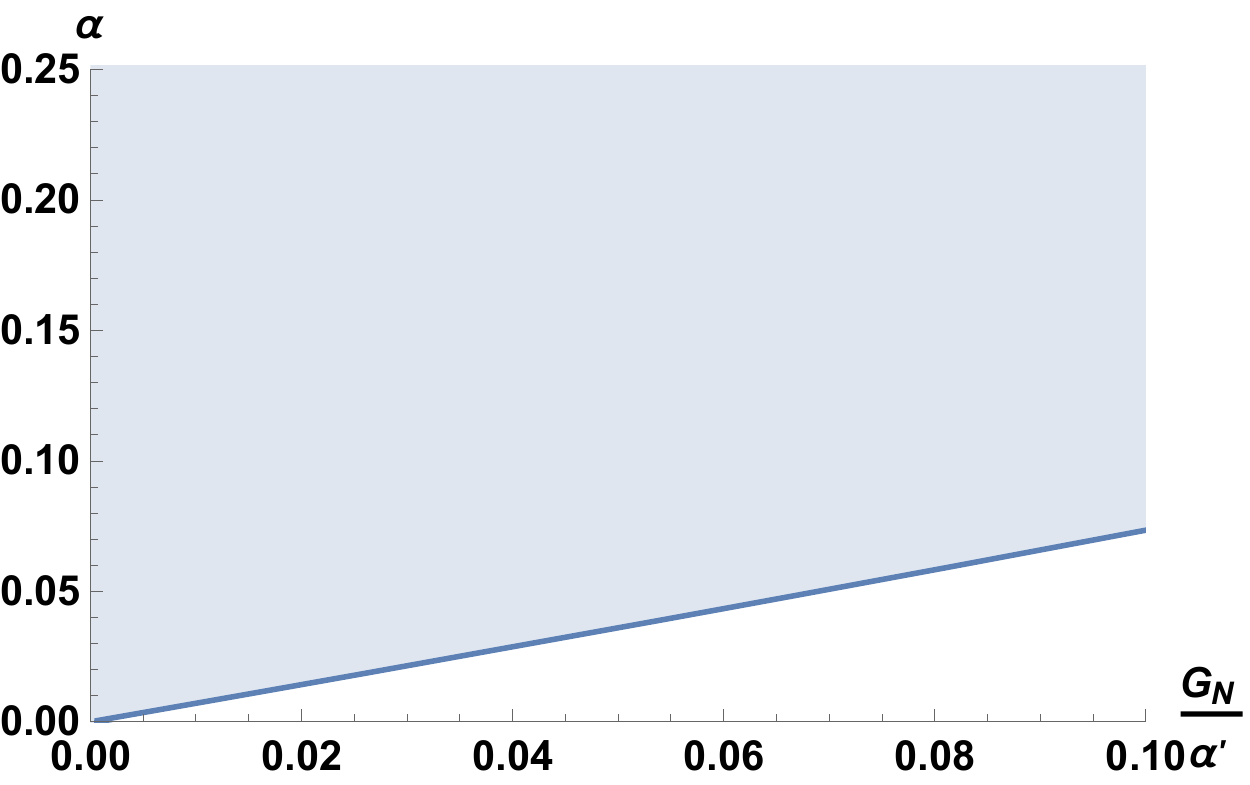}
\caption{The behavior of the parameter $\alpha$ of the harmonics' distribution  as a function of ${G_N\over\alpha'}$.  }
\end{figure}
giving the possibility to tune $\alpha'$ and the parameter (or in general the parameters), associated to the distribution of the coherent state harmonics. Notice that, even if this condition can be satisfied, the profound physical reason why the state is compact and behaves like a (small) BH is not completely obvious \cite{Horowitz:1996nw, Maldacena:1996ds, Damour:1999aw, Chialva:2009pf, Brustein:2016msz, Mathur:2005zp}. 

\subsection{Interactions}

In \cite{Bianchi:2019ywd}, amplitudes with coherent states for open bosonic string interacting with massless vector bosons were shown to expose the expected soft factor at tree level. Very much as for the amplitudes with heterotic string mass eigenstates studied in Section \ref{StringMemoSect}, we expect the amplitudes with heterotic string coherent interacting with graviton to expose Weinberg's soft factor and a more involved Shapiro-Virasoro dressing of the 3-point amplitude for the coherent states. To this purpose, we need the (non-vanishing on-shell) 3-point amplitude of coherent states. 

For the R-movers (bosonic string) one can borrow the result from the open bosonic string \cite{Bianchi:2019ywd}, dropping the integrations over $X_0$ and $\beta$ one has
\begin{equation}
\begin{split}
&{\cal A}^R_3(\tilde\zeta^{(1)}_{n_1},K_1;\tilde\zeta^{(2)}_{n_2}, K_2; \tilde\zeta^{(3)}_{n_3}, K_3) =  
\exp\left\{\sum_{a<b}^{1,3}\sum_{n_a,n_b}\tilde\zeta^{(a)}_{n_a}\tilde\zeta^{(b)}_{n_b}
{\cal L}_{n_a{-}1,n_b{-}1}(Q_a,Q_b)\right\} \\
&\hspace*{4cm}\exp\Bigg\{\sum_{a=1}^3 \Bigg[\sum_{n_a}\tilde\zeta^{(a)}_{n_a}(K_{a{+}1}{-}K_{a{-}1})
{\cal R}_{n_a{-}1}(Q_a)\Bigg\} \\
&\hspace*{2cm}\exp\Bigg\{\sum_{a=1}^3\sum_{r_a,s_a}\tilde\zeta^{(a)}_{r_a}\tilde\zeta^{(a)}_{s_a}{r_a s_a(Q_a^2{-}1)\over 2(r_a{+}s_a)}
{\cal R}_{r_a{-}1}(Q_a){\cal R}_{s_a{-}1}(Q_a)\Bigg]\Bigg\}
\end{split}
\end{equation}
where
\be
Q_a= q_a(K_{a{+}1}{-}K_{a{-}1}) \,, \quad {\cal R}_{n{-}1}(Q)= {(-)^n \over 2\, n!} {\Gamma\left( {n\over 2}(Q{+}1)\right) \over 
\Gamma\left( {n\over 2}(Q{-}1){+}1\right)}
\ee
and 
\be
\!\!\!\! \!\!\!\! \!\!\!\! \!\!\!\!\!\!\!\!\!\!\!\! \!\!\!\!\!\!\!\!\!\!\!\!\!\!\!\! {\cal L}_{n_1{-}1,n_2{-}1}(Q_1,Q_2) = {(-)^{n_1{+}1} \over n_1n_2} \sum_{h_1=1}^{n_1}\sum_{h_2=1}^{n_2}
{\Gamma(h_1+h_2) \over \Gamma(h_1)\Gamma(h_2)} \nonumber\\
\ee
\be
 \qquad \qquad \qquad \qquad \qquad \qquad {\cal Z}_{n_1{-}h_1}\left( {n_1\over 2}(Q_1{-}1)\right) {\cal Z}_{n_2{-}h_2}\left( {n_2\over 2}(Q_2{-}1)\right)\,.
\ee
The expression drastically simplifies when $\tilde\zeta^{(a)}_{r_a}\tilde\zeta^{(a)}_{s_a} = 0$ (for instance, if $\tilde\zeta^{(a)\mu}_{r_a} = C_{r_a} V^\mu_a$, with $V^\mu_a$ some null vector and $C_{r_a}$ `arbitrary' constants) and even more so if $\tilde\zeta^{(a)}_{r_a}\tilde\zeta^{(b)}_{s_b} = 0$ (as before with $V_a{\cdot}V_b=0$). States in the first Regge trajectory correspond to considering only $\tilde\zeta^{(a)}_1\neq 0$ and null.

For the L-movers (superstring), under the assumption that we consider only bosonic excitations over the BPS ground state and modulo some subtleties \cite{Aldi:2019osr}, we can also borrow from the open strings. Dropping the integrations over $X_0$ and $\beta$, the relevant amplitude is 
$$
{\cal A}^L_3(A_1,K_1;A_2,K_2;\zeta^{(3)}_{n},K_3) = \left\{\sum_n A_1\zeta^{(3)}_{n}{\cal M}_{n{-}1}(Q)+
\sum_m A_2\zeta^{(3)}_{m}{\cal M}_{m{-}1}(Q) + \right.
$$ 
\be 
\left. +A_1A_2 - A_1K_2\,A_2K_1+\sum_n [A_1K_2 \zeta^{(3)}_{n}A_2- A_2K_1 \zeta^{(3)}_{n}A_1]
{\cal M}_{n{-}1}(Q)\right\}
\ee
$$
\exp\left\{\sum_{n}\zeta^{(3)}_{n}(K_{1}{-}K_{2})
{\cal R}_{n{-}1}(Q) +\sum_{r,s}\zeta^{(3)}_{r}\zeta^{(3)}_{s}{rs(Q^2{-}1)\over 2(r{+}s)}
{\cal R}_{r{-}1}(Q){\cal R}_{s{-}1}(Q)\right\} \,,
$$
where $Q=q_3(K_1{-}K_2)$ and 
\be
{\cal M}_{n{-}1}(Q)=  {(-)^{n{+}1} \over n!} {\Gamma\left( {n\over 2}(Q{+}1){+}1\right) \over 
\Gamma\left( {n\over 2}(Q{-}1){+}2\right)} =  - {2n(Q{+}1) \over n(Q{-}1){+}2}
 {\cal R}_{n{-}1}(Q)
\ee
As above, there are major simplifications if $\zeta^{(3)}_{n}\zeta^{(3)}_{m}=0$ and/or $A_{a}\zeta^{(3)}_{n}=0$ for $a=1,2$. Combining the two expressions one gets the complete closed-string 3-point amplitude
\be \label{3pt_coherent}
\begin{split}
&{\cal M}_3(A_1,\tilde\zeta^{(1)}_{n_1},K_1;A_2,\tilde\zeta^{(2)}_{n_2}, K_2;\zeta^{(3)}_{n}, \tilde\zeta^{(3)}_{n_3}, K_3) = \int_0^{2\pi} \prod_a {d\beta_a\over 2\pi} \int d^4X_0 e^{\imath \sum_ap_aX_0}\\
&\quad \quad{\cal A}^L_3(\widehat{A}^L_1,\widehat{K}^L_1;\widehat{A}^L_2,\widehat{K}^L_2;\widehat\zeta^{(3)}_{n}, \widehat{K}^L_3)
{\cal A}^R_3(\widehat{\tilde\zeta}^{(a)}_{n_a},\widehat{K}^R_a) \prod_i \delta_{\sum_a P^i_{a,L}}
\prod_A \delta_{\sum_a P^A_{a,R}}\,\,,
\end{split}
\end{equation}
where $\widehat{A}^L_a = A^L_a e^{-\imath q_aX_0+ \imath \beta_a}$ ($a=1,2$), \, $\widehat\zeta^{(3)}_{n} = \zeta^{(3)}_{n} e^{- \imath nq_3X_0+ \imath n\beta_3}$ and $\widehat{\tilde\zeta}^{(a)}_{n} = \tilde\zeta^{(a)}_{n} e^{- \imath nq_aX_0- \imath n\beta_a}$. It is crucial to recall that $q_aq_b=0$ since the $q$'s are indeed all collinear, with only $q^-_a=-1/p_a^+\neq 0$. Moreover, the two integrations over $\beta_1$ and $\beta_2$ simply project the R-movers onto the level $N_R= 1 + {1\over 2} ({{\bf{P}}}_L^2-{{\bf{P}}}_R^2)$ for $a=1,2$ (BPS states) for level-matching. Level-matching for the non{-}BPS state gives an infinite number of states with $N_R-(N_L-\delta_L) = 1 + {1\over 2} ({{\bf{P}}}_L^2-{{\bf{P}}}_R^2)$.

It is instructive to study more explicitly the amplitude (\ref{3pt_coherent}) in the very simple case where $\zeta_n^{(3)}\zeta_n^{(3)}=0=\tilde{\zeta}_{n_{a}}^{(a)} \tilde{\zeta}_{n_{b}}^{(b)}$ and compute the level-matching integrals. In fact one can use the following relation
\begin{equation}
\begin{split}
&\int_{0}^{2\pi}{d\beta \over 2\pi} \, e^{\imath \beta k} \,e^{\sum_{n=1} e^{-\imath \beta n} f_n}= \left(\prod_{n=1}\sum_{\ell_{n}=0}^{\infty} {f_n^{\ell_{n}}\over \ell_{n}!}\right) \int_{0}^{2\pi}{d\beta \over 2\pi} \, e^{\imath \beta \left(k-\sum_{n=1} n \ell_{n}\right) }
\end{split}
\end{equation}
where implementing the $\delta$-function integral as $k-\sum_{n=2}n \ell_n=\ell_1$
one obtains the following polynomial
\begin{equation}
{\cal B}_k=\sum_{\ell_{n}: \sum_{n=2}n\ell_n\leq k}^{\infty}  {f_1^k \over \left(k{-}\sum_{n=2}n\ell_n\right)!} \prod_{n=2} {f_n^{\ell_{n}} \over f_1^{n \ell_{n}} \ell_{n}!}
\end{equation}
that for $k=1$ gives simply ${\cal B}_1=f_1$, and in general is a function of the elements $f_1,..,f_n$ i.e ${\cal B}_k={\cal B}_k(\{f_n\})$ . Following a similar strategy, starting from the integral
\begin{equation}
\int_{0}^{2\pi}{d\beta \over 2\pi} \, e^{\imath \beta m} \,e^{\sum_{n=1} e^{\imath \beta n} f_n} e^{\sum_{\bar{n}=1} \,e^{-\imath \beta \bar{n}} \bar{f}_{\bar{n}}}
\end{equation}
one obtains the following polynomial
\begin{equation}
\begin{split}
\mathfrak{B}_m=\sum_{(\ell_{n},\bar{\ell}_{\bar{n}}): \sum_{\bar{n}=1}\bar{n}\bar{\ell}_{\bar{n}} {-}\sum_{n=2}n\ell_{n}\geq m  } &{f_1^{-m} \over \left(\sum_{\bar{n}=1}\bar{n}\bar{\ell}_{\bar{n}} {-}\sum_{n=2}n\ell_{n}{-} m\right)!} 
\prod_{\bar{n}=1}\prod_{n=2} { \bar{f}_{\bar{n}}^{\bar{\ell}_{\bar{n}}}  f_1^{\bar{n} \bar{\ell}_{\bar{n}}{-}n\ell_{n}} f_n^{\ell_n}      \over \ell_{n}! \, \bar{\ell}_{\bar{n}}!  }
\end{split}
\end{equation}
where now $\mathfrak{B}_m=\mathfrak{B}_m(\{f_n\},\{\bar{f}_{\bar{n}}\})$. In terms of these two polynomials it is possible to represent (\ref{3pt_coherent}) as
\begin{equation}
\begin{split}
&{\cal M}_3(A_1,\tilde\zeta^{(1)}_{n_1},K_1;A_2,\tilde\zeta^{(2)}_{n_2}, K_2;\zeta^{(3)}_{n}, \tilde\zeta^{(3)}_{n_3}, K_3) =\int d^4X_0 \,e^{\imath \sum_ap_aX_0}  \prod_i \delta_{\sum_a P^i_{a,L}}
\prod_A \delta_{\sum_a P^A_{a,R}}\\
&\hspace{1cm}\Big\{(A_1A_2 {-} A_1K_2 A_2K_1 ) e^{-\imath(q_1{+}q_2)X_0} \bar{f}_1^{(1)}(\tilde{\zeta}_1,Q_1)\bar{f}_1^{(2)}(\tilde{\zeta}_2,Q_2) \mathfrak{B}_{0}(f_n^{(3)},\bar{f}_{n}^{(3)}) +  \\
&\hspace{1cm}+ \sum_{m=1} {\cal M}_{m{-}1}(Q) \mathfrak{B}_{m}(f_n^{(3)},\bar{f}_{n}^{(3)}) \,e^{-i(q_1{+}mq_3)X_0} A_1\zeta_m^{(3)} \bar{f}_{1}^{(1)}(\tilde{\zeta}_1,Q_1) \\
&\hspace{1cm}+\sum_{m=1} {\cal M}_{m{-}1}(Q) \mathfrak{B}_{m}(f_n^{(3)},\bar{f}_{n}^{(3)}) \,e^{-i(q_2{+}mq_3)X_0} A_2\zeta_m^{(3)}\bar{f}_{2}^{(2)}(\tilde{\zeta}_2,Q_2) \\
&\hspace{1cm}+\sum_{m=1} {\cal M}_{m{-}1}(Q) \mathfrak{B}_{m}(f_n^{(3)},\bar{f}_{n}^{(3)}) \,e^{-i(q_1{+}q_2{+}mq_3)X_0} \\
&\hspace{1cm}\hspace{1.2cm} (A_1K_2\zeta_m^{(3)}A_2{-}A_2K_1\zeta_m^{(3)}A_1)\bar{f}_{1}^{(1)}(\tilde{\zeta}_1,Q_1)\bar{f}_{2}^{(2)}(\tilde{\zeta}_2,Q_2) \Big\}
\end{split}
\end{equation}
where the arguments of the polynomials are given by
\begin{equation}
\begin{split}
&\bar{f}_r^{(a)}=e^{-\imath rq_a X_0}\tilde{\zeta}_r^{(a)}(K^R_{a{+}1}{-}K^R_{a{-}1}){\cal R}_{r{-}1}(Q_a)\,, \qquad f_n^{(3)}=e^{-\imath nq_3 X_0}\zeta_n^{(3)}(K^L_{1}{-}K^L_{2}){\cal R}_{n{-}1}(Q)\,.
\end{split}
\end{equation}
One can plug this non-vanishing result into an inelastic heterotic string amplitude, such as the one computed with mass eigenstates in Sect. \ref{StringMemoSect}, and obtain the GW profile. We will not perform this laborious analysis here but we expect the result to be similar, since this is largely determined by the Shapiro-Virasoro dressing of the GR result\footnote{We thank P.~Di~Vecchia for stressing this point.}.


\section{Conclusions and final comments}
\label{ConcSect}

We investigated the $\ap$ stringy corrections to the GW emitted during the merging of two BPS BHs. For this purpose, we used a toy model whereby small BH's are described by vertex operators in heterotic string or coherent state thereof. This allowed to compute the exact amplitude at tree level (sphere).

As expected, we found that the leading order corrections to the GW signal calculated in GR are of the order $(\ap)^{3}$. Although the suppression is cubic in $\ap$, which as such would produce a signal decaying too fast with the distance, using the full tree-level scattering amplitude, we find an imprinting in the GW signals due to the infinite tower of massive string resonances that we dub `string memory'. This string footprint contributes to the falsifiability of this scenario, laying within the sensitivity region of aLIGO/VIRGO and future interferometers. 

The effect of string resonances tends to be partly lost due to loop effects that broaden and shift the poles, providing a {\it lost memory effect} that can be partly regained in the GW signal both in the merging and in the ring-down phase, the latter governed by QNM \cite{Berti:2009kk, Bianchi:2020des} of the stringy version of the BH \cite{Bena:2020see, Bianchi:2020bxa, Bena:2020uup, Bianchi:2020Comp}. Indeed, we found that GWs can carry information on string resonances and that the signal, being polynomially rather than exponentially decaying in (retarded) time, does actually enable for the search of such string memory effect.

Our present work can be extended in several directions. In particular it would be very interesting to study the case of spinning BHs as well as to refine the analysis in the coherent state description.


\section*{Acknowledgements}
\noindent
We would like to thank G.~Di~Russo for collaboration at an early stage of this project. We wish to thank A.~Aldi, D.~Consoli, A.~Grillo, P.~Di~Vecchia, J.~F.~Morales, P.~Pani, G.~Raposo, R.~Russo, A.~Sen, D.~Skliros, G.~Veneziano,  
for discussions and their valuable comments. 

A.M. wishes to acknowledge support by the Shanghai Municipality, through the grant No. KBH1512299, by Fudan University, through the grant No. JJH1512105, and by NSFC, through the grant No. 11875113.

\appendix
\section*{Appendix}

\section{Note on 2-body decay kinematics}
\label{KinematicsApp}

In the soft limit $\omega = 0$, the resulting kinematics is the one of a 2-body decay / production.  

In the rest frame of the produced BH the 10-d momenta (barring R-moving components) read
\be
K_3 = (M_3, 0; {\bf P}_1{+}{\bf P}_2)\,, \qquad \eta_1 K_1 = (E_1, \vec{p}; {\bf P}_1)\,, \qquad \eta_2 K_2 = (E_2,-\vec{p}; {\bf P}_2)\,,
\ee
where
\be
E_1 = {M_3^2+M_1^2-M_2^2 \over 2M_3} \,, \qquad E_2 = {M_3^2+M_2^2-M_1^2 \over 2M_3} \,, \qquad
|\vec{p}| = {\sqrt{{\cal F}(M^2_1,M^2_2,M^2_3)}\over 2M_3} \,,
\ee
with 
${\cal F}(x,y,z)= x^2+y^2+z^2-2xy-2yz-2zx$ the ubiquitous `fake square'.  

Setting $\mu_1=M_1^2/M_3^2$ and $\mu_2=M_1^2/M_2^2$, the physical domain is 
\be
0<\mu_1<1 \quad , \quad 0<\mu_2<1\quad , \quad (\mu_1-\mu_2)^2 - 2(\mu_1+\mu_2) + 1>0
\ee
that represents a triangular region in the first quadrant bounded by the two axis $\mu_2=0$ and $\mu_1=$ and the oblique parabola $(\mu_1-\mu_2)^2 - 2(\mu_1+\mu_2) + 1=0$ (${\cal F}(\mu_1,\mu_2,1)=0$) where $|\vec{p}|=0$. See Fig.~\ref{PhysDomFig}. Along the bisectrix $0<\mu_1=\mu_2<1/4$, $E_1=E_2=M_3/2$ and $|\vec{p}|=M_3\sqrt{1- 4\mu_{1=2}}/2$. In particular at the vertex $\mu_1=\mu_2=1/4$, $E_1=E_2=M_3/2$ and $|\vec{p}|=0$. While at the origin $\mu_1=\mu_2=0$$E_1=E_2=M_3/2=|\vec{p}|$.

For $M_1=M_2=M$, the equality holds ${\cal F}(M^2,M^2,M^2_3)=M_3^2(M_3^2-4M^2)$. 

Notice that the amount of `internal' non-collinearity is constrained by the condition  ${\cal F}(M^2_1,M^2_2,M^2_3)= (M^2_1+M^2_2-M^2_3)^2 - 4M^2_1M^2_2\ge 0$ that, using ${\bf P}_3={\bf P}_1+{\bf P}_2$, $M^2_{1/2}= |{\bf P}_{1/2}|^2$ and $M^2_{3}= |{\bf P}_1+{\bf P}_2|^2 + 2N_3 = M_1^2 + M_2^2 + 2{\bf P}_1+{\bf P}_2 + N_3$ yields 
\be
(2{\bf P}_1{\cdot}{\bf P}_2  + 2N_3)^2 - 4 |{\bf P}_1|^2|{\bf P}_2|^2\ge 0\,,
\ee 
so that 
\be
N_3  \ge |{\bf P}_1||{\bf P}_2| - {\bf P}_1{\cdot}{\bf P}_2 =  |{\bf P}_1||{\bf P}_2|(1- \cos\gamma_{1,2})\,,
\ee
or
\be
 1\ge \cos\gamma_{1,2} \ge 1 - {N_3\over M_1M_2}\,.
\ee

Obviously for $N_3=0$ also the third BH is BPS, and ${\bf P}_1$, ${\bf P}_2$, and thus ${\bf P}_3={\bf P}_1+{\bf P}_2$, must be collinear, very much as the three space-time momenta. For $N_3 = 2 M_1M_2$, $\cos\gamma_{1,2}=-1$, and for larger $N_3>2 M_1M_2$ the condition on $\cos\gamma_{1,2}$ is always satisfied.  

In this frame $np_3=-M_3=-|\ell_3|$, while the other two scalar invariants read 
\be
np_1= E_1 - |\vec{p}|\cos\theta = {M_3^2{+}M_1^2{-}M_2^2 \over 2M_3} - \cos\theta {\sqrt{{\cal F}(M^2_1,M^2_2,M^2_3)}\over 2M_3} \,,
\ee
\be
np_2= E_2 + |\vec{p}|\cos\theta = {M_3^2{-}M_1^2{+}M_2^2 \over 2M_3} + \cos\theta {\sqrt{{\cal F}(M^2_1,M^2_2,M^2_3)}\over 2M_3}\,.
\ee
Their ratios are given by
\be 
\lambda_{1/2,3} = {\ell_{1/2}\over\ell_3} = {M_3^2{\pm}M_1^2{\mp}M_2^2 \over 2M^2_3} \mp \cos\theta {\sqrt{{\cal F}(M^2_1,M^2_2,M^2_3)}\over 2M^2_3}\,,
\ee
with
 \be
 \lambda_{1,3}+\lambda_{2,3} = 1\,.
\ee

The expressions simplify for $M_1=M_2=M$, since $E_1=E_2=M_3/2$ and $|\vec{p}| = \sqrt{{M_3^2\over 4}-M^2}$, so that (for $\ap =2$) one finds 
\be
\lambda_{1/2,3} = {1\over 2}\left[1\mp \cos\theta \sqrt{1-4{M^2\over M_3^2}}\right] = 
{1\over 2}\left[1\mp \cos\theta \sqrt{ 1- {4|{\bf P}_{1/2}|^2\over |{\bf P}_3|^2 + 2N_3}}\right]\,.
\ee

The phase space gets modified by the emission of the massless graviton when $\omega \neq 0$. In the CoM frame of the system
\be
p_1'=(E_1', \vec{p})\quad , \quad p_2'=(E_2', -\vec{p})\quad , \quad p_3'=(E_3', -\vec{k}) \quad , \quad k = (\omega, \vec{k})
\ee
with $\omega=|\vec{k}|$, $E_3' = \sqrt{M_3^2 + \omega^2}$, $E_1'=\sqrt{M_1^2 + p^2}$, and $E_2'=\sqrt{M_2^2 + p^2}$. Moreover since
\be
E_3' + \omega = E_1'+E_2' 
\ee
the equations look identical to the ones for $\omega=0$ after the replacement 
$M_3\rightarrow \widetilde{M}_3 = E_3' + \omega = \omega + \sqrt{M_3^2 + \omega^2}$ so that the solution is 
\be
E_1' = {\widetilde{M}_3^2+M_1^2-M_2^2 \over 2\widetilde{M}_3} \,, \qquad E_2' = {\widetilde{M}_3^2+M_2^2-M_1^2 \over 2\widetilde{M}_3} \,, \qquad
|\vec{p}| = {\sqrt{{\cal F}(M^2_1,M^2_2,\widetilde{M}^2_3)}\over 2\widetilde{M}_3} \,,
\ee
setting
\be
\tilde\mu_1 = {M_1^2\over \widetilde{M}_3^2} = 
{M_1^2\over (\omega + \sqrt{M_3^2 + \omega^2})^2} \quad , \quad 
\tilde\mu_2 = {M_2^2\over \widetilde{M}_3^2} = 
{M_2^2\over (\omega + \sqrt{M_3^2 + \omega^2})^2}
\ee
one finds the same kinematical domain as before, where however the denominator $\omega + \sqrt{M_3^2 + \omega^2}\ge M_3$, so much so that $M_1$ and $M_2$ can be larger than $M_3$. 
In the main text we denote 
$\tilde\mu_1$ and $\tilde\mu_2$ by $\mu_1$ and $\mu_2$ to avoid too cumbersome notations. 

\section{Vanishing of 3-point amplitude for BPS states}
\label{no3BPSApp}


One of the ingredients of the 4-point scattering amplitude that we consider is the 3-point amplitude of the massive states. We would like to show that this is zero in the case of 3 BPS states.

Indeed the L-moving part is 
\begin{equation}
{\cal V}^{(int),L}_{3{-}YM}(\{\zeta_L\},\{p_L\})=\sqrt{{\alpha' \over 2}}\Big(\zeta_{3,L}{\cdot}\zeta_{4,L} \,\zeta_{2,L}{\cdot}p_{3,L}{+}\zeta_{2,L}{\cdot}\zeta_{3,L} \,\zeta_{4,L}{\cdot}p_{2,L}{+} \zeta_{2,L}{\cdot}\zeta_{4,L} \,\zeta_{3,L}{\cdot}p_{4,L}\Big)\,,
\end{equation}
while the second is the three-point like coupling of higher spin states\footnote{The most general representation of this coupling can be constructed by coherent states, representing the full superposition of all the possible string states.} with only internal momenta and polarisations [see BF].
Quite remarkably,  
\begin{equation}
{\cal V}^{(int),L}_{3{-}YM}(\{\zeta_L\},\{p_L\})= 0 \, .
\end{equation}
This is a consequence of the BPS nature of the 3 BH states that requires collinearity of their full 10-dim left `massless' momenta \ie $K_a^2 = p_a^2 + {{\bf{P}}}_a^2 = 0$, since $M_a^2={{\bf{P}}}_a^2=-p_a^2$. As a consequence 
$K_1+K_2= -K_3$ implies $K_1K_2=0$ and cyclic. But two light-like momenta in any dimension are `orthogonal' only if they are `parallel' \ie collinear. Indeed (assuming both `in' or both `out')
\be
K_1K_2 = - E_1E_2 +\vec{K}_1{\cdot}\vec{K}_2 = - |\vec{K}_1||\vec{K}_2| + |\vec{K}_1||\vec{K}_2|\cos\theta_{12} = |\vec{K}_1||\vec{K}_2|(\cos\theta_{12} -1) = 0 \ee 
so that $\theta_{12}=0$ as expected. 

Whichever left `polarisations' $A_a$ (dropping the $L$ subscript) one chooses for the 3 BPS BHs, one gets 
\begin{equation}
{\cal V}^{YM}_{3}(\{A\},\{K\})=
A_1{\cdot}A_2 A_{3}{\cdot}K_{1} + A_2{\cdot}A_3 A_{1}{\cdot}K_{2} + A_3{\cdot}A_1 A_{2}{\cdot}K_{3} = 0\,, 
\end{equation}
since $A_aK_b = A_a K_a \rho_{b,a} = 0$ due to the BRST condition $A_aK_a = 0$ and collinearity, namely  
$K_b = \rho_{b,a} K_a $ with $\rho_{b,a}$ some constants such that 
$\sum_{b\neq a} \rho_{b,a} = -1$.

\section{Generalized Hypergeometric Functions and Meijer G-function}
\label{GenHypFunApp}

For rational values of $\lambda_{a,b}$ integration over $\omega$ produces combinations of generalised hypergeometric functions 
\be
\label{seriesFgen}
{}_pF_q(a_1,..., a_p; b_1,...,b_q;z) = {\prod_{j=1}^q \Gamma(b_j)\over 
\prod_{i=1}^p \Gamma(a_i)} \sum_{n=0}^\infty{\prod_{i=1}^p \Gamma(n{+}a_i) \over \prod_{j=1}^q \Gamma(n{+}b_j)} {z^n\over n!}
\ee
For generic values of $a_i$ and $b_j$, the series expansion is well defined: for $p\le q$ for all $z$, for $p=q{+}1$ for $|z|<1$ and for $p\ge q{+}2$ only for $z=0$. 

An analytic continuation valid for $p\ge q{+}2$ for all complex $z\neq 0$ and  for $p=q{+}1$ for $|z| > 1$ can be given in terms of Meijer G-functions \cite{KSST} defined by the contour integral of the Mellin-Barnes type
\be 
G_{p,q}^{m,n}\left(^{a_1,..., a_p} _{b_1,...,b_q} \vert z\right) = \int_{\cal L} {z^s ds\over 2\pi i}{\prod_{j=1}^m \Gamma(b_j{-}s) \prod_{i=1}^n\Gamma(1{-}a_i{+}s)\over 
\prod_{j'=m{+}1}^q \Gamma(1{-}b_{j'}{+}s) \prod_{i'=n{+}1}^p\Gamma(a_{i'}{+}s)} 
\ee
Notice the somewhat confusing historical notation for which the first top index $m$ is related to the second bottom index $q$ by $0\le m\le q$, while  the second top index $n$ is related to the first bottom index $p$ by $0\le n\le p$. It is understood that $z\neq 0$ and that $a_i{-}b_j$ is not an integer for $i=1,..n$ and $j=1, ...m$ (to avoid double poles in $s$). By judicious choice of the contour ${\cal L}$ one can make sense of the integral for any $p,q$ and $z$ with $m,n$ in the allowed ranges\footnote{According to \cite{KSST} the Meijer G-function is a special case of the H-function. For a recent application in string amplitudes see e.g. \cite{Bianchi:2020cfc}.}. 

By analytic continuation, under the conditions of Theorem 3.1 in \cite{KSST}, the generalized hypergeometric function ${}_pF_q$  admits a representation as a G-function of the form 
\be
\label{GrepofF} 
{}_pF_q(a_1,..., a_p; b_1,...,b_q;z) = 
{\prod_{j=1}^q \Gamma(b_j)\over 
\prod_{i=1}^p \Gamma(a_i)} G_{p,q{+}1}^{1,p}\left(^{1{-}a_1,..., 1{-}a_p} _{0,1{-}b_1,...,1{-}b_q} \vert {-}z\right)
\ee
According to Remark 3.5 in \cite{KSST}, this representation can be considered as an
extension of the generalized hypergeometric function defined by the series in (\ref{seriesFgen}) from the usual range
of the parameters $p, q$ and of the complex variable $z$. The representation (\ref{GrepofF}) for the Gauss hypergeometric function $_2F_1$ is well known. This allows to define ${}_2F_1(a,b;c;z)$ for $|z|>1$ including the cases $a-b\in {\bf Z}$ that gives rise to $\log(-z)$
\be
{}_2F_1(a,a;c;z) = {(-z)^{-a} \Gamma(c) \over \Gamma(a)\Gamma(c{-}a)}\sum_{n=0}^\infty {(a)_n (1{-}c{+}a)_n\over n!^2 {} z^n}  
[\log(-z)+2\psi(n{+}1)-\psi(a{+}n)- \psi(c{-}a{-}n)]
\ee

\section{Resummation for `Rational' kinematics}
\label{RationalKinApp}
Starting from the sum
\begin{equation}
\sum_{n=1}^\infty 
{(-)^{n} \over n!^2} {\Gamma(1{-}n\lambda_{1,3})\Gamma(1{-}n_3\lambda_{2,3}) \over \Gamma(1{+}n\lambda_{1,3})\Gamma(1{+}n_3\lambda_{2,3})}  \,x^n = \sum_{n=1}^{\infty} {(-)^n\over n!^2}\, (1+n\lambda_{1,3})_n (1+n\lambda_{2,3})_n \,x^n
\end{equation}
with $(1{+}n\lambda)_n= \Gamma(1{+}n{+}n\lambda)/ \Gamma(1{+}n\lambda)$ a generalised Pochhammer symbol and performing some simple manipulations one can arrive at the representation 
 \begin{equation}\label{MasterSum}
 {1+\lambda \over \lambda}\,\sum_{n=1}^{\infty} \left( {(n\lambda)_n \over n!} \right)^2 x^n\,,
 \end{equation} 
 with $\lambda=\lambda_{2,3}$, where the relation (in the physical regime)  $\lambda_{1,3}+\lambda_{2,3}+1=0$ was used. The parameter $\lambda$, as explained in Appendix A, has a continuous range of variability between $-1/2$ and $1/2$. Thus one can study the sum for the discrete values of $\lambda$ using the following parametrization
\begin{equation}\label{lambParam}
\lambda=\pm {1\over \ell}\,,\qquad \ell >1\,,\qquad \ell\in\mathbb{N}\,,
\end{equation}
and without loss of generality one can consider only the positive values. Nonetheless, it is straightforward verify that the following analysis holds also for negative values of $\lambda$. 
 Following this parametrization the starting is 
 \begin{equation}
 \sum_{n=1}^{\infty} \left( (n/\ell )_n \over n! \right)^2 x^n\,,
 \end{equation}
 one can fragment the range of the sum in $\ell-1$ ranges using the replacements $n \rightarrow n\ell-j$ as follows
 \begin{equation}
 \sum_{n=1}^{\infty} f\left({n\over\ell}\right)= \sum_{j=0}^{\ell-1} \sum_{n=1}^{\infty} f\left({n\ell{-}j\over \ell}\right)\,,
 \end{equation}
 yields, in the specific case, 
 \begin{equation}
 \sum_{j=0}^{\ell-1}\sum_{n=1}^{\infty} \left( \Gamma\big[(\ell{+}1)(n-j/\ell)\big] \over \Gamma(n-j/\ell) \,\Gamma\big[\ell\big(n- (j-1)/\ell \big) \big] \right)^2 x^{\ell n -j + \ell}\,.
 \end{equation}
Finally, using the Gauss's multiplication formula
\begin{equation}
 \prod_{j=0}^{b-1} \Gamma\left(h+{j\over b}\right)= \sqrt{b\,2^{b{-}1} \pi^{b{-}1}}\,\Gamma(b\, h)\, b^{-b\,h}\,,
 \end{equation} 
 after some manipulations the sum can be represented as
 \begin{equation}
 \begin{split}
{1+\ell^{-1} \over \ell^{-1}} &\sum_{j=0}^{\ell-1} { \ell^{\,2j{-}2\ell{-}1}  \over 2\pi}   (\ell+1)^{2(\ell{+}1)(1{-}j/\ell)-1}  \prod_{t=1}^{\ell} \left(     { \Gamma\big[1{-}j/\ell {+} t/(\ell{+}1)\big]  \over \Gamma\big[ 1- (j-1)/\ell + (t{-}1)/\ell \big]  }\right)^2 x^{\ell-j} \\
&\sum_{n=0}^{\infty}{ (1)_n \over n!} \prod_{t=1}^{\ell} \left(  {\big( 1+t/(\ell{+}1) - j/\ell\big)_n \over \big( 1+(t{-}1)/\ell - (j{-}1)/\ell  \big)_n}   \right)^2 \left( {(\ell{+}1)\over \ell^{\,2 \ell}}^{2(\ell{+}1)}  \,x^\ell \right)^n\,,
\end{split}
 \end{equation}
which, in a more compact way, becomes
\begin{equation}
\begin{split}
 & \sum_{n=1}^{\infty} \left( (n/\ell )_n \over n! \right)^2 x^n=\\
 &\ \ \ \ \sum_{j=0}^{\ell-1} { \ell^{\,2j{-}2\ell{-}1}  \over 2\pi}   (\ell+1)^{2(\ell{+}1)(1{-}j/\ell)-1}  \prod_{t=1}^{\ell} \left(     { \Gamma\big[1{-}j/\ell {+} t/(\ell{+}1)\big]  \over \Gamma\big[ 1- (j-1)/\ell + (t{-}1)/\ell \big]   }\right)^2 x^{\ell-j} \\
 &\ \ \ \ \pFq{2\ell{+}1}{2\ell}{1,\{1{-}j/\ell{+}t/(\ell{+}1)\}_{t=1}^{\ell},\{1{-}j/\ell{+}t/(\ell{+}1)\}_{t=1}^{\ell}}{\{1{-}(j{-}1)/\ell{+}(t{-}1)/\ell\}_{t=1}^{\ell},\{1{-}(j{-}1)/\ell{+}(t{-}1)/\ell\}_{t=1}^{\ell}}{{(\ell{+}1)\over \ell^{2\ell}}^{2(\ell{+}1)} x^{\ell}} \,.
 \end{split}
 \end{equation}
 This is the final expression, which holds for any positive values of the parameter $\lambda=1/\ell$,  $\forall\, \ell >1$ with $\ell \in \mathbb{N}$.
 
 Pretty much similar to the previous manipulations, one can study the sum for integer and positive values of the kinematical parameter $\lambda$ arriving to the following result
 \begin{equation}
 \begin{split}
  &\sum_{n=1}^{\infty} \left( (n \lambda )_n \over n! \right)^2 x^n=\\
  &{\prod_{f=0}^{\lambda} \Gamma\big[ f/(\lambda{+}1) \big]^2 \over 2\pi \prod_{j=0}^{\lambda{-}1} \Gamma(j/\lambda)^2} \Bigg\{ -1 +        
   \pFq{2\lambda}{2\lambda{-}1}{  \{ f/(\lambda{+}1) \}_{f=1}^{\lambda},\{ f/(\lambda{+}1) \}_{f=1}^{\lambda}}{\{j/\lambda\}_{j=1}^{\lambda{-}1},\{j/\lambda\}_{j=1}^{\lambda{-}1}}{{(\lambda{+}1)\over \lambda^{2\lambda}}^{2(\lambda{+}1)} x}  \Bigg\}
  \end{split}
 \end{equation}

\end{document}